\documentclass[11pt,a4paper]{article}
\usepackage[T1]{fontenc}
\usepackage{lmodern}
\usepackage[margin=27mm]{geometry}
\usepackage{amsmath,amssymb,amsthm,mathtools,microtype}
\usepackage{mathrsfs}
\usepackage{graphicx}
\usepackage{aliascnt}
\usepackage[hidelinks]{hyperref}
\usepackage[nameinlink,noabbrev]{cleveref}
\numberwithin{equation}{section}
\newtheorem{theorem}{Theorem}[section]
\newaliascnt{lemma}{theorem}
\newtheorem{lemma}[lemma]{Lemma}
\aliascntresetthe{lemma}
\newaliascnt{proposition}{theorem}
\newtheorem{proposition}[proposition]{Proposition}
\aliascntresetthe{proposition}
\newaliascnt{corollary}{theorem}
\newtheorem{corollary}[corollary]{Corollary}
\aliascntresetthe{corollary}
\newcommand{\C}{\mathbb C}
\newcommand{\tr}{\operatorname{tr}}
\newcommand{\diag}{\operatorname{diag}}
\newcommand{\Id}{\mathrm{Id}}
\newcommand{\trans}{\mathsf T}
\newcommand{\BA}{Baker-Akhiezer}

\title{Baker-Akhiezer realization and reconstruction for three-wave
interactions in canonical multiwave Hamiltonians}
\author{Graham Hesketh\\
Independent Researcher, UK\\
\href{mailto:gdh1e10@gmail.com}{\texttt{gdh1e10@gmail.com}}}
\date{}

\begin{document}
\maketitle

\begin{abstract}
We revisit the canonical multiwave Hamiltonians with three-wave
interactions studied by Menyuk, Chen and Lee, and by Jiang,
Rauch-Wojciechowski and Bullough. With compatible linear phase terms, we construct
their general solution on the generic spectral locus from multipoint
Baker-Akhiezer functions. Conversely, we give a finite interpolation
criterion and completion procedure that realize prescribed
Baker-Akhiezer evaluations on arbitrary smooth algebraic curves in
the same Hamiltonian system. We identify the prescribed curve with
the normalized spectral curve and show that commuting polynomial
Hamiltonians span every Abel direction on each birational spectral
level, including the accompanying mode phases. Together, these
results establish a correspondence between Baker-Akhiezer data and
canonical multiwave dynamics, supporting our view of this system as
a fundamental canonical Hamiltonian setting for Baker-Akhiezer
functions. We make explicit its connections with other established
integrable systems and validate hyperelliptic, trigonal and
multipuncture examples numerically.
\end{abstract}

\clearpage
\begingroup
\makeatletter
\renewcommand{\@pnumwidth}{2.5em}
\renewcommand{\@tocrmarg}{3.5em}
\makeatother
\tableofcontents
\endgroup
\clearpage

\section{Introduction}

The \BA\ (BA) function is a central object in the algebro-geometric
integration of nonlinear equations. Krichever's construction, building
on earlier work of Baker and Akhiezer, associates such functions with
an algebraic curve, marked points and prescribed divisor data
\cite{Krichever1977,KricheverShiota2011}. They solve the auxiliary
linear problems underlying the Kadomtsev--Petviashvili hierarchy and,
with the appropriate spectral data, finite-gap Korteweg--de Vries
solutions. The same functions enter the integration of
finite-dimensional systems, including Neumann and Garnier systems;
Mumford's polynomial coordinates describe the associated moving
divisors \cite{Kostov1999,Eilbeck2000,Mumford1984,Vanhaecke2001}.

These constructions connect functions and dynamics in two directions.
A Lax representation associates spectral curves and eigenvector data
with a Hamiltonian system. Conversely, prescribed curve and divisor
data produce BA functions from which solutions can be constructed.
Adler--Kostant--Symes theory explains the commuting Hamiltonians in
many matrix realizations, while spectral line bundles describe their
algebraic integration \cite{AdlerVanMoerbeke2018,Beauville1990}.
Together these methods provide a broad framework, within which
particular systems select particular curves, marked points and
functional realizations.

Finite-dimensional multiwave equations provide a direct setting for
this relationship between functions and dynamics. In genus one,
elliptic sigma ratios describe individual modes of integrable
three-wave and four-wave interactions
\cite{Hesketh2015,HeskethFWM2026}; their shifts and exponential factors
encode both powers and phases. Multiwave systems also have a history
of Hamiltonian and spectral constructions that extends beyond these
elliptic examples.

Three-wave interaction equations arise in models of ocean waves,
plasma waves and quadratic nonlinear optical media. Restricted
multiple three-wave systems couple several triads through a common
mode \cite{KostovTsiganov2008}; continuous-wave optical interactions
also lead to finite-dimensional mode equations \cite{Hesketh2015}.
Menyuk, Chen and Lee investigated integrable couplings through
Hamiltonian, Lax and Painlev\'e analyses
\cite{MenyukChenLee1982,MenyukChenLee1983Integrable,
MenyukChenLee1983Painleve}. They obtained a complete commuting
integral set for the equal-coupling shared-mode family, constructed
broader wave systems, and integrated the equal-mismatch family
using elliptic functions and phase quadratures.
Rauch-Wojciechowski, Jiang and Bullough gave an alternative matrix
Lax representation and constructed a broader class of multiwave
ordinary differential equations \cite{WJB1986}. Jiang and
Rauch-Wojciechowski then developed this approach in a graded Lie
algebra framework, constructing canonical cubic systems with
rectangular mode arrays and studying the completeness of their
commuting integrals \cite{JRW1991}.
Kostov and Tsiganov studied the restricted shared-mode family through
a two-component Dirac problem. They constructed its hyperelliptic
spectral curve and finite-gap eigenfunction representation, together
with a classical $r$-matrix and bi-Hamiltonian structure; the
single-triad case was evaluated explicitly in Weierstrass functions,
including its phase \cite{KostovTsiganov2008}.
Geng, Guan and Du constructed finite-dimensional Lie--Poisson
systems from a rational $\mathfrak{gl}_3$ Gaudin matrix, with separated
variables and Jacobi inversion on non-hyperelliptic spectral curves
\cite[Sections~3--5]{GengGuanDu2022}.

Adams, Harnad and Previato developed a moment-map
construction of isospectral Hamiltonian systems. Adams, Harnad and
Hurtubise (AHH) integrated both
the matrix flows and their canonical lifts in generalized BA functions
\cite{AdamsHarnadPreviato1988,AdamsHarnadHurtubise1990II}.
This work complements the matrix Lax approach to multiwave equations
by constructing canonical realizations through moment maps and
relating their evolution to spectral curves and BA functions.

\subsection{Our contribution}

Our motivation comes from nonlinear optics, where elegant canonical
Hamiltonians describe continuous-wave and quasi-continuous-wave
interactions, with individual modes expressed by genus-one
Baker-Akhiezer functions \cite{Hesketh2015,HeskethFWM2026}.
Familiar integrable systems, including classical tops, can in turn
be obtained as projections of such canonical mode Hamiltonians.
In the higher-genus constructions motivating this work,
Hamiltonians with a single mixing product likewise admit BA solutions,
with conservation laws associated with \^Onishi's determinant
identities \cite{HeskethHyperelliptic2026}, but the Abelian argument $z$,
as well as the elementary shifts, is restricted to an Abel lift of
a point on the curve. Composite modes carry sums of elementary
shifts. This suggests a further question: what canonical Hamiltonian
dynamics accommodates these functions when $z$ varies freely on the
Jacobian?

Pursuing this question leads to the cubic multiwave constructions
of Menyuk, Chen and Lee and of Jiang and Rauch-Wojciechowski.
The moment-map framework of Adams, Harnad and Hurtubise provides
a related quartic canonical realization with generalized BA solutions.
Connections with other BA-solvable systems are therefore to be
expected. We make these connections explicit through a correspondence
with the quartic moment-map systems of Adams, Harnad and Hurtubise
and a constrained-KP operator realization. These correspondences also
highlight the simplicity of the canonical multiwave formulation:
its Hamiltonian is cubic, its canonical brackets are constant, and
its nonlinear evolution terms contain products of only two modes,
the lowest possible polynomial nonlinearity. The dependent variables
and Poisson structures differ between these descriptions. We therefore
advocate the cubic system as a fundamental canonical setting for BA
functions.

The BA correspondence also works in both directions. Prescribed BA
evaluations on arbitrary smooth curves can
be accommodated by compatible canonical mode systems, while generic
canonical initial modes determine their spectral BA solutions. The
commuting Hamiltonians also generate the Abel directions, with
explicit formulas for the accompanying mode phases. The curve changes
the compatible parameters and initial data, while the cubic interaction
law retains its form. When the prescribed genus is below the regular
spectral genus, the realization occupies the special family described
in \cref{prop:BA-prescribed-spectral-curve}. Taken together, these results support viewing
this classical multiwave system as the canonical Hamiltonian setting
for Baker-Akhiezer dynamics and motivate its further study from that
perspective.

Here $L,M,N$ are positive integers specifying three
mode matrices $\boldsymbol u\in\C^{L\times M}$,
$\boldsymbol v\in\C^{M\times N}$ and
$\boldsymbol w\in\C^{N\times L}$.
Each entry has an independent Hamiltonian partner, giving
$LM+MN+NL$ canonical mode pairs; when $N=1$, the latter two
matrices are vectors.
The paper develops the correspondence in the following stages.

\noindent\textbf{Hamiltonian and spectral foundations
(\cref{sec:canonical}).}
We begin with the canonical rectangular multiwave formulation of
Jiang and Rauch-Wojciechowski and its earlier wave-system setting.
Established Lax, AKS and spectral-line-bundle methods provide the
framework. Our treatment supplies the following ingredients needed
by the BA construction:
\begin{itemize}
\item The symmetric phase formulation and exact source dictionaries,
with the independent rates determined from prescribed BA data in
\cref{thm:general-BA-realization}.
Such rates are compatible with the earlier multiwave framework; their
curve-point compatibility is central here.
\item A uniform complete Liouville count for every positive $L,M,N$
and every fixed phase profile, including repeated phases and their
additional moment integrals
(\cref{thm:canonical-independence}). For distinct phases,
Bolsinov's established criterion also gives completeness
\cite{Bolsinov1992}.
\item A concrete characterization of regular spectral curves at fixed
matrix sizes, their genus and covering degrees, together with the
special levels needed for lower-genus realizations
(\cref{thm:geometry-classification,prop:geometry-genus,prop:geometry-hyperelliptic}).
\end{itemize}
The fixed-size classification and the arbitrary-curve construction
answer complementary questions: the first describes the curves at
fixed matrix sizes, while the second chooses compatible matrix sizes
and a spectral level for a prescribed curve.

\noindent\textbf{The BA correspondence and commuting flows
(\cref{sec:BA-correspondence}).}
This section contains the principal results. We use ordinary and dual
BA functions, their two-point kernel and meromorphic interpolation to
establish three parts of the correspondence:
\begin{itemize}
\item \textbf{Main theorem, part I: Canonical BA realization.}
\Cref{thm:general-BA-realization} gives the residue and phase criterion,
including necessity for the normalized ansatz on a nonstationary open
Jacobian family. The test consists of two $g$-dimensional linear
systems for arbitrary $N$. \Cref{cor:BA-finite-list} completes
prescribed finite evaluation lists, and kernel convolution supplies
the functional identities behind Hamilton's equations.
\item \textbf{Main theorem, part II: Generic BA reconstruction.}
The explicit curve map in \cref{prop:BA-prescribed-spectral-curve}
identifies birational realizations with the normalized Lax curve;
for $N=1$ birationality is automatic.
\Cref{thm:BA-spectral-converse} then reconstructs the BA bases, kernel
and all modes from generic canonical initial data. Repeated phases
use constant block bases and active reduction, after which original
coordinates can be linear combinations of scalar BA evaluations.
\item \textbf{Main theorem, part III: The spectral BA hierarchy.}
\Cref{thm:BA-higher-flows} computes the puncture times and constant
endpoint phase rates of the commuting polynomial Hamiltonians.
Their flows span the Jacobian directions on each fixed birational
spectral level; the required combinations depend on its curve data.
This identifies the BA translations and mode gauges within the
Hamiltonian hierarchy.
\end{itemize}
For $N=1$, one ordinary/dual BA pair supplies two elementary vectors
and their pair-mode matrix. For $N>1$, the additional index labels
puncture-normalized sections sharing one curve, divisor and kernel;
it does not describe independent copies of the one-point system.

\noindent\textbf{Correspondences and explicit specializations
(\cref{sec:correspondences-specializations}).}
We make the general construction concrete and relate it to other
functional and Hamiltonian descriptions:
\begin{itemize}
\item An explicit all-size dictionary connects a quartic canonical
AHH lift to the cubic equations. The two matrix brackets and BA
convolution explain the shared evolution and interaction factorization.
\item The weighted BA identities give a matrix constrained-KP
interpretation for arbitrary $N$, including the normalized source
evolution and its levelwise canonical generators. The one-point
specialization gives the elementary-mode notation. For hyperelliptic curves,
polynomial interpolation gives explicit sigma-mode weights and
phase coefficients, including half-period al functions and elliptic
realizations with arbitrarily many elementary modes.
\item An explicit mode-power map identifies the hyperelliptic Mumford
subsystem and reconstructs the modes from its divisor data, explaining
how the canonical description retains phase information alongside
the established projected dynamics.
\end{itemize}

\noindent\textbf{Numerical validation
(\cref{sec:numerical-validation}).}
The checks test both directions and the commuting evolution:
\begin{itemize}
\item Hyperelliptic and trigonal sigma solutions are compared with
independent integration of Hamilton's equations.
\item Multipuncture examples in genera one and two test the full
arrays, spectral identities and higher-flow phases; the genus-two
case follows two independent Abel directions from the same initial modes.
\item A further genus-two example starts from canonical initial modes
and reconstructs their curve and BA solution, testing the inverse
construction separately from prescribed-curve examples.
\end{itemize}
The discussion then brings these results together and identifies
further questions concerning the hierarchy, reductions and BA
interaction identities.

\section{The Hamiltonian system and its spectral geometry}
\label{sec:canonical}

We establish the Hamiltonian and spectral framework needed for the BA
correspondence. The phase parameters specify the rates to be matched,
the commuting integrals provide Liouville integrability, and the pencil
identifies the curve carried by a solution.

We use the canonical cubic multiwave family developed in the earlier
wave and matrix constructions
\cite{MenyukChenLee1983Integrable,WJB1986,JRW1991}.
Independent constant diagonal phases accommodate the rates determined
by prescribed BA data in \cref{thm:general-BA-realization}; power
charges redistribute them into a symmetric convention. We first give
the full three-matrix system, which supplies the pencil and commuting
Hamiltonians used by its BA realization for all three group sizes. The coordinate
comparisons with the earlier systems are in
Appendices~\ref{sec:MCL-dictionary} and~\ref{sec:JRW-dictionary}.

\subsection{Canonical modes and phase parameters}
\label{sec:canonical-phase-setup}

We fix the modes, brackets and phase conventions so that the BA
formulas in \cref{sec:BA-correspondence} can be compared directly with
Hamilton's equations.

\subsubsection{Modes, brackets and Hamilton equations}
\label{sec:canonical-modes}

The mode matrices are $\boldsymbol u\in\C^{L\times M}$,
$\boldsymbol v\in\C^{M\times N}$ and
$\boldsymbol w\in\C^{N\times L}$, for positive integers $L,M,N$.
Each entry has an independent barred Hamiltonian partner.
Their common cubic interaction has the matrix form
\begin{equation}\label{eq:intro-mixing}
 H_{\mathrm{mix}}=\sqrt2\,\tr\!\left(
 \boldsymbol u\boldsymbol v\boldsymbol w
 +\bar{\boldsymbol u}\bar{\boldsymbol v}\bar{\boldsymbol w}\right).
\end{equation}
Each term joins one mode from each matrix; the constants in
\cref{eq:canonical-H} supply its linear phase rates.

There are $p=LM+MN+NL$ canonical mode pairs. Writing $i$ for the
imaginary unit, their nonzero basic brackets are
\begin{equation}\label{eq:canonical-brackets}
\begin{aligned}
 \{u_{\ell m},\bar{u}_{\hat\ell\hat m}\}&=-i\delta_{\ell\hat\ell}\delta_{m\hat m},
 &&1\leq\ell,\hat\ell\leq L,\quad1\leq m,\hat m\leq M,\\
 \{v_{mn},\bar{v}_{\hat m\hat n}\}&=-i\delta_{m\hat m}\delta_{n\hat n},
 &&1\leq m,\hat m\leq M,\quad1\leq n,\hat n\leq N,\\
 \{w_{n\ell},\bar{w}_{\hat n\hat\ell}\}&=-i\delta_{n\hat n}\delta_{\ell\hat\ell},
 &&1\leq n,\hat n\leq N,\quad1\leq\ell,\hat\ell\leq L.
\end{aligned}
\end{equation}
All other basic brackets vanish, apart from antisymmetry. Bars label
independent Hamiltonian partners, with the same array dimensions as the
unbarred modes. Stars mean actual complex conjugation; $\trans$ means
ordinary transpose, and $\dagger$ means actual conjugate transpose.
A dot denotes differentiation with respect to time $t$, and
$\dot h=\{h,H\}$. Thus
\[
 i\dot u=\frac{\partial H}{\partial\bar u},\qquad
 i\dot{\bar u}=-\frac{\partial H}{\partial u}.
\]
This is the usual complex canonical convention: writing
$u=(q+ip)/\sqrt2$ and $\bar u=(q-ip)/\sqrt2$ gives $\{q,p\}=1$.
The variables $q,p$ are complex unless a real form is imposed.

For real phase coefficients and real coupling, the real forms
$\bar u=\epsilon_u u^*$, $\bar v=\epsilon_v v^*$ and
$\bar w=\epsilon_w w^*$, with $\epsilon_u,\epsilon_v,\epsilon_w\in\{1,-1\}$,
are invariant precisely when $\epsilon_u\epsilon_v\epsilon_w=1$.
The all-positive choice is the explosive three-wave form; the three
choices with two negative signs are decay forms and admit definite
combinations of the power charges. The Hamiltonian is real on each
form, with $i\dot u=\epsilon_u\partial H/\partial u^*$ and the
corresponding equations for $v,w$. The general complex system retains
independent partners; Liouville integrability is understood in the
complex Hamiltonian sense.

Let $b_\ell^{L}$, $b_m^{M}$ and $b_n^{N}$ be constants, with
$1\leq\ell\leq L$, $1\leq m\leq M$, $1\leq n\leq N$.
The superscripts distinguish the three groups of phase coefficients.

\begin{samepage}
The Hamiltonian is
\begin{align}\label{eq:canonical-H}
H={}&\sum_{\ell=1}^{L}\sum_{m=1}^{M}
 (b_\ell^{L}+b_m^{M})u_{\ell m}\bar{u}_{\ell m}
 +\sum_{m=1}^{M}\sum_{n=1}^{N}
 (b_m^{M}+b_n^{N})v_{mn}\bar{v}_{mn}\nonumber\\
&+\sum_{n=1}^{N}\sum_{\ell=1}^{L}
 (b_n^{N}+b_\ell^{L})w_{n\ell}\bar{w}_{n\ell}\nonumber\\
&+\sqrt2\sum_{\ell=1}^{L}\sum_{m=1}^{M}\sum_{n=1}^{N}
 \left(u_{\ell m}v_{mn}w_{n\ell}
       +\bar{u}_{\ell m}\bar{v}_{mn}\bar{w}_{n\ell}\right).
\end{align}
\end{samepage}
Each mode's linear phase coefficient is the sum of the constants at its two
endpoint groups. The Hamiltonian is invariant in form under cyclic
permutation of the groups and their mode matrices.
The three mode powers are
\begin{equation}\label{eq:canonical-powers}
 I^u_{\ell m}=u_{\ell m}\bar{u}_{\ell m},\qquad
 I^v_{mn}=v_{mn}\bar{v}_{mn},\qquad
 I^w_{n\ell}=w_{n\ell}\bar{w}_{n\ell},
 \quad1\leq\ell\leq L,\ 1\leq m\leq M,\ 1\leq n\leq N.
\end{equation}
The superscripts $u,v,w$ label the associated mode families and are
not exponents. These powers become squared moduli when the partners are actual complex
conjugates.

The Hamilton equations are, for
$1\leq\ell\leq L$, $1\leq m\leq M$, $1\leq n\leq N$,
\begin{equation}\label{eq:canonical-rates}
\begin{aligned}
 i\dot u_{\ell m}&=(b_\ell^L+b_m^M)u_{\ell m}
            +\sqrt2\sum_{k=1}^{N}\bar{v}_{mk}\bar{w}_{k\ell},\\
 i\dot{\bar{u}}_{\ell m}&=-(b_\ell^L+b_m^M)\bar{u}_{\ell m}
            -\sqrt2\sum_{k=1}^{N}v_{mk}w_{k\ell},\\
 i\dot v_{mn}&=(b_m^{M}+b_n^N)v_{mn}
       +\sqrt2\sum_{\ell=1}^{L}\bar{u}_{\ell m}\bar{w}_{n\ell},\\
 i\dot{\bar{v}}_{mn}&=-(b_m^{M}+b_n^N)\bar{v}_{mn}
       -\sqrt2\sum_{\ell=1}^{L}u_{\ell m}w_{n\ell},\\
 i\dot w_{n\ell}&=(b_n^N+b_\ell^{L})w_{n\ell}
       +\sqrt2\sum_{j=1}^{M}\bar{u}_{\ell j}\bar{v}_{jn},\\
 i\dot{\bar{w}}_{n\ell}&=-(b_n^N+b_\ell^{L})\bar{w}_{n\ell}
       -\sqrt2\sum_{j=1}^{M}u_{\ell j}v_{jn}.
\end{aligned}
\end{equation}

\Cref{eq:canonical-H,eq:canonical-rates} define the canonical multiwave
system at the centre of this work. In \cref{sec:BA-correspondence},
we show that its simple cubic Hamiltonian provides a common setting
for BA functions on arbitrary smooth algebraic curves, and reconstruct
its generic solution from the associated spectral data.

\subsubsection{Power symmetries and phase conventions}
\label{sec:canonical-phase-gauges}

The linear terms cancel from the mode-power derivatives. The conserved
power combinations generating diagonal mode gauges are
\begin{equation}\label{eq:canonical-charges}
\begin{aligned}
 Q_\ell^L&=\sum_{n=1}^{N}I^w_{n\ell}
             -\sum_{m=1}^{M}I^u_{\ell m},&&1\leq\ell\leq L,\\
 Q_m^M&=\sum_{n=1}^{N}I^v_{mn}
             -\sum_{\ell=1}^{L}I^u_{\ell m},&&1\leq m\leq M,\\
 Q_n^N&=\sum_{\ell=1}^{L}I^w_{n\ell}
             -\sum_{m=1}^{M}I^v_{mn},&&1\leq n\leq N.
\end{aligned}
\end{equation}
They commute, and their only linear relation is
\begin{equation}\label{eq:canonical-charge-relation}
 \sum_{\ell=1}^{L}Q_\ell^L-\sum_{m=1}^{M}Q_m^M
                         -\sum_{n=1}^{N}Q_n^N=0.
\end{equation}
Hence there are $L+M+N-1$ independent power charges on a generic
mode chart. The mixed signs in their definitions specify the
orientation of each gauge action.

Adding a constant linear combination of these charges to $H$
redistributes the linear mode rates without changing any power
derivative. We distinguish the symmetric-phase Hamiltonian $H$ from
a second convention $H_0$, in which the $u$ modes have zero linear
rate. This convention is convenient for comparison with Jiang and
Rauch-Wojciechowski. Define
\begin{align}\label{eq:canonical-H0}
 H_0&=H+\sum_{\ell=1}^{L}b_\ell^LQ_\ell^L
        +\sum_{m=1}^{M}b_m^MQ_m^M\nonumber\\
 &=\sum_{n=1}^{N}\sum_{\ell=1}^{L}
       (2b_\ell^L+b_n^N)w_{n\ell}\bar{w}_{n\ell}
   +\sum_{m=1}^{M}\sum_{n=1}^{N}
       (2b_m^M+b_n^N)v_{mn}\bar{v}_{mn}+H_{\mathrm{mix}},
\end{align}
where $H_{\mathrm{mix}}$ is \cref{eq:intro-mixing}. Let
$(\tilde{\boldsymbol u},\tilde{\boldsymbol v},\tilde{\boldsymbol w})$
and their barred partners denote solutions of Hamilton's equations
generated by $H_0$. The corresponding solutions of $H$ are
\begin{equation}\label{eq:canonical-phase-gauge}
\begin{aligned}
 u_{\ell m}(t)&=e^{-i(b_\ell^L+b_m^M)t}\tilde u_{\ell m}(t),
 &&1\leq\ell\leq L,\quad1\leq m\leq M,\\
 v_{mn}(t)&=e^{ib_m^Mt}\tilde v_{mn}(t),
 &&1\leq m\leq M,\quad1\leq n\leq N,\\
 w_{n\ell}(t)&=e^{ib_\ell^Lt}\tilde w_{n\ell}(t),
 &&1\leq n\leq N,\quad1\leq\ell\leq L.
\end{aligned}
\end{equation}
The barred partners receive the
inverse factors. This time-dependent change is canonical at each
fixed time. Its exponents cancel in every cubic mixing product, so
both those products and all mode powers are identical in the two
descriptions. Appendix~\ref{sec:JRW-dictionary} identifies $H_0$ with the
constant-diagonal extension of the 1991 Hamiltonian \cite{JRW1991}.

The independent diagonal parameters and this change of phase convention
have distinct roles. In each mixing product $u_{\ell m}v_{mn}w_{n\ell}$,
the sum of the three linear phase coefficients is
$2(b_\ell^L+b_m^M+b_n^N)$, for
$1\leq\ell\leq L$, $1\leq m\leq M$, $1\leq n\leq N$.
The phase gauge \cref{eq:canonical-phase-gauge} preserves this sum.
The outer diagonal parameters therefore supply interaction-dependent
linear rates that remain present in either convention. Their values
are fixed by the marked curve data in the BA realization.

\subsection{Commuting Hamiltonians and complete integrability}
\label{sec:canonical-integrable-framework}

We construct the Lax pencil and its commuting Hamiltonians, then give
the complete integral count needed to place the BA solutions in an
integrable canonical system.

\subsubsection{The matrix pencil and its Lax equation}
\label{sec:canonical-lax}

A Lax representation packages the mode equations into a matrix
commutator, from which conserved quantities follow through the
characteristic polynomial. We obtain the representation from the
three-block construction of Jiang and Rauch-Wojciechowski
\cite[equation (2.7)]{JRW1991}, using the constant-diagonal extension
and the symmetric phase convention. The coordinate and phase changes
are given explicitly in Appendix~\ref{sec:JRW-dictionary}.

First write the equations using the mode matrices
$\boldsymbol u\in\C^{L\times M}$,
$\boldsymbol v\in\C^{M\times N}$,
$\boldsymbol w\in\C^{N\times L}$ and the constant phase matrices
\begin{equation}\label{eq:canonical-diagonals}
 \mathsf B_L=\diag(b_1^L,\ldots,b_L^L),\qquad
 \mathsf B_M=\diag(b_1^M,\ldots,b_M^M),\qquad
 \mathsf B_N=\diag(b_1^N,\ldots,b_N^N).
\end{equation}
Then \cref{eq:canonical-rates} is equivalently
\begin{equation}\label{eq:canonical-matrix-rates}
\begin{aligned}
 i\dot{\boldsymbol u}
 &=\mathsf B_L\boldsymbol u+\boldsymbol u\mathsf B_M
       +\sqrt2\bar{\boldsymbol w}^{\trans}\bar{\boldsymbol v}^{\trans},\\
 i\dot{\bar{\boldsymbol u}}^{\trans}
 &=-\mathsf B_M\bar{\boldsymbol u}^{\trans}-\bar{\boldsymbol u}^{\trans}\mathsf B_L
       -\sqrt2\boldsymbol v\boldsymbol w,\\
 i\dot{\boldsymbol v}
 &=\mathsf B_M\boldsymbol v+\boldsymbol v\mathsf B_N
       +\sqrt2\bar{\boldsymbol u}^{\trans}\bar{\boldsymbol w}^{\trans},\\
 i\dot{\bar{\boldsymbol v}}^{\trans}
 &=-\mathsf B_N\bar{\boldsymbol v}^{\trans}-\bar{\boldsymbol v}^{\trans}\mathsf B_M
       -\sqrt2\boldsymbol w\boldsymbol u,\\
 i\dot{\boldsymbol w}
 &=\mathsf B_N\boldsymbol w+\boldsymbol w\mathsf B_L
       +\sqrt2\bar{\boldsymbol v}^{\trans}\bar{\boldsymbol u}^{\trans},\\
 i\dot{\bar{\boldsymbol w}}^{\trans}
 &=-\mathsf B_L\bar{\boldsymbol w}^{\trans}-\bar{\boldsymbol w}^{\trans}\mathsf B_N
       -\sqrt2\boldsymbol u\boldsymbol v.
\end{aligned}
\end{equation}
These same blocks assemble into the $d\times d$ coefficient matrix
$\mathsf X(t)$, where $d=L+M+N$:
\begin{equation}\label{eq:canonical-X}
 \mathsf X=\begin{pmatrix}
 -2\mathsf B_L&\sqrt2\boldsymbol u&\bar{\boldsymbol w}^{\trans}\\
 \sqrt2\bar{\boldsymbol u}^{\trans}&-2\mathsf B_M&\boldsymbol v\\
 -\boldsymbol w&-\bar{\boldsymbol v}^{\trans}&\mathsf B_N
 \end{pmatrix}.
\end{equation}
The Lax matrix $\mathsf L(\nu,t)$ forms the affine matrix pencil
\begin{equation}\label{eq:canonical-pencil}
 \mathsf L(\nu,t)=\mathsf X(t)+\nu\mathsf D,\qquad
 \mathsf D=\diag(\Id_L,-\Id_M,0_N).
\end{equation}
Here $\nu$ is a constant spectral parameter, $\mathsf D$ is fixed,
and $\mathsf L(0,t)=\mathsf X(t)$. The time dependence of
$\mathsf X$ comes entirely from the mode matrices; its diagonal
blocks are constant. Suppressing $t$ in the notation, block
multiplication gives
\begin{equation}\label{eq:canonical-lax}
 i\dot{\mathsf L}=[\mathsf L,\mathsf T],\qquad
 \mathsf T(\nu)=\begin{pmatrix}
 -\mathsf B_L&0&\bar{\boldsymbol w}^{\trans}\\
 0&\mathsf B_M&-\boldsymbol v\\
 -\boldsymbol w&\bar{\boldsymbol v}^{\trans}&-\nu\Id_N
 \end{pmatrix}.
\end{equation}
Equivalently, $\dot{\mathsf L}=[\mathsf L,-i\mathsf T]$. Its off-diagonal blocks reproduce \cref{eq:canonical-matrix-rates},
while its diagonal blocks have zero derivatives.
The spectral determinant is
\begin{equation}\label{eq:canonical-spectral}
 F(\lambda,\nu)=\det(\lambda\Id_d-\mathsf L(\nu)).
\end{equation}
Its coefficients are polynomials in the modes and are conserved by
\cref{eq:canonical-lax}.

\subsubsection{The AKS framework and commuting integrals}
\label{sec:canonical-AKS}

To establish integrability, the conserved quantities must also commute
under the canonical bracket. The AKS framework explains this
commutativity by expressing the mode bracket in terms of the same
matrices $\mathsf X,\mathsf D$ used in
\cref{eq:canonical-X,eq:canonical-pencil}.
Write $X_{hr},D_{hr}$ for their entries. The canonical mode brackets give
\begin{equation}\label{eq:canonical-frozen}
 \{X_{hr},X_{jk}\}
 =-i(D_{hh}-D_{rr})\delta_{hk}\delta_{rj},
 \qquad1\leq h,r,j,k\leq d.
\end{equation}
Its coefficients depend only on the fixed matrix $\mathsf D$.
Fixing the three diagonal blocks of $\mathsf X$ leaves precisely the
off-diagonal canonical mode coordinates; these form a symplectic
affine leaf of the matrix bracket.

For scalar functions $f(\mathsf X),g(\mathsf X)$, define their
matrix gradients $P,Q$ by
\begin{equation}\label{eq:canonical-matrix-gradients}
 \mathrm df=\tr(P\,\mathrm d\mathsf X),\qquad
 \mathrm dg=\tr(Q\,\mathrm d\mathsf X).
\end{equation}
These gradients can depend on the mode values. Contracting
\cref{eq:canonical-frozen} gives
\begin{equation}\label{eq:canonical-gradient-bracket}
 \{f,g\}
 =i\tr(\mathsf D[P,Q]).
\end{equation}
This constant bracket and the shifted invariants of
$\mathsf X+\nu\mathsf D$ give the argument-shift realization of
the AKS construction \cite{MishchenkoFomenko1978,Bolsinov1992,
AdlerVanMoerbeke2018}.

The Hamiltonian belongs to this construction through the trace identity
\begin{align}\label{eq:canonical-trace-H}
 H={}&-\tfrac13\tr\mathsf X^3
      +3\tr\!\left[\mathsf B_L
        (\bar{\boldsymbol w}^{\trans}\boldsymbol w
                     -\boldsymbol u\bar{\boldsymbol u}^{\trans})\right]
        \nonumber\\
 &+3\tr\!\left[\mathsf B_M
        (\boldsymbol v\bar{\boldsymbol v}^{\trans}
                     -\bar{\boldsymbol u}^{\trans}\boldsymbol u)\right]
      -\tfrac83\tr\mathsf B_L^3
      -\tfrac83\tr\mathsf B_M^3
      +\tfrac13\tr\mathsf B_N^3.
\end{align}
The two quadratic trace terms are weighted combinations of the power
charges in \cref{eq:canonical-charges}; the traces of the fixed phase
matrices are constants. Thus $H$ combines a cubic spectral invariant
with commuting gauge generators.

\begin{proposition}\label{prop:canonical-commutation}
Every coefficient of $F$ in \cref{eq:canonical-spectral} commutes
with every other coefficient and with the charges in
\cref{eq:canonical-charges}.
\end{proposition}
\begin{proof}
Using the gradient definition \cref{eq:canonical-matrix-gradients},
let $P$ and $Q$ be the matrix gradients of conjugation-invariant polynomials
of $\mathsf X+\nu\mathsf D$ and $\mathsf X+\mu\mathsf D$,
respectively. They commute with their respective matrix arguments.
Cyclicity of the trace gives
\[
 \tr((\mathsf X+\nu\mathsf D)[P,Q])
 =\tr((\mathsf X+\mu\mathsf D)[P,Q])=0.
\]
Subtracting proves $\tr(\mathsf D[P,Q])=0$ for $\nu\ne\mu$;
polynomial continuation includes $\nu=\mu$. This proves spectral
commutativity under \cref{eq:canonical-frozen}. Each power charge
acts on $\mathsf X$ by constant diagonal conjugation, which also
preserves $F$.
\end{proof}

\subsubsection{The complete count for Liouville integrability}
\label{sec:canonical-independence}

The spectral invariants are complete when the phase coefficients are
distinct within each group. Equal phases allow additional continuous
changes of mode basis, whose conserved quantities supply any missing
integrals. We state the complete count here; Appendix~\ref{sec:integrability-proof}
gives its independence proof for arbitrary phase coefficients.

Write $s_h=D_{hh}$ and $e_h=X_{hh}$, $1\leq h\leq d$.
Partition each original group into maximal sets $B$ with a common
intercept $e_h$. We call these sets equal-phase blocks; their slopes
$s_h$ are also constant, with common value $s_B$.
Define the block moment matrix
\begin{equation}\label{eq:canonical-block-moment}
 (\mathsf J_B)_{ij}
 =\sum_{\substack{h=1\\s_h\ne s_B}}^d
       \frac{X_{ih}X_{hj}}{s_B-s_h},
 \qquad i,j\in B.
\end{equation}
For a constant $|B|\times|B|$ matrix $\mathsf Z$, let $\widehat{\mathsf Z}$ denote its
embedding by zeros outside $B$. Then
\[
 \{\mathsf X,\tr(\mathsf Z\mathsf J_B)\}=-i[\widehat{\mathsf Z},\mathsf X].
\]
Thus its Hamiltonian flow is conjugation by $e^{-i\epsilon\widehat{\mathsf Z}}$.
It preserves the Hamiltonian, since $\widehat{\mathsf Z}$ commutes with the fixed phase
matrices, so the entries of $\mathsf J_B$ are Noether integrals.
Different blocks act independently. Within each block the moment
entries have $-i$ times the matrix Lie--Poisson bracket.
These are the standard quadratic moment maps for rectangular canonical
matrix pairs \cite[Section~2]{Harnad1994}.

Fix an ordering of the indices in $B$, and let $\mathsf J_B^{[t]}$
denote its leading $t\times t$ principal submatrix. The functions
\begin{equation}\label{eq:canonical-moment-chain}
 I_{B,t,k}=\frac1k\tr\!\left((\mathsf J_B^{[t]})^k\right),
 \qquad 1\leq k\leq t\leq |B|,
\end{equation}
commute. For two nested principal blocks, the gradient of an
invariant of the larger block commutes with that block; the
Lie--Poisson pairing with the smaller-block gradient is therefore zero.
This is the Gelfand--Zeitlin construction
\cite{GuilleminSternberg1983}. These functions also commute with every
spectral coefficient, since block conjugation preserves $F$.
Let $\mathcal A$ be the polynomial algebra generated by these functions
and the coefficients of $F$. It is a commuting algebra of first
integrals of $H$.

\begin{theorem}\label{thm:canonical-independence}
For every positive $L,M,N$ and every fixed choice of phase coefficients,
$\mathcal A$ has generic differential rank $p=LM+MN+NL$.
Thus \cref{eq:canonical-H} is generically Liouville integrable.
A complete local set can include $H$ wherever $\mathrm dH\ne0$.
\end{theorem}
The proof in Appendix~\ref{sec:integrability-proof} proceeds by
increasing one matrix dimension at a time. At each insertion, the
spectral coefficients and principal moments together supply one new
independent integral per new canonical pair. This includes arbitrary
coincident phase coefficients. The same argument quantifies the
spectral deficit in \cref{eq:canonical-spectral-deficit}.

When all phases within each original group are distinct, the spectral
family alone is complete; an explicit selection by
line intersections and a spanning tree is given in
\cref{sec:distinct-phase-integrals}. Retaining the $d-1$ power charges
leaves
\begin{equation}\label{eq:canonical-genus-preview}
 G=p-d+1=LM+MN+NL-L-M-N+1
\end{equation}
spectral integrals on the regular locus used in
\cref{sec:spectral-geometry}.

Bolsinov's argument-shift completeness criterion
\cite[Theorem~1.1]{Bolsinov1992} includes the constant bracket
\cref{eq:canonical-frozen}. Its conditions are that
$\mathsf X+\nu\mathsf D$ is regular for every finite $\nu$ and that
the projected kernel has the required dimension; the latter becomes
distinctness of the phases within each original group.
A smooth spectral curve supplies the first condition, so completeness
in the distinct-phase case is a known consequence of this criterion.
\Cref{thm:canonical-independence} gives a uniform direct proof and
completes the count for repeated phases using the established
moment-map and Gelfand--Zeitlin constructions.
Jiang and Rauch-Wojciechowski prove completeness for row counts two
and three (our $L+M$), and leave four rows unresolved
\cite[Sections II.B--II.C, especially p.~1723]{JRW1991}. The insertion proof covers
arbitrary row and column counts and every fixed phase profile.

\begin{samepage}
\subsubsection{Commuting spectral flows and higher-order interactions}
\label{sec:spectral-flows}

The spectral invariants provide Hamiltonians as well as conserved
quantities. Choosing another invariant changes the evolution of the
same mode matrices while preserving their spectral curve. This is
the commuting-flow construction of the AKS framework
\cite{AdlerVanMoerbeke2018}, and applies to arbitrary positive
$L,M,N$.\par
\end{samepage}

For each integer $r\geq2$, define the trace Hamiltonian
\begin{equation}\label{eq:hierarchy-traces}
 H_r=\frac1r\tr\mathsf X^r.
\end{equation}
It is a spectral invariant at $\nu=0$ and is a polynomial of degree
at most $r$ in the modes. By
\cref{prop:canonical-commutation,eq:canonical-trace-H}, these
Hamiltonians commute with one another, with $H$ and $H_0$, and with
all the power charges and spectral coefficients. The complete
integral set in \cref{thm:canonical-independence} is therefore
preserved by every such flow.

For the $H_r$ evolution, the bracket
\cref{eq:canonical-frozen} gives an explicit Lax representation:
\begin{equation}\label{eq:hierarchy-lax}
 i\dot{\mathsf X}=[\mathsf D,\mathsf X^{r-1}],\qquad
 i\dot{\mathsf L}=[\mathsf L,\mathsf T_r],\qquad
 \mathsf T_r(\nu)=
       \frac{\mathsf X^{r-1}-\mathsf L(\nu)^{r-1}}{\nu}.
\end{equation}
The numerator vanishes at $\nu=0$, so $\mathsf T_r$ is polynomial
in $\nu$. Indeed, $[\mathsf L,\mathsf L^{r-1}]=0$ and
$[\mathsf X,\mathsf X^{r-1}]=0$ reduce its commutator to
$[\mathsf D,\mathsf X^{r-1}]$. The block-diagonal entries of
$\mathsf X$ remain fixed because $\mathsf D$ is scalar on each
of its three blocks.

The fourth member $H_4=\tr\mathsf X^4/4$ includes quartic
interactions, with lower-degree terms supplied by the fixed diagonal
entries. Appendix~\ref{sec:quartic-expansion} gives its expanded
quartic part.

For a constant $\alpha$, the combined Hamiltonian and time matrix are
\begin{equation}\label{eq:hierarchy-combination}
 H_\alpha=H+\alpha H_4,\qquad
 \mathsf T_\alpha=\mathsf T+\alpha\mathsf T_4.
\end{equation}
The same initial state and phase coefficients give the same
conserved polynomial $F(\lambda,\nu)$ under either evolution.
In a Jacobian or Prym linearization, if the geometric generators $i\partial_t$ have translation
vectors $\boldsymbol U_H,\boldsymbol U_4$, the divisor's Abel
coordinate evolves as
\begin{equation}\label{eq:hierarchy-direction}
 z(t)=z(0)-it(\boldsymbol U_H+\alpha\boldsymbol U_4).
\end{equation}
This is the spectral interpretation of choosing a different scalar
evolution within the commuting family
\cite{Beauville1990,Vanhaecke2001}.

The hyperelliptic realization in \cref{sec:BA-realization} selects
the scalar Schr\"odinger direction $z(t)=z_0-it e_g$ for $H_0$.
More generally, \cref{thm:BA-higher-flows} identifies the spectral
flows with multipoint BA times and constant reciprocal mode phases
for arbitrary group sizes. On each birational spectral level the
full family supplies every Abel direction.

\subsection{Regular spectral curves at fixed matrix sizes}
\label{sec:spectral-geometry}

The Lax pencil associates a spectral curve to each trajectory. For
fixed matrix sizes $L,M,N$, the phase coefficients and initial modes
determine a family of marked curves. We describe its regular members,
their genus, and how special spectral levels accommodate curves of
lower genus.

This fixed-size question complements the BA construction of
\cref{sec:general-BA-realization}: any prescribed smooth curve admits
a compatible BA realization, enlarging the mode lists when needed.
The classification here describes which curves occur at fixed matrix
sizes, while that construction chooses compatible matrix sizes,
parameters and initial modes to accommodate the curve.

\subsubsection{From the Lax pencil to a fixed curve}

For each spectral parameter $\nu$, the $d\times d$ Lax matrix
$\mathsf L(\nu,t)$ in \cref{eq:canonical-pencil} has eigenvalue
$\lambda$ and nonzero eigenvector $\psi\in\C^d$ when
\begin{equation}\label{eq:geometry-eigenvalue}
 \mathsf L(\nu,t)\psi=\lambda\psi,\qquad \psi\ne0.
\end{equation}
Its solvability condition is precisely
\begin{equation}\label{eq:geometry-affine}
 F(\lambda,\nu)=\det\bigl(\lambda\Id_d-\mathsf X(t)-\nu\mathsf D\bigr)=0,
 \qquad d=L+M+N.
\end{equation}
Thus the spectral curve records the eigenvalues as $\nu$ varies.
At a generic value of $\nu$ there are $d$ eigenvalues, so projection
to $\nu$ gives a degree-$d$ covering of the projective line.
Its branch values occur when eigenvalues coincide; they can be
located by eliminating $\lambda$ from $F=\partial_\lambda F=0$.

The equation is fixed along the Hamiltonian flow. Indeed,
\cref{eq:canonical-lax} implies $\partial_tF=0$, since
the trace of the adjugate of $\lambda\Id_d-\mathsf L$ multiplied
by $[\mathsf L,\mathsf T]$ vanishes. The phase coefficients and
initial modes determine the coefficients of $F$, which are the
spectral first integrals. Different integral values can therefore
give different curves, while a single trajectory preserves its curve.
This is the standard spectral-curve mechanism for matrix Lax systems
\cite{JRW1991,Beauville1990,AdlerVanMoerbeke2018}.

To obtain a compact smooth curve, first take the projective closure
of \cref{eq:geometry-affine}, then its normalization. Normalization
separates the local branches at singular points. We denote this smooth
curve by $C$; its genus, rather than the degree of the determinant,
is the genus relevant to the associated Abelian functions.

\subsubsection{Branches at infinity and genus}

Introduce a homogeneous coordinate $\tau$, with affine chart $\tau=1$.
The projective equation and its restriction to the line at infinity are
\begin{align}
 \widehat F(\lambda,\nu,\tau)
 &=\det\bigl(\lambda\Id_d-\nu\mathsf D-\tau\mathsf X\bigr)=0,
 \label{eq:geometry-projective}\\
 \widehat F(\lambda,\nu,0)
 &=(\lambda-\nu)^L(\lambda+\nu)^M\lambda^N.
 \label{eq:geometry-infinity}
\end{align}
The plane curve has degree $d$ and meets infinity at
$[1:1:0]$, $[-1:1:0]$ and $[0:1:0]$. The three points contain,
respectively, $L$, $M$ and $N$ branches. Their affine expansions are
\begin{equation}\label{eq:geometry-branches}
\begin{aligned}
 \lambda&=\nu-2b_\ell^L+O(\nu^{-1}),&&1\leq\ell\leq L,\\
 \lambda&=-\nu-2b_m^M+O(\nu^{-1}),&&1\leq m\leq M,\\
 \lambda&=b_n^N+O(\nu^{-1}),&&1\leq n\leq N.
\end{aligned}
\end{equation}
In particular, the phase coefficients specify the branch intercepts.
Distinct intercepts within each list give distinct tangent lines at
the corresponding plane point. Such a point is called an
\emph{ordinary multiple point}: all its branches are smooth and
their tangent lines are distinct. Multiplicity one means a smooth
point. Every branch in \cref{eq:geometry-branches} meets the infinity
line transversely.

We call a spectral curve \emph{regular} when the phase coefficients
are pairwise distinct within each of the three lists, its equation is
irreducible, and it has no affine singularities. Its only possible
plane singularities are then the prescribed ordinary points at infinity.

\begin{proposition}\label{prop:geometry-genus}
The normalization of a regular spectral curve has genus
\begin{equation}\label{eq:geometry-genus}
 G=LM+MN+NL-L-M-N+1.
\end{equation}
It has $d$ marked points at infinity, one for each branch in
\cref{eq:geometry-branches}.
\end{proposition}
\begin{proof}
For example, near $[1:1:0]$, use $\nu=1$ and write
$\lambda=1+\tau h$. The first nonzero coefficient in
$\widehat F(1+\tau h,1,\tau)$ is a nonzero constant times
\[
 \tau^L\prod_{\ell=1}^{L}(h+2b_\ell^L).
\]
Its distinct roots give the $L$ tangent lines. The other two points
give the factors
$\prod_{m=1}^{M}(h+2b_m^M)$ and
$\prod_{n=1}^{N}(h-b_n^N)$ in the same way.
An ordinary point of multiplicity $q$ reduces the genus of a degree-$d$
plane curve by $q(q-1)/2$. The plane genus formula
\cite[Chapter~IV, \S1]{Hartshorne1977} therefore gives
\[
 G=\frac{(d-1)(d-2)}2
   -\frac{L(L-1)+M(M-1)+N(N-1)}2,
\]
which simplifies to \cref{eq:geometry-genus}.
\end{proof}

This also explains the count in \cref{eq:canonical-genus-preview}:
the number of mode pairs is $p=G+(d-1)$, with $d-1$ independent
power charges. The spectral curve supplies the remaining geometric
part of that count.

The regular locus is nonempty when phases and initial modes vary;
Appendix~\ref{sec:geometry-details} proves this by smoothing a
union of lines.

\subsubsection{Marked-curve characterization}

The infinity branches provide a characterization in terms of marked
points and meromorphic functions. Write their divisors as
\begin{equation}\label{eq:geometry-markings}
 \mathcal D_L=\sum_{\ell=1}^{L}P_\ell^L,\qquad
 \mathcal D_M=\sum_{m=1}^{M}P_m^M,\qquad
 \mathcal D_N=\sum_{n=1}^{N}P_n^N.
\end{equation}
Here $P_\ell^L$, $P_m^M$ and $P_n^N$ are the points of $C$
corresponding to the infinity branches in \cref{eq:geometry-branches}.
Each point occurs once, and the three sets are disjoint.
For a meromorphic function $f$, the pole divisor $(f)_\infty$
lists its poles with their orders. Its degree is the number of
preimages of a generic value of $f$, counted with multiplicity.

\begin{theorem}\label{thm:geometry-classification}
Fix positive $L,M,N$, and let $G$ be given by
\cref{eq:geometry-genus}. For a smooth connected complex curve $C$
of genus $G$, the following statements are equivalent.
\begin{enumerate}
 \item $C$ is the normalization of a regular spectral curve of
 \cref{eq:canonical-pencil}, for some phase coefficients and initial modes.
 \item $C$ has a birational plane model of degree $d=L+M+N$
 with three distinct collinear ordinary points of multiplicities
 $L,M,N$, no other singularities, and with their common line
 transverse to every branch.
 \item $C$ has divisors as in \cref{eq:geometry-markings} and
 meromorphic functions $\lambda,\nu$ generating its function field
 $\C(C)$, with exact pole divisors
 \begin{equation}\label{eq:geometry-poles}
 \begin{aligned}
  (\lambda-\nu)_\infty&=\mathcal D_M+\mathcal D_N,\\
  (\lambda+\nu)_\infty&=\mathcal D_L+\mathcal D_N,\\
  (\lambda)_\infty&=\mathcal D_L+\mathcal D_M.
 \end{aligned}
 \end{equation}
\end{enumerate}
\end{theorem}
The equivalence is proved in Appendix~\ref{sec:geometry-details}.
The marked functions specify the regular curve family at these sizes.
Varying the phase coefficients and spectral integral values explores
this family; each Hamiltonian trajectory preserves its own spectral
curve. A prescribed curve of lower genus can occur at a special level,
as described in \cref{sec:special-spectral-levels}.

\subsubsection{Covering degrees and hyperelliptic cases}

The three cancellations in \cref{eq:geometry-poles} give maps to
the projective line of degrees
\begin{equation}\label{eq:geometry-degrees}
 \deg\lambda=L+M,\qquad
 \deg(\lambda-\nu)=M+N,\qquad
 \deg(\lambda+\nu)=L+N.
\end{equation}
Any two of these functions generate $\C(C)$, since they recover
$\lambda,\nu$ by linear combinations. These are concrete
restrictions on the curve, additional to its genus.

\begin{proposition}\label{prop:geometry-hyperelliptic}
A regular spectral curve of genus $G\geq2$ is hyperelliptic if and
only if at least two of $L,M,N$ equal one. The case $L=M=N=1$
has genus one.
\end{proposition}
This criterion classifies the regular spectral curves at fixed matrix
sizes. Hyperelliptic BA solutions also occur at special spectral levels
when fewer than two of $L,M,N$ equal one. In particular, the
genus-$g$ construction in \cref{sec:BA-realization} permits
$L,M\geq g$ and $N=1$; for $g\geq2$, these hyperelliptic curves
are smooth normalizations of special spectral levels of systems whose
regular curves are non-hyperelliptic
(\cref{sec:special-spectral-levels}).

Two singleton groups give a degree-two map in
\cref{eq:geometry-degrees}. Appendix~\ref{sec:geometry-details}
proves the converse using the covering degrees.
For example, in the regular $(1,M,1)$ case the degree-two function
$\lambda+\nu$ has pole divisor $Q_1+R_1$: its involution exchanges
these points, so the BA puncture is not a branch point. The
odd-degree specialization of \cref{sec:BA-realization} instead marks
a Weierstrass point as its single puncture. Its conditions $L,M\geq g$
come from interpolation at that marking, not from a restriction on
all hyperelliptic spectral curves. The general residue criterion
covers both settings; in the singleton group, the principal divisor
of $\lambda+\nu$ supplies the required relation between the two
holomorphic evaluation vectors.

\noindent\begin{minipage}{\textwidth}
Examples of the covering degrees from \cref{eq:geometry-degrees},
listed in increasing order, are
\begin{center}
\begin{tabular}{ccll}
 $(L,M,N)$ & $G$ & Covering degrees & Regular curve\\\hline
 $(1,1,1)$ & $1$ & $2,2,2$ & elliptic\\
 $(1,1,2)$ & $2$ & $2,3,3$ & hyperelliptic\\
 $(1,2,2)$ & $4$ & $3,3,4$ & non-hyperelliptic\\
 $(1,2,3)$ & $6$ & $3,4,5$ & non-hyperelliptic\\
 $(2,2,2)$ & $7$ & $4,4,4$ & non-hyperelliptic\\
 $(2,2,3)$ & $10$ & $4,5,5$ & non-hyperelliptic
\end{tabular}
\end{center}
\end{minipage}\par\medskip
In particular, the $(1,2,2)$ family is trigonal: it has a
degree-three map and no degree-two map. The $(2,2,2)$ family is
tetragonal. A hypothetical degree-three map and one of its
degree-four maps would generate the function field, since their
degrees are coprime; Castelnuovo--Severi would then give $G\leq6$.

There is also a useful product-surface model when $N=1$.
Using $\lambda-\nu$ and $\lambda+\nu$ as coordinates gives a
smooth curve of bidegree $(L+1,M+1)$ in
$\mathbb P^1\times\mathbb P^1$, with genus $LM$.
Here bidegree records the degrees in the first and second coordinates
of its defining equation; the corresponding projection degrees are
$M+1$ and $L+1$. For $N>1$, the same change gives bidegree
$(L+N,M+N)$, but its $N$ branches at the common infinity corner
share a tangent. Their total defect is $N(N-1)$: each pair has
intersection multiplicity two, as follows from their distinct
intercepts in \cref{eq:geometry-branches}. Consequently its genus is
\[
 (L+N-1)(M+N-1)-N(N-1)=G.
\]

\subsection{Special spectral levels and reductions}
\label{sec:special-spectral-levels}

We explain how special integral values and mode constraints change the
regular geometry, so that lower-genus BA curves and symmetric
reductions can be located within the same Hamiltonian family.

\subsubsection{Special spectral levels}

Keep the phase coefficients distinct within each group and suppose
the plane curve remains irreducible. If particular integral values
create affine singularities, the standard genus--delta formula for
normalization gives
\begin{equation}\label{eq:geometry-special-genus}
 g(C)=G-\sum_{P\in\operatorname{Sing}_{\mathrm{aff}}(F)}\delta_P.
\end{equation}
Here $C$ is the smooth normalization of the plane spectral curve
$F(\lambda,\nu)=0$, $g(C)$ is its genus, and $G$ is the regular genus
in \cref{eq:geometry-genus}. The set
$\operatorname{Sing}_{\mathrm{aff}}(F)$ consists of its affine singular
points, and $\delta_P$ is the local genus defect at $P$, equal to one
at an ordinary node.
Thus the sizes determine the regular genus, while the spectral
integrals determine whether a particular level has lower genus.
Coincident phase coefficients can also change the infinity
singularities and require their defects to be recomputed.

For birational prescribed BA realizations,
\cref{prop:BA-prescribed-spectral-curve} identifies the prescribed smooth
curve with the normalization and gives the total excess defect $G-g$.
For example, the $(L,M,N)=(3,3,1)$ system has regular genus nine,
whereas the genus-three trigonal realization in
\cref{sec:numerical-validation} has excess defect six. The additional
singularities belong to its plane spectral image; the prescribed BA
curve is its smooth normalization.

\subsubsection{A symmetric reduction and its Prym}

A symmetric reduction with $L=M=s$, $N=1$ gives an involution
$\iota(\lambda,\nu)=(\lambda,-\nu)$. The moving divisor then
lies in a translate of its Prym, the connected part of the Jacobian
reversed by $\iota$. On a regular level the curve has genus $s^2$,
and the Prym has dimension $s(s+1)/2$.
Appendix~\ref{sec:geometry-prym-proof} gives the mode constraints,
the dimension count and the eigenvector-divisor proof.

\section{The Baker-Akhiezer correspondence and its commuting flows}
\label{sec:BA-correspondence}

\Cref{sec:canonical} introduced the canonical Hamiltonian system,
established its Liouville integrability and spectral properties, and
described the associated algebraic curves. We now focus on the general
solution of the system in terms of Baker-Akhiezer functions.

We establish the correspondence in both directions: prescribed BA data
produce canonical mode solutions, and generic canonical initial modes
recover their BA data. We then identify the commuting evolution and
its mode phases, making the Abel directions explicit.

\subsection{From prescribed BA data to canonical dynamics}
\label{sec:general-BA-realization}

We construct solutions for arbitrary positive $L,M,N$ directly from a
smooth curve. The two incident arrays come from ordinary and dual BA
bases normalized at $N$ punctures; their two-point kernel supplies the
$L\times M$ array. Meromorphic interpolation determines the weights
and phase coefficients. Its residue identities give Hamilton's equations
without first reconstructing a matrix spectral problem.
For $N=1$ the basis has a single ordinary/dual pair, giving two
mode vectors and their pair-mode matrix. We develop this specialization
after the general reconstruction and Hamiltonian comparison.

\subsubsection{Scalar BA bases and their kernel}
\label{sec:multipoint-BA-data}

Fix a smooth compact connected curve $\mathcal C$ of genus $g$,
distinct punctures $R_k$ with local parameters $s_k$, $1\leq k\leq N$,
and an ordinary BA divisor $\mathscr D$ of degree $g+N-1$, disjoint
from the punctures. Put $\mathscr R=\sum_{k=1}^{N}R_k$.
For a divisor $E$, $H^0(\mathcal O(E))$ denotes the vector space of
global sections of its associated line bundle: meromorphic functions
$f$ satisfying $(f)+E\geq0$, together with the zero function.
Regularity at the reference time means
$H^0(\mathcal O(\mathscr D-\mathscr R))=0$, so no nonzero function
with poles bounded by $\mathscr D$ vanishes at every puncture: the $N$
leading values then determine a BA section uniquely.

The dual divisor is determined by these data. Write $K_{\mathcal C}$
for a canonical divisor. The map from
$H^0(K_{\mathcal C}+2\mathscr R)$ to the $N$ double-pole coefficients and the
$g+N-1$ restriction values along $\mathscr D$ (including jets at
multiple points) has kernel
$H^0(K_{\mathcal C}+\mathscr R-\mathscr D)$, which vanishes by Serre duality
and Riemann--Roch, since $\deg(\mathscr D-\mathscr R)=g-1$ and its $H^0$
vanishes. The source and target both have dimension $g+2N-1$, so the map is an isomorphism.
There is therefore a unique differential $d\Omega_0$ vanishing at
$\mathscr D$ with double-pole leading parts $-ds_k/s_k^2$.
Its remaining zero divisor is $\mathscr D^\vee$, of degree $g+N-1$,
and
\[
 \mathcal O(\mathscr D^\vee-\mathscr R)
 \simeq\mathcal O(K_{\mathcal C}+\mathscr R-\mathscr D)
\]
shows that the dual leading-value problem is regular too.
Zero divisors are counted with multiplicity. The simple-pole residues
of $d\Omega_0$ are
allowed and are determined by this construction. The first-order
local parameters fix its double-pole coefficients and hence the
dual divisor. The normalized bases satisfy
\begin{equation}\label{eq:general-BA-asymptotics}
\begin{aligned}
 \psi_k(P,t)&=e^{-it/s_m}(\delta_{km}+O(s_m)),\\
 \psi_k^\vee(P,t)&=e^{it/s_m}(\delta_{km}+O(s_m))
 \quad(P\to R_m),\quad1\leq k,m\leq N.
\end{aligned}
\end{equation}
These are scalar functions on one curve with shared divisor data:
$k$ selects a normalized section, rather than an independent one-point
solution \cite{Krichever1977,KricheverShiota2011}.
At the reference time both bases sum to one: the constant function
has these leading values, so uniqueness gives
$\sum_{k=1}^{N}\psi_k(P,0)=\sum_{k=1}^{N}\psi_k^\vee(P,0)=1$.
The number $N$ specifies these BA punctures independently of the genus;
they need not be infinities in another plane equation for the curve.
The zero divisor of the fixed duality differential is
$\mathscr D+\mathscr D^\vee$.

Choose normalized holomorphic differentials $\boldsymbol\omega$.
For each puncture $R_k$, define its Abel translation vector
$\boldsymbol U_k\in\C^g$ by the local expansion
\begin{equation}\label{eq:multipoint-Abel-direction}
 \boldsymbol\omega=(-\boldsymbol U_k+O(s_k))ds_k\quad(P\to R_k),
 \qquad \boldsymbol U=\sum_{k=1}^{N}\boldsymbol U_k,\qquad
 z(t)=z_0-it\boldsymbol U.
\end{equation}
Their sum $\boldsymbol U$ is the common Abel direction of the selected
time evolution. The normalized second-kind integral $\Xi$ has expansions
$\Xi=s_k^{-1}+\Xi_k+O(s_k)$. Its principal parts give this common time.

For a holomorphic basis $du$ and compatible second-kind basis $dr$,
our half-period convention is $2\omega=\oint_a du$,
$2\omega'=\oint_b du$, $2\eta=-\oint_a dr$,
$2\eta'=-\oint_b dr$. Thus $\mathbb B=\omega^{-1}\omega'$ and
$\varkappa=\eta\omega^{-1}$, with $\varkappa$ symmetric.
Write $\theta=\theta[\delta](\,\cdot\mid\mathbb B)$ for a fixed
characteristic. The theta-exponential sigma convention is
\cite{BEL1997,KorotkinShramchenko2012}
\[
 \sigma(u)=C\exp\!\left(\tfrac12u^{\trans}\varkappa u\right)
 \theta((2\omega)^{-1}u),\qquad C\ne0.
\]
The bases and characteristic fix this convention. Here
$du=\boldsymbol\omega$ and $2\omega=\Id_g$; the explicit examples
state their unnormalized bases. These conventions agree with Generapy,
the Python library used for our numerical validation in
\cref{sec:numerical-validation} \cite{Generapy}.

Base the Abel map $A$ at $R_1$, and put
$\Delta_k(P)=A(P)-A(R_k)$ and $J_k(P)=\Xi(P)-\Xi_k$.
The theta formulas are
\begin{equation}\label{eq:multipoint-BA-theta}
\begin{aligned}
 \psi_k(P,t)&=a_k(P)e^{-itJ_k(P)}
       \frac{\theta(z(t)+\Delta_k(P))}{\theta(z(t))},\\
 \psi_k^\vee(P,t)&=b_k(P)e^{itJ_k(P)}
       \frac{\theta(z(t)-\Delta_k(P))}{\theta(z(t))},
       \qquad1\leq k\leq N.
\end{aligned}
\end{equation}
The fixed factors are determined by the reference sections:
$a_k=\psi_k(P,0)\theta(z_0)/\theta(z_0+\Delta_k)$ and
$b_k=\psi_k^\vee(P,0)\theta(z_0)/\theta(z_0-\Delta_k)$.
They encode the divisors and puncture normalizations. The time-dependent
formulas are evaluated off $\theta(z)=0$ and extend meromorphically
in the BA times, with possible poles on that divisor.
Equivalently, put $J_{\sigma,k}=J_k-\boldsymbol U^{\trans}\varkappa\Delta_k$,
$a_{\sigma,k}=a_k e^{-z_0^{\trans}\varkappa\Delta_k-\Delta_k^{\trans}\varkappa\Delta_k/2}$
and
$b_{\sigma,k}=b_k e^{z_0^{\trans}\varkappa\Delta_k-\Delta_k^{\trans}\varkappa\Delta_k/2}$.
The sigma versions are
$\psi_k=a_{\sigma,k}e^{-itJ_{\sigma,k}}\sigma(z+\Delta_k)/\sigma(z)$ and
$\psi_k^\vee=b_{\sigma,k}e^{itJ_{\sigma,k}}\sigma(z-\Delta_k)/\sigma(z)$.
For necessity, an open Jacobian family means that $z_0$ ranges over
an open set while the curve, punctures, evaluation points, weights and
constant mode gauges remain fixed; the dot denotes the selected first time.

The scalar Cauchy--BA kernel $\Omega(P,Q;t)$ has the fixed ordinary
and dual divisors and diagonal normalization
\begin{equation}\label{eq:general-kernel-normalization}
 \Omega(P,Q)\sim\frac1{\beta_Q(s_Q(P)-s_Q(Q))},\qquad
 d\Omega_0=(\beta_Q+O(s_Q))ds_Q.
\end{equation}
At each puncture,
\begin{equation}\label{eq:general-kernel-infinity}
\begin{aligned}
 \Omega(P,Q)&\sim s_k(P)e^{-it/s_k(P)}\psi_k^\vee(Q)
                                      &&(P\to R_k),\\
 \Omega(P,Q)&\sim-s_k(Q)\psi_k(P)e^{it/s_k(Q)}
                                      &&(Q\to R_k),\quad1\leq k\leq N.
\end{aligned}
\end{equation}
We use the differential-kernel characterization of
\cite[Definition~3.2 and (3.22)]{Bertola2021}, converted to this sign.
Differentiation removes the diagonal principal part, and the leading
values at all punctures give
\begin{equation}\label{eq:general-kernel-rate}
 i\dot\Omega(P,Q)=\sum_{k=1}^{N}\psi_k(P)\psi_k^\vee(Q).
\end{equation}
Subtracting the sum leaves a BA section with zero leading vector, hence
zero. For $N=1$ this is the usual BA-product primitive
\cite{GrinevichOrlov1989,AratynNissimovPacheva1997}.

\subsubsection{Mode arrays and meromorphic interpolation}
\label{sec:general-BA-interpolation}

We assemble the BA evaluations into the three canonical mode arrays
and show that two finite linear residue conditions determine the
admissible weights for meromorphic interpolation, from which the
phase rates are obtained.

Put $K=L+M$. Choose distinct ordinary points $Q_a$, $1\leq a\leq K$,
away from the punctures, fixed poles and zeros of $d\Omega_0$.
For nonzero weights $\kappa_a$ and reciprocal gauges $c_a,e_k$ define
\begin{equation}\label{eq:multipoint-mode-arrays}
\begin{aligned}
 u_{\ell j}^{\rm BA}&=\frac{\sqrt2\kappa_{\ell}\kappa_{L+j}}{c_{\ell}c_{L+j}}
                  \Omega(Q_{\ell},Q_{L+j}),&
 \bar u_{\ell j}^{\rm BA}&=-\sqrt2c_{\ell}c_{L+j}\Omega(Q_{L+j},Q_{\ell}),\\
 v_{jk}^{\rm BA}&=\frac{c_{L+j}}{e_k}\psi_k(Q_{L+j}),&
 \bar v_{jk}^{\rm BA}&=\frac{\kappa_{L+j}e_k}{c_{L+j}}\psi_k^\vee(Q_{L+j}),\\
 w_{k\ell}^{\rm BA}&=c_{\ell}e_k\psi_k^\vee(Q_{\ell}),&
 \bar w_{k\ell}^{\rm BA}&=\frac{\kappa_{\ell}}{c_{\ell}e_k}\psi_k(Q_{\ell}),
 \quad1\leq\ell\leq L,\ 1\leq j\leq M,\ 1\leq k\leq N.
\end{aligned}
\end{equation}
The bars are independent partner labels. The fixed normalizations have
one redundancy: for any $s\ne0$, replacing $c_{\ell}$ by $sc_{\ell}$ in the
first group, $c_{L+j}$ by $s^{-1}c_{L+j}$ in the second, and $e_k$
by $s^{-1}e_k$ leaves every mode unchanged. This is the single
relation among the $d$ vertex gauge generators, leaving $d-1$
independent actions as in \cref{eq:canonical-charge-relation}.
The kernel rate already gives
the $u$ equations; the other arrays require weighted sums over the
opposite ordinary-point group.

At $Q_a$ choose a vanishing coordinate and write
\begin{equation}\label{eq:general-residue-data}
 \boldsymbol\omega=(E_a+O(s_a))ds_a,\qquad
 d\Omega_0=(\beta_a+O(s_a))ds_a,\qquad r_a=2\kappa_a/\beta_a.
\end{equation}
Seek $h_L,h_M$ with finite simple poles only at their respective
groups, of principal coefficients $r_a$, and expansions
\begin{equation}\label{eq:general-h-normalization}
 h_A=s_k^{-1}+a_{A,k}+O(s_k)\quad(P\to R_k),
 \qquad A\in\{L,M\},\quad1\leq k\leq N.
\end{equation}
One additive constant per function may be fixed. Their local constants
$a_{A,k}$ determine the third-group phase rates.

\begin{lemma}[Meromorphic interpolation]\label{lem:general-BA-interpolation}
The functions exist if and only if
\begin{equation}\label{eq:general-residue-criterion}
 \sum_{\ell=1}^{L}r_{\ell}E_{\ell}=\boldsymbol U,\qquad
 \sum_{j=L+1}^{K}r_jE_j=\boldsymbol U.
\end{equation}
They are unique up to additive constants, and
\begin{equation}\label{eq:general-residue-weights}
 \kappa_{\ell}=\tfrac12\operatorname{Res}_{Q_{\ell}}(h_Ld\Omega_0),\quad1\leq\ell\leq L,
 \qquad
 \kappa_j=\tfrac12\operatorname{Res}_{Q_j}(h_Md\Omega_0),\quad L+1\leq j\leq K.
\end{equation}
\end{lemma}
\begin{proof}
The finite residues of $h\omega_\alpha$ are $r_a(E_a)_\alpha$;
their sum at the punctures is $-U_\alpha$. Their vanishing is exactly
the compact-curve Mittag--Leffler obstruction \cite{Hartshorne1977}.
Two solutions differ by a global holomorphic function, hence a constant.
\end{proof}
The test is two $g$-dimensional linear systems. Generic groups of
$n\geq g$ points have rank $g$ and $n-g$ free principal coefficients.
Every coefficient must be nonzero to retain its modes. Special smaller
groups obey the same consistency test.

\subsubsection{The kernel convolution identity}
\label{sec:kernel-verification}

The interpolation functions turn the kernel into weighted BA identities,
coupling the three arrays. The underlying Szeg\H{o}-kernel composition
identity is given by Ben-Zvi and Biswas
\cite[Proposition~5.2.1]{BenZviBiswas2003}; we write its weighted BA form
with the puncture contributions explicit and give the residue proof in
our normalization.

\begin{lemma}[Convolution and weighted BA equations]
\label{lem:general-kernel-convolution}
For $h=h_A$ let $(a_0,a_1)=(1,L)$ or $(L+1,K)$ respectively. Then
\begin{equation}\label{eq:general-kernel-convolution}
 \sum_{k=1}^{N}\psi_k(P)\psi_k^\vee(Q)
 =(h(P)-h(Q))\Omega(P,Q)
 -2\sum_{a=a_0}^{a_1}\kappa_a\Omega(P,Q_a)\Omega(Q_a,Q).
\end{equation}
For $1\leq k\leq N$,
\begin{equation}\label{eq:general-weighted-BA}
\begin{aligned}
 i\dot\psi_k(P)&=(h(P)-a_{A,k})\psi_k(P)
       -2\sum_{a=a_0}^{a_1}\kappa_a\psi_k(Q_a)\Omega(P,Q_a),\\
 i\dot\psi_k^\vee(P)&=-(h(P)-a_{A,k})\psi_k^\vee(P)
       -2\sum_{a=a_0}^{a_1}\kappa_a\psi_k^\vee(Q_a)\Omega(Q_a,P).
\end{aligned}
\end{equation}
\end{lemma}
\begin{proof}
Apply the residue theorem in $T$ to
$h(T)\Omega(P,T)\Omega(T,Q)d\Omega_0(T)$.
The zeros of $d\Omega_0$ cancel its fixed divisor poles. The residues
at $P,Q,Q_a$ are $-h(P)\Omega(P,Q)$, $h(Q)\Omega(P,Q)$ and
$2\kappa_a\Omega(P,Q_a)\Omega(Q_a,Q)$. Each $R_k$ contributes
$\psi_k(P)\psi_k^\vee(Q)$, proving the convolution.
For the first weighted equation the finite principal parts cancel
by \cref{eq:general-residue-data}. At $R_m$ the residual leading
value is $-a_{A,m}\delta_{km}$ before the $a_{A,k}$ correction.
After that correction every leading value vanishes, so BA uniqueness
applies. The dual proof has the reversed diagonal sign.
\end{proof}

\subsubsection{Realization and finite completion}

For $1\leq\ell\leq L$, $1\leq j\leq M$, $1\leq k\leq N$,
define the phase coefficients and auxiliary phase rates by
\begin{equation}\label{eq:general-phase-values}
\begin{aligned}
 b_{\ell}^L&=-\tfrac12h_M(Q_{\ell}),&
 b_j^M&=\tfrac12h_L(Q_{L+j}),&
 b_k^N&=\tfrac12(a_{M,k}-a_{L,k}),\\
 \widetilde b_{\ell}^L&=-\tfrac12h_M(Q_{\ell}),&
 \widetilde b_j^M&=-\tfrac12h_L(Q_{L+j}),&
 \widetilde b_k^N&=-\tfrac12(a_{M,k}+a_{L,k}).
\end{aligned}
\end{equation}
Use the cycle-preserving transformation
\begin{equation}\label{eq:multipoint-phase-gauge}
 u_{\ell j}=e^{-it(\widetilde b_{\ell}^L-\widetilde b_j^M)}u_{\ell j}^{\rm BA},\quad
 v_{jk}=e^{-it(\widetilde b_j^M-\widetilde b_k^N)}v_{jk}^{\rm BA},\quad
 w_{k\ell}=e^{-it(\widetilde b_k^N-\widetilde b_{\ell}^L)}w_{k\ell}^{\rm BA},
\end{equation}
with reciprocal partner factors. Its three rates sum to zero in
every triad, preserving the mixing terms.

Regular multipoint data means the smooth-curve and divisor
conditions of \cref{sec:multipoint-BA-data}, distinct evaluation points
avoiding both divisors and the punctures, and nonzero weights and gauges.

\begin{theorem}[Main theorem, part I: Canonical BA realization]\label{thm:general-BA-realization}
\emph{Sufficiency.} For regular multipoint data, \cref{eq:general-residue-criterion} gives
a meromorphic solution of \cref{eq:canonical-H} through
\cref{eq:multipoint-mode-arrays,eq:general-phase-values,eq:multipoint-phase-gauge}.
\emph{Necessity.} For $g\geq1$ and $\boldsymbol U\ne0$, these residue
conditions are necessary for the displayed BA ansatz to solve the
canonical equations on an open Jacobian family.
\emph{Phase compatibility.} For preassigned $b^L,b^M,b^N$, such a BA solution
can be matched by constant-rate reciprocal mode gauges if and only if
\begin{equation}\label{eq:multipoint-phase-compatibility}
 2(b_{\ell}^L+b_j^M+b_k^N)
 =h_L(Q_{L+j})-h_M(Q_{\ell})+a_{M,k}-a_{L,k},
 \quad1\leq\ell\leq L,\ 1\leq j\leq M,\ 1\leq k\leq N.
\end{equation}
\end{theorem}
\begin{proof}
The kernel and weighted identities give the raw equations
\begin{equation}\label{eq:multipoint-raw-rates}
\begin{aligned}
 i\dot u_{\ell j}^{\rm BA}&=\sqrt2\sum_{k=1}^{N}\bar v_{jk}^{\rm BA}\bar w_{k\ell}^{\rm BA},\\
 i\dot v_{jk}^{\rm BA}&=(h_L(Q_{L+j})-a_{L,k})v_{jk}^{\rm BA}
              +\sqrt2\sum_{\ell=1}^{L}\bar u_{\ell j}^{\rm BA}\bar w_{k\ell}^{\rm BA},\\
 i\dot w_{k\ell}^{\rm BA}&=(-h_M(Q_{\ell})+a_{M,k})w_{k\ell}^{\rm BA}
              +\sqrt2\sum_{j=1}^{M}\bar u_{\ell j}^{\rm BA}\bar v_{jk}^{\rm BA},
\end{aligned}
\end{equation}
and negative partner equations. The phase transformation gives
\cref{eq:canonical-rates}. Comparing sums of the linear rates around
each triad proves \cref{eq:multipoint-phase-compatibility}; admissible
redistributions are the constant gauges of \cref{sec:canonical-phase-gauges}.

For necessity let $\mathsf H_\theta$ be the Hessian of $\log\theta$
and $\partial_v=\sum_{\alpha=1}^{g}v_\alpha\partial_{z_\alpha}$
for a constant direction $v$.
Fay's two-kernel identity is
\begin{equation}\label{eq:general-Fay-power}
 \Omega(P,Q)\Omega(Q,P)d\Omega_0(P)d\Omega_0(Q)
 =-B(P,Q)-\boldsymbol\omega(P)^{\trans}\mathsf H_\theta\boldsymbol\omega(Q),
\end{equation}
where $B$ is the normalized fundamental bidifferential
\cite[p.~26]{Fay1973}; see also \cite[(3.12)]{Korotkin2003}.
To obtain the diagonal BA product, use the theta form of the kernel.
With $\Delta=A(P)-A(Q)$ it has the reference-normalized evolution
\[
 \Omega(P,Q;t)=\Omega(P,Q;0)e^{-it(\Xi(P)-\Xi(Q))}
 \frac{\theta(z(t)+\Delta)\theta(z_0)}
      {\theta(z(t))\theta(z_0+\Delta)}.
\]
In a local coordinate $s$, write $\boldsymbol\omega=E_P\,ds$ and
$\partial_s\Xi=\Xi_s$. Its expansion as $Q\to P$ is
\[
\begin{aligned}
 \Omega(P,Q;t)\frac{d\Omega_0(Q)}{ds(Q)}
 &=\frac1{s(P)-s(Q)}+a(P)-it\Xi_s(P)\\
 &\quad+E_P^{\trans}\bigl(\nabla\log\theta(z)-\nabla\log\theta(z_0)\bigr)
 +O(s(P)-s(Q)),
\end{aligned}
\]
where $a(P)$ is the time-independent regular coefficient at $t=0$.
Differentiate by $i\partial_t$ and take the diagonal limit.
The kernel rate \cref{eq:general-kernel-rate} and
$i\dot z=\boldsymbol U$ give
\begin{equation}\label{eq:general-elementary-power}
 \left(\sum_{k=1}^{N}\psi_k(P)\psi_k^\vee(P)\right)d\Omega_0(P)
 =d\Xi(P)+\boldsymbol\omega(P)^{\trans}\mathsf H_\theta\boldsymbol U.
\end{equation}
Thus the diagonal identity follows from the kernel's regular part;
the off-diagonal products are given separately by
\cref{eq:general-Fay-power}.

Let $\mathcal Q_{\ell}=Q_{\ell}^L/\kappa_{\ell}$, where $Q_{\ell}^L$ is the first-group
power charge of \cref{eq:canonical-charges}. The mode definitions give
\[
 \mathcal Q_{\ell}=\sum_{k=1}^{N}\psi_k(Q_{\ell})\psi_k^\vee(Q_{\ell})
 +2\sum_{j=L+1}^{K}\kappa_j\Omega(Q_{\ell},Q_j)\Omega(Q_j,Q_{\ell}).
\]
Put $V_M=\sum_{j=L+1}^{K}r_jE_j-\boldsymbol U$ and let
$B_{\ell j}$ denote the coefficient of $B(Q_{\ell},Q_j)$ in the chosen
coordinates at $Q_{\ell},Q_j$. Substituting the two product identities,
with $r_j=2\kappa_j/\beta_j$, yields
\[
 \beta_{\ell}\mathcal Q_{\ell}
 =\Xi_s(Q_{\ell})-\sum_{j=L+1}^{K}r_jB_{\ell j}
       -\partial_{E_{\ell}}\partial_{V_M}\log\theta(z).
\]
The first two terms are independent of $z$. Since $\mathcal Q_{\ell}$ is
conserved and $\dot z=-i\boldsymbol U$, the ansatz on an open
Jacobian family implies the meromorphic identity
\[
 \partial_{\boldsymbol U}\partial_{E_{\ell}}\partial_{V_M}\log\theta=0.
\]
At a smooth point of the theta divisor its leading cubic pole has
coefficient
$2(\partial_{\boldsymbol U}\theta)(\partial_{E_{\ell}}\theta)
(\partial_{V_M}\theta)$. The theta divisor of a smooth connected
curve is irreducible, reduced and ample \cite{Mumford1984}.
The product therefore vanishes on its smooth locus only if one
factor vanishes identically along the divisor. A constant direction
everywhere tangent to this divisor lies in the Lie algebra of its
translation stabilizer, which is finite. The first factor cannot
vanish identically because $\boldsymbol U\ne0$, nor can the second:
holomorphic differentials have no common zero, so $E_{\ell}\ne0$.
It follows that $V_M=0$. This is the second residue equation.
A column charge gives the other residue equation. The linear rates
in \cref{eq:multipoint-raw-rates} then force the triad sums in
\cref{eq:multipoint-phase-compatibility}.
\end{proof}
Sufficiency applies also to genus zero and stationary Abel directions;
the nonzero-direction hypothesis is used only for necessity.

\begin{corollary}[Including prescribed evaluations]\label{cor:BA-finite-list}
Fix regular multipoint data and any finite list of ordinary evaluation
points in the two groups with nonzero principal coefficients.
Adjoining $g+1$ generic points to each group permits both residue
equations to be solved with all new coefficients nonzero.
If all $N$ regular constants of each interpolation function are also
prescribed, $g+N$ generic additional points per group suffice.
The enlarged arrays have the realization of
\cref{thm:general-BA-realization}.
\end{corollary}
\begin{proof}
For the residue equations choose $g+1$ columns such that every $g$
are independent. The solution set for any residual right side is an
affine line whose kernel direction has no zero coordinate. Avoid its
finitely many zero-coordinate parameter values.
For prescribed regular constants the obstruction space is
$H^1(\mathcal O(-\sum_{k=1}^{N}R_k))$, of dimension $g+N-1$.
Its dual consists of holomorphic and third-kind differentials with
at most simple puncture poles. Generic $g+N$ evaluation columns have
the same full-rank and nonzero-kernel property. Apply the same argument
to each group. In genus zero the residue system is empty.
\end{proof}

With $a_L,a_M$ added evaluation points the resulting system has
\[
 (L+a_L)(M+a_M)+N(L+M+a_L+a_M)
\]
canonical mode pairs. The original evaluations are retained as
entries of these enlarged arrays; their equations include the new
modes. Completion constructs a coupled enlargement and generally
changes $h_L,h_M$ and the original modes' linear phase coefficients.

Here $a_L=a_M=g+1$ for the basic completion, and
$a_L=a_M=g+N$ when all regular puncture constants are prescribed.

\subsection{From canonical initial modes to BA solutions}
\label{sec:BA-spectral-converse}
\label{sec:BA-reconstruction-flows}

We identify the curve carried by a prescribed realization and recover
BA solutions from generic canonical initial modes, completing the
reverse direction of the correspondence.

The interpolation functions $h_L,h_M$ give spectral coordinates on
the prescribed curve. With several punctures, identifying that curve
with the normalized characteristic curve requires the birationality
condition in \cref{prop:BA-prescribed-spectral-curve}.

\subsubsection{Identifying the prescribed spectral curve}

\begin{proposition}[The spectral curve of a prescribed realization]
\label{prop:BA-prescribed-spectral-curve}
For the phase coefficients \cref{eq:general-phase-values}, put
\begin{equation}\label{eq:BA-prescribed-spectral-map}
 \nu(P)=-\tfrac12(h_L(P)+h_M(P)),\qquad
 \lambda(P)=\tfrac12(h_M(P)-h_L(P)).
\end{equation}
The characteristic polynomial of the canonical pencil vanishes on
this map. If the map is birational onto its image (it induces an isomorphism
of function fields), $\mathcal C$ normalizes the full characteristic curve. A sufficient condition is
that one puncture has a finite limiting $\lambda$ distinct from the
other punctures, in particular pairwise distinct $b_k^N$.
For $N=1$ this condition is automatic.

Let $\Gamma$ be the projective plane characteristic curve. Write
$d=L+M+N$ and $G=LM+MN+NL-d+1$. On this birational locus
the total singularity defect beyond the ordinary three infinity
multiple points is
\begin{equation}\label{eq:BA-prescribed-genus-defect}
 \sum_{q\in\operatorname{Sing}(\Gamma)}\delta_q
 -\binom L2-\binom M2-\binom N2=G-g.
\end{equation}
For distinct phases within groups the additional defect is affine;
coincident phases can place it at infinity, as in \cref{sec:BA-al-functions}.
\end{proposition}
\begin{proof}
Let $\mathsf X_{\rm BA}$ be the block matrix in
\cref{eq:canonical-pencil} with the raw arrays
\cref{eq:multipoint-mode-arrays} and the phase diagonals
\cref{eq:general-phase-values}. The explicit vector is
\begin{equation}\label{eq:BA-prescribed-eigenvector}
 r(P)=\begin{pmatrix}
 (-\kappa_{\ell}\Omega(Q_{\ell},P)/c_{\ell})_{\ell=1}^{L}\\
 (c_{L+j}\Omega(Q_{L+j},P))_{j=1}^{M}\\
 (e_k\psi_k^\vee(P))_{k=1}^{N}
 \end{pmatrix}.
\end{equation}
The first two block rows of
$(\mathsf X_{\rm BA}+\nu(P)\mathsf D)r=\lambda(P)r$
are the convolution identities for $h_M,h_L$. The third follows by
subtracting their dual weighted equations:
\[
 \sum_{\ell=1}^{L}\kappa_{\ell}\psi_k^\vee(Q_{\ell})\Omega(Q_{\ell},P)
 -\sum_{j=L+1}^{K}\kappa_j\psi_k^\vee(Q_j)\Omega(Q_j,P)
 =(\lambda(P)-b_k^N)\psi_k^\vee(P).
\]
The phase transformation \cref{eq:multipoint-phase-gauge} conjugates
the pencil and vector by the same diagonal matrix, preserving this
spectral equation. Adding a scalar origin $\rho\Id_d$ to the
pencil shifts $\lambda$ by $\rho$ and preserves the eigenvectors.

The poles of $\nu$ are simple, at all $Q_a,R_k$. Their branches obey
\[
 \lambda-\nu\to-2b_{\ell}^L\ (P\to Q_{\ell}),\qquad
 \lambda+\nu\to-2b_j^M\ (P\to Q_{L+j}),\qquad
 \lambda\to b_k^N\ (P\to R_k).
\]
If one $R_k$ has a unique finite limiting ordinate, it is the sole
point over its image in the normalization. Its simple pole of $\nu$
makes this point unramified, so the covering degree is one.
The same argument applies to a uniquely separated outer branch.
For a birational map, $\deg\nu=d$ implies that the minimal polynomial
of $\lambda$ over $\C(\nu)$ has degree $d$. The monic characteristic
polynomial has that degree and vanishes, so it is this minimal
polynomial. The plane degree is $d$; subtracting the ordinary
infinity defects gives \cref{eq:BA-prescribed-genus-defect}.
\end{proof}

\paragraph{Prescribed curves and spectral fibres.}
Fix the marked spectral curve and its functions $\lambda,\nu$,
with distinct phase intercepts. On the regular framed spectral
correspondence, a smooth genus-$G$ level has $G$ divisor-class
parameters and $d-1$ independent mode gauges, hence dimension
$G+d-1=p$. A prescribed genus-$g$ normalization instead supplies
$g+d-1$ parameters after identifying equivalent meromorphic frames.
Relative to an expected-dimensional spectral fibre, this realized
family has codimension $G-g$. It is an invariant normalization-based
family on a special level when $g<G$, and is open in the regular
fibre when $g=G$.

This family has a concrete spectral-sheaf interpretation. Separate
the forced ordinary infinity points, and let
$n:\mathcal C\to\Gamma_{\mathrm{sp}}$ normalize the remaining
integral spectral model. At the reference time the components of
\cref{eq:BA-prescribed-eigenvector} are sections of the line bundle
\[
 \mathscr L=\mathcal O\left(\mathscr D^\vee+
                         \sum_{a=1}^{K}Q_a\right).
\]
Here $\mathcal O(E)$ denotes the line bundle associated with a divisor
$E$; thus $\mathscr L$ allows poles along $\mathscr D^\vee$ and at
the evaluation points $Q_a$, with the indicated multiplicities.
Subtracting the full pole fibre of $\nu$ gives
$\mathscr L(-\nu^{-1}(\infty))=\mathcal O(\mathscr D^\vee-
\sum_{k=1}^{N}R_k)$, with both cohomology groups zero.
Its leading-value sequence shows that the $d$ components form a
basis and that $\nu_*\mathscr L\simeq\mathcal O^d$.
Multiplication by $\lambda$ induces a map
$\mathcal O^d\to\mathcal O(1)^d$, represented by the pencil.
The component spectral sheaf on $\Gamma_{\mathrm{sp}}$ is $n_*\mathscr L$.
Thus the realized family is the stratum obtained by pushing line
bundles forward from the normalization, with the mode framings
retained. Its divisor-class dimension is $g$.
The full compactified spectral fibre can also contain sheaves with
singularity gluing data \cite{AdamsHarnadHurtubise1990II,Beauville1990}.
The general solution of
\cref{thm:BA-spectral-converse} reconstructs the curve belonging to
generic initial modes. For repeated phases its active dimensions,
moment data and restored block bases are accounted for separately
in Appendix~\ref{app:BA-active-converse}.

With several punctures, the interpolation functions can instead factor
through a covering; curve identification then requires birationality.
The stronger completion in \cref{cor:BA-finite-list} permits distinct
differences $a_{M,k}-a_{L,k}$ to be prescribed, giving the sufficient
branch separation used in the proposition.
For $N=1$, the product coordinates $\lambda-\nu,\lambda+\nu$ have
degrees $M+1,L+1$; their image has bidegree $(L+1,M+1)$ and arithmetic
genus $LM$, recovering the one-point defect $LM-g$.

\subsubsection{Reconstruction from generic canonical data}

The spectral eigenvectors now recover every mode, so that the BA
formulas give solutions from canonical initial data rather than only
from data chosen in advance on a curve.

\begin{theorem}[Main theorem, part II: Generic BA reconstruction]\label{thm:BA-spectral-converse}
Fix positive $L,M,N$ and arbitrary phase coefficients. Zariski-generic
canonical initial modes reconstruct scalar multipoint BA bases and
their kernel on a smooth normalized active spectral curve.
For distinct phases within groups all directions are active and the
genus is $G=LM+MN+NL-L-M-N+1$; the last group supplies $N$
punctures, and the ordinary and dual divisors have degree $G+N-1$.
For repeated phases, use constant canonical bases within equal-phase
blocks and remove inactive directions (coordinates with zero incoming
and outgoing couplings) as in
Appendix~\ref{app:BA-active-converse}. Apply the construction with
active dimensions $L_a,M_a,N_a$ and genus
\cref{eq:BA-active-genus}, then restore the zero directions and bases.
The original coordinates can then be constant linear combinations
of the active scalar BA evaluations.
\end{theorem}
\begin{proof}
Work in the $H_0$ phase gauge of \cref{eq:canonical-phase-gauge},
omitting its tildes during this proof. The pencil retains the diagonal
blocks $-2\mathsf B_L,-2\mathsf B_M,\mathsf B_N$, and its time matrix is
\begin{equation}\label{eq:multipoint-H0-time}
 \mathsf T_0(\nu)=
 \begin{pmatrix}0&0&\bar{\boldsymbol w}^{\trans}\\
 0&0&-\boldsymbol v\\
 -\boldsymbol w&\bar{\boldsymbol v}^{\trans}&-\nu\Id_N
 \end{pmatrix}.
\end{equation}
For distinct phases the regular normalized curve has branches
\begin{equation}\label{eq:BA-converse-infinity}
\begin{aligned}
 \lambda&=\nu-2b_{\ell}^L+O(\nu^{-1})&&\text{at }Q_{\ell},\quad1\leq\ell\leq L,\\
 \lambda&=-\nu-2b_j^M+O(\nu^{-1})&&\text{at }Q_{L+j},\quad1\leq j\leq M,\\
 \lambda&=b_k^N+O(\nu^{-1})&&\text{at }R_k,\quad1\leq k\leq N.
\end{aligned}
\end{equation}
The nonempty regular locus, including repeated-phase active
normalizations, is proved in \cref{prop:BA-active-regularity}.
Put $s_k^{-1}=-\nu$ at $R_k$. The functions
\begin{equation}\label{eq:BA-converse-h}
 h_L=-\lambda-\nu,\qquad h_M=\lambda-\nu
\end{equation}
have the required poles and satisfy
$h_M(Q_{\ell})=-2b_{\ell}^L$, $h_L(Q_{L+j})=2b_j^M$,
$a_{L,k}=-b_k^N$, $a_{M,k}=b_k^N$.
The interpolation and phase conditions thus come from the initial
spectral curve, rather than restricting initial modes further.

Let $f_k$ be the last $N$ coordinate vectors, $\mathsf J$ their
coordinate projection, and $a=\sum_{k=1}^{N}f_k$.
Choose right and left spectral eigenvectors $r,r^{\vee}$ with
$i\dot r=-\mathsf T_0r$, $i\dot r^{\vee}=r^{\vee}\mathsf T_0$ and reference
normalizations $a^{\trans}r(P,0)=r^{\vee}(P,0)a=1$. Define
\begin{equation}\label{eq:BA-converse-functions}
 \psi_k(P)=r^{\vee}(P)f_k,\qquad
 \psi_k^\vee(P)=f_k^{\trans}r(P),\qquad
 \Omega(P,Q)=-\frac{r^{\vee}(P)r(Q)}{\nu(P)-\nu(Q)}.
\end{equation}
Since $\mathsf T_0(\nu(P))-\mathsf T_0(\nu(Q))
=-(\nu(P)-\nu(Q))\mathsf J$, this kernel satisfies
\cref{eq:general-kernel-rate}. The eigenvalue equations remove apparent
poles at distinct points of an equal-$\nu$ fiber. At $R_k$, the leading
eigenvectors are its coordinate direction and the temporal equation
gives exactly \cref{eq:general-BA-asymptotics}.

For the divisor count use the positive-degree spectral component
line bundle $\mathscr L$, whose pushforward under $\nu$ is
$\mathcal O^d$ \cite{Beauville1990}. Its degree is $G+d-1$.
The last $N$ components vanish simply at the $L+M$ outer infinity
points. Their remaining common BA divisor has degree $G+N-1$.
More explicitly, put $\mathscr E=\mathscr L(-\nu^{-1}(\infty))$.
Here $H^0$ and $H^1$ denote zeroth and first sheaf-cohomology
groups, respectively, whereas $H_0$ denotes the Hamiltonian in
\cref{eq:canonical-H0}.
Then $H^0(\mathscr E)=H^1(\mathscr E)=0$, since its pushforward is
$\mathcal O(-1)^d$. Adding the $N$ punctures gives an exact
leading-value isomorphism
$H^0(\mathscr E(\sum_{k=1}^{N}R_k))\simeq\C^N$.
This proves the required BA uniqueness. The transpose module gives
the dual statement; the analytic infinity justification for repeated
phases is given in Appendix~\ref{app:BA-active-converse}.
The duality differential is
\begin{equation}\label{eq:BA-converse-duality}
 d\Omega_0=-\frac{d\nu}{r^{\vee}(P)r(P)}.
\end{equation}
It is regular at the outer branches, has the required double poles
at $R_k$, and its apparent projection-branch poles cancel with $d\nu$.

The reference normalization gives a concrete recovery chart. Set
\begin{equation}\label{eq:BA-converse-constants}
\begin{aligned}
 c_{\ell}&=\sum_{k=1}^{N}w_{k\ell}(0),&
 \bar c_{\ell}&=\sum_{k=1}^{N}\bar w_{k\ell}(0),&
 \kappa_{\ell}&=c_{\ell}\bar c_{\ell},\quad1\leq\ell\leq L,\\
 c_{L+j}&=\sum_{k=1}^{N}v_{jk}(0),&
 \bar c_{L+j}&=\sum_{k=1}^{N}\bar v_{jk}(0),&
 \kappa_{L+j}&=c_{L+j}\bar c_{L+j},\quad1\leq j\leq M.
\end{aligned}
\end{equation}
The barred constants are independent partner normalizations, not
complex conjugates; the weights are products of initial sums.
On the open chart where these sums are nonzero, outer eigenvector
expansions give, for $1\leq\ell\leq L$, $1\leq j\leq M$ and $1\leq k\leq N$,
\begin{equation}\label{eq:BA-converse-recovery}
\begin{aligned}
 \psi_k^\vee(Q_{\ell})&=w_{k\ell}/c_{\ell},&
 \psi_k(Q_{\ell})&=\bar w_{k\ell}/\bar c_{\ell},\\
 \psi_k(Q_{L+j})&=v_{jk}/c_{L+j},&
 \psi_k^\vee(Q_{L+j})&=\bar v_{jk}/\bar c_{L+j},\\
 \Omega(Q_{\ell},Q_{L+j})&=\frac{u_{\ell j}}{\sqrt2\bar c_{\ell}\bar c_{L+j}},&
 \Omega(Q_{L+j},Q_{\ell})&=-\frac{\bar u_{\ell j}}{\sqrt2c_{\ell}c_{L+j}}.
\end{aligned}
\end{equation}
For example $r_{\ell}\sim-\nu/c_{\ell}$ at $Q_{\ell}$, and the last eigenvector
rows give $f_k^{\trans}r=w_{k\ell}/c_{\ell}$. The opposite outer block then
gives the kernel recovery, with slope difference two. At an outer
point, in coordinate $1/\nu$, the relevant $h$ has principal
coefficient $-2$ and $d\Omega_0$ has coefficient $-\kappa_a$, so
the recovered weights agree with \cref{eq:general-residue-weights}.
Thus \cref{eq:multipoint-mode-arrays} recovers every mode, with
$e_k=1$. Weighted reference anchors give other generic charts.
The forward theorem and uniqueness of the polynomial ODE identify
the evolution. Applying the phase transformation
\cref{eq:canonical-phase-gauge} recovers the solutions generated by
the Hamiltonian $H$ in \cref{eq:canonical-H}.

For repeated phases apply this argument in the constant active bases
of Appendix~\ref{app:BA-active-converse}. Their resolved infinity
directions supply the same normalized leading values. Restore
inactive zero directions and invert the canonical bases afterwards.
\end{proof}

For comparison with a prescribed realization, its fixed $e_k$ gauges
are set to one by the puncture-vertex canonical gauge transformations.
The common rescaling of \cref{sec:general-BA-interpolation} accounts
for the single redundancy among all vertex normalization constants.
The reference partition of unity in \cref{sec:multipoint-BA-data}
then makes \cref{eq:BA-converse-constants} recover the original
weights and gauges directly.

\subsection{Commuting Hamiltonians, Abel directions and mode phases}
\label{sec:BA-higher-flows}

The commuting spectral Hamiltonians evolve the same BA data in
different directions. We compute their puncture times and constant
mode phases by meromorphic interpolation. Jacobian linearization
and its principal-part description are classical
\cite{Griffiths1985,Beauville1990,AdamsHarnadHurtubise1993}; the normalization here gives the
rates for these canonical arrays. With several punctures, a fixed
choice of time representatives replaces the one-point gap convention.
Thus, on each fixed spectral level, the commuting Hamiltonians capture
the full multidimensional BA evolution: their flows span every
direction of the Abelian vector $z$, with the corresponding mode
phases, rather than only the single direction selected by our original
Hamiltonian.

\subsubsection{BA times and endpoint phases}

Recall that $\mathsf L(\nu)=\mathsf X+\nu\mathsf D$ is the Lax
matrix, with mode-dependent coefficient matrix $\mathsf X$ and fixed
slope matrix $\mathsf D$ from \cref{eq:canonical-X,eq:canonical-pencil}.
For a fixed finite spectral value $\nu_0$ and an integer trace order
$r\geq2$, define
\begin{equation}\label{eq:BA-shifted-flow}
\begin{aligned}
 H_{r,\nu_0}&=\tfrac1r\tr(\mathsf X+\nu_0\mathsf D)^r,&
 \mathsf G&=(\mathsf X+\nu_0\mathsf D)^{r-1},\\
 \mathsf T_{r,\nu_0}(\nu)&=
 \frac{\mathsf G-\mathsf L(\nu)^{r-1}}{\nu-\nu_0},&
 q(P)&=\frac{\lambda(P)^{r-1}}{\nu(P)-\nu_0}.
\end{aligned}
\end{equation}
Here $H_{r,\nu_0}$ is the spectral Hamiltonian, $\mathsf G$ its matrix
gradient, and $\mathsf T_{r,\nu_0}$ its Lax time matrix.
The scalar meromorphic function $q$ uses the spectral coordinates
$\lambda(P),\nu(P)$ of a point $P$ on the normalized curve and
determines the local BA exponential rates.
The matrix quotient is polynomial, and
$i\dot{\mathsf L}=[\mathsf L,\mathsf T_{r,\nu_0}]$.
All coefficients of these Hamiltonians in $\nu_0$ commute.

Let $\operatorname{pp}_P$ denote the principal part at $P$.
Choose a puncture divisor $D_R=\sum_{k=1}^{N}m_kR_k$ with
$H^1(\mathcal O(D_R))=0$. For $g\geq1$ a safe choice is
$m_1=2g-1$, $m_k=1$ for $2\leq k\leq N$.
The space $\mathcal V_R$ of these puncture principal parts has
quotient
\[
 \mathcal V_R/
 \operatorname{pp}H^0(\mathcal O(D_R))\simeq H^1(\mathcal O),
\]
of dimension $g$. Fix a $g$-dimensional complement $\mathcal G$
to the meromorphic image. It selects representatives of the Abel
times. When $\boldsymbol U\ne0$, it may be chosen to contain the
original vector $(s_1^{-1},\ldots,s_N^{-1})$, retaining our first
time among those representatives. For $N=1$, the ordinary
Weierstrass gap powers give this complement. The local parameters
and $\mathcal G$ are fixed throughout the hierarchy.

\begin{lemma}[Meromorphic time normalization]\label{lem:BA-higher-normalization}
There is a unique meromorphic function $\mathfrak f$ with outer
principal parts
\[
 \operatorname{pp}_{Q_a}\mathfrak f=-\operatorname{pp}_{Q_a}q,
 \qquad1\leq a\leq K,
\]
whose further poles lie in $D_R$, whose puncture principal-part
vector belongs to $\mathcal G$, and whose regular constant at $R_1$
is zero. Its coefficients are obtained by a finite linear
meromorphic interpolation problem.
For $g=0$ take no puncture poles and normalize $\mathfrak f(R_1)=0$.
\end{lemma}
\begin{proof}
The vanishing of $H^1(\mathcal O(D_R))$ extends arbitrary outer
principal parts with extra poles in $D_R$
\cite[Chapter~IV]{Hartshorne1977}. Subtract a meromorphic function
supported on $D_R$ to remove the principal-part component outside
$\mathcal G$, then subtract the regular constant at $R_1$.
A difference has principal parts in both the meromorphic image and
its complement, hence has no poles. Its normalized constant makes
it zero. On the sphere the outer parts extend without extra poles.
\end{proof}

Define the time polynomial and vertex phase rates by
\begin{equation}\label{eq:BA-higher-time-phase}
\begin{aligned}
 p_k&=-\operatorname{pp}_{R_k}\mathfrak f,&
 \varphi_k^N&=\operatorname{reg}_{R_k}(q+\mathfrak f),&&1\leq k\leq N,\\
 \varphi_{\ell}^L&=(q+\mathfrak f)(Q_{\ell}),&&&1\leq\ell\leq L,\\
 \varphi_j^M&=(q+\mathfrak f)(Q_{L+j}),&&&1\leq j\leq M.
\end{aligned}
\end{equation}
Here $\operatorname{reg}$ means the constant Laurent coefficient.
The outer sums are regular by construction. At $R_k$,
$q=O(\nu^{-1})$, since $\lambda$ is finite there, and
$\varphi_1^N=0$. These numbers depend only on the fixed marked
curve and the chosen time normalization. Changing the complement
$\mathcal G$ changes the mode phases by power-charge flows while
preserving the Abel direction.

\begin{theorem}[Main theorem, part III: The spectral BA hierarchy]\label{thm:BA-higher-flows}
Every prescribed realization in \cref{thm:general-BA-realization}
evolves under $H_{r,\nu_0}$ by the same multipoint BA bases and
kernel, with ordinary and dual puncture exponentials
$e^{-itp_k}$ and $e^{itp_k}$.
Use \cref{eq:multipoint-mode-arrays} with reciprocal gauges, for
$1\leq\ell\leq L$, $1\leq j\leq M$ and $1\leq k\leq N$,
\begin{equation}\label{eq:multipoint-higher-gauges}
 c_{\ell}(t)=c_{\ell} e^{it\varphi_{\ell}^L},\quad
 c_{L+j}(t)=c_{L+j}e^{-it\varphi_j^M},\quad
 e_k(t)=e_k e^{-it\varphi_k^N}.
\end{equation}
Thus the explicit mode factors, relative to the unphased BA formulas,
are
\begin{equation}\label{eq:multipoint-higher-phases}
 u_{\ell j}:e^{-it(\varphi_{\ell}^L-\varphi_j^M)},\qquad
 v_{jk}:e^{-it(\varphi_j^M-\varphi_k^N)},\qquad
 w_{k\ell}:e^{-it(\varphi_k^N-\varphi_{\ell}^L)}.
\end{equation}
Partner factors are inverse. Finite constant linear combinations
of Hamiltonians add their $p_k$ and phase rates.

On a birational prescribed spectral level, the family with
$2\leq r\leq d=L+M+N$ and varying $\nu_0$ spans every Jacobian
direction. Finitely many polynomial spectral coefficients therefore
suffice to generate all its Abel translations.
\end{theorem}
\begin{proof}
The right eigenvector is \cref{eq:BA-prescribed-eigenvector}.
A compatible left vector is
\[
 r^{\vee}(P)=\bigl((-c_{\ell}\Omega(P,Q_{\ell}))_{\ell=1}^{L},
 (\kappa_{L+j}\Omega(P,Q_{L+j})/c_{L+j})_{j=1}^{M},
 (\psi_k(P)/e_k)_{k=1}^{N}\bigr).
\]
Their convolution identities give the spectral equations and
\begin{equation}\label{eq:BA-higher-initial-kernel}
 r^{\vee}(P)r(Q)=-(\nu(P)-\nu(Q))\Omega(P,Q).
\end{equation}
Evolve the vectors under
$i\dot r_H=-\mathsf T_{r,\nu_0}r_H$ and
$i\dot r_H^{\vee}=r_H^{\vee}\mathsf T_{r,\nu_0}$.
Near every infinity branch the eigenvalue equation gives the
convergent local factorization
\begin{equation}\label{eq:BA-higher-local-factorization}
 r_H=e^{-itq(P)}\mathsf V(t,\nu(P))r(P,0),\qquad
 i\dot{\mathsf V}=-\frac{\mathsf G}{\nu-\nu_0}\mathsf V,\qquad
 \mathsf V(0)=\Id.
\end{equation}
Since $\mathsf V=\Id+O(\nu^{-1})$, every leading coordinate direction
is preserved, including coincident-phase branches. At finite points
the original polynomial time matrix preserves the fixed divisor
poles; the local factorization is used only near infinity.

Dress the vectors reciprocally by
$\widetilde r=e^{-it\mathfrak f}r_H$,
$\widetilde r^{\vee}=e^{it\mathfrak f}r_H^{\vee}$.
At $Q_{\ell},Q_{L+j}$ the polar exponents cancel and the right leading
coefficients acquire $e^{-it\varphi_{\ell}^L}$,
$e^{-it\varphi_j^M}$. Their initial coefficients
$-\nu/c_{\ell}$ and $c_{L+j}\nu/\kappa_{L+j}$ give the first two gauges.
At $R_k$ the leading dual factor is
$e^{itp_k}e^{-it\varphi_k^N}$, giving $e_k(t)$.
After these gauges the leading vectors are normalized at all
punctures. BA uniqueness identifies every ordinary and dual basis
function. Each outer component has its fixed divisor, its one
additional simple outer pole, the calculated residue and zero
leading vector at all punctures, so the same argument identifies
the kernel components.

No finite equal-$\nu$ poles are introduced: the derivative of
$r_H^{\vee}(P)r_H(Q)$ is divisible by $\nu(P)-\nu(Q)$ because the
time matrix is polynomial. Initially removable poles remain
removable. Reciprocal dressing is regular there and leaves the
diagonal pairing unchanged. Thus $d\Omega_0$ and the weights are
fixed. The spectral equations recover all arrays as in
\cref{eq:BA-converse-recovery}, proving their higher-flow formulas.

For spanning choose an unramified finite fiber $\nu=\nu_0$
with $d$ distinct $\lambda$ values. Birationality guarantees such
fibers. For a holomorphic differential $du=w(P)d\nu$, the residue
theorem for $(q+\mathfrak f)du$ gives
\[
 \sum_{k=1}^{N}\operatorname{Res}_{R_k}(p_kdu)
 =\sum_{h=1}^{d}\lambda(P_h)^{r-1}w(P_h).
\]
A differential annihilating all these directions has zero right side
for $2\leq r\leq d$. The $r=1$ relation follows from residues of
$du/(\nu-\nu_0)$. Vandermonde inversion forces $w$ to vanish on
every such fiber, hence $du=0$. Residue duality proves spanning;
the normalized Jacobian has dimension $g$, so $g$ independent
directions can be selected from this family (none are needed for
$g=0$). Expanding in $\nu_0$ also permits the selection from $g$
polynomial spectral coefficients. All these directions belong to
the Jacobian of the normalization.
\end{proof}

On each fixed spectral level the chosen time coefficients and phase
rates are constants. Linear combinations selecting a normalized Abel
direction depend on that level; this is not a globally fixed
polynomial generator for every separately normalized puncture time.
If the prescribed curve map has covering degree greater than one,
the distinct-$\lambda$ fiber argument does not apply; the spanning
statement concerns the birational curve identified in
\cref{prop:BA-prescribed-spectral-curve}.

The vertex phases are generated by the diagonal power charges.
Explicitly,
\[
\begin{aligned}
 H_{\rm gauge}={}&
 \sum_{\ell=1}^{L}\varphi_{\ell}^L
 \left(\sum_{j=1}^{M}I^u_{\ell j}-\sum_{k=1}^{N}I^w_{k\ell}\right)
 +\sum_{j=1}^{M}\varphi_j^M
 \left(\sum_{k=1}^{N}I^v_{jk}-\sum_{\ell=1}^{L}I^u_{\ell j}\right)\\
 &+\sum_{k=1}^{N}\varphi_k^N
 \left(\sum_{\ell=1}^{L}I^w_{k\ell}-\sum_{j=1}^{M}I^v_{jk}\right).
\end{aligned}
\]
Subtracting this fixed-coefficient Hamiltonian removes the framing
phases on that level. Treating its coefficients as globally variable
spectral functions would introduce derivatives of those coefficients.

\begin{corollary}[Higher flows from generic initial modes]
\label{cor:BA-higher-reconstruction}
The reconstructed data of \cref{thm:BA-spectral-converse} solve the
commuting flows with \cref{eq:BA-higher-time-phase,eq:multipoint-higher-gauges}.
For repeated phases use the same active bases and restore the original
coordinates after BA evolution.
\end{corollary}
\begin{proof}
Every spectral Hamiltonian is invariant under equal-phase block
conjugations, so their moment matrices and adapted bases remain
constant. Inactive summands stay uncoupled under every matrix power.
Apply the forward theorem to the active data and restore the bases.
\end{proof}

\subsubsection{Gauge symmetries and Abel translations}
\label{sec:gauge-Abel-symmetries}

We interpret the conserved charges and spectral flows as symmetries,
so that their roles in the integral count and the BA phase data can
be distinguished. A conserved function $Q$ generates a canonical
symmetry because $\{Q,H\}=0$ implies that its Hamiltonian flow
commutes with the time evolution. This is the Hamiltonian form of
Noether's correspondence.
For the resonant three-wave system, reduction by mode-phase symmetries
and reconstruction of the geometric and dynamical phases were developed
by Alber, Luther, Marsden and Robbins
\cite[Sections~2--4]{AlberLutherMarsdenRobbins1998}.

For arbitrary $L,M,N$, the power charges in
\cref{eq:canonical-charges} generate the constant reciprocal gauges
\begin{equation}\label{eq:general-mode-gauge-action}
\begin{aligned}
 u_{\ell m}&\longmapsto e^{\theta_\ell^L-\theta_m^M}u_{\ell m},\\
 v_{mn}&\longmapsto e^{\theta_m^M-\theta_n^N}v_{mn},\\
 w_{n\ell}&\longmapsto e^{\theta_n^N-\theta_\ell^L}w_{n\ell},
 \qquad1\leq\ell\leq L,\quad1\leq m\leq M,\quad1\leq n\leq N.
\end{aligned}
\end{equation}
The barred partners have inverse factors. The $\theta$ parameters
are constant complex numbers; their common shift acts trivially,
leaving $d-1$ independent gauges. Each cubic interaction is invariant
because its three exponents sum to zero.

On the regular spectral locus, the ambient integral count separates these gauge generators
from the remaining spectral integrals:
\begin{equation}\label{eq:gauge-translation-count}
 p=(d-1)+G,\qquad d=L+M+N,
 \qquad p=LM+MN+NL,
\end{equation}
where $G$ is the regular spectral genus in
\cref{eq:canonical-genus-preview}. Locally, fixing the $d-1$ independent
charges and quotienting their gauge actions leaves $G$ Hamiltonian
degrees of freedom. Under eigenvector-divisor linearization, the
commuting isospectral flows act as Abel translations
\cite{Beauville1990,Vanhaecke2001}. The autonomous Hamiltonian selects
the time direction within this commuting family; it is included in the
complete integral set of \cref{thm:canonical-independence}.

When phases coincide, the moment matrices
\cref{eq:canonical-block-moment} extend the diagonal gauges to
continuous changes of basis within equal-phase groups.
Their principal-block invariants provide the additional integrals
counted in \cref{eq:canonical-spectral-deficit}.

On a prescribed birational BA realization,
\cref{thm:BA-higher-flows} supplies the translations of its Jacobian
and the accompanying endpoint phases. The power-charge flows change
the multiplicative mode normalizations. These give the two symmetry
actions visible in the BA formulas; the hyperelliptic divisor lift in
\cref{sec:mumford-gauge-lift} makes their distinction explicit.

\section{Correspondences and explicit specializations}
\label{sec:correspondences-specializations}

\Cref{sec:BA-correspondence} established the general BA solution of
the canonical systems defined in \cref{sec:canonical}. We now place
these systems in context by comparing them with other known
integrable systems admitting BA solutions. We develop the Hamiltonian
and operator correspondences, followed by particularly explicit
one-point and hyperelliptic specializations of the general theory.

\subsection{Correspondence with quartic moment-map systems}
\label{sec:AHH-comparison}

We compare the cubic modes with a quartic canonical lift to explain
how the same matrix evolution can have different interaction orders.
The moment-map construction of Adams, Harnad and Previato pulls
matrix Hamiltonians back through bilinear maps; Adams, Harnad and
Hurtubise integrate
both the matrix flows and their canonical lifts using spectral line
bundles and generalized BA functions
\cite{AdamsHarnadPreviato1988,AdamsHarnadHurtubise1990II}.
Wu and Geng obtain a related quartic canonical system by nonlinearizing
the ordinary and adjoint $3\times3$ AKNS spectral problems, prove its
Liouville integrability and reconstruct three-wave fields through
bilinear eigenfunction pairings \cite[Sections~III--V]{WuGeng1999}.
We relate a quartic canonical realization to our cubic system through
a shared spectral matrix evolution, while identifying the different
mode variables and matrix Poisson brackets underlying the two
descriptions. A BA convolution identity supplies the functional
connection: weighted bilinear sums of intermediate evaluations
collapse to individual endpoint BA functions, explaining the change
from quartic to cubic interaction expressions.

\subsubsection{A quartic canonical parent}

The moment-map framework of Adams, Harnad and Previato
\cite{AdamsHarnadPreviato1988} generates many Hamiltonian systems. Because
its matrix entries are bilinear in the canonical modes, polynomial
matrix Hamiltonians pull back to expressions of even degree in those
modes. The cubic Hamiltonian considered here therefore requires a
different canonical realization.

Let $\mathsf F(s),\mathsf G(s)\in\C^{q\times d}$ be matrices of
canonical modes evolving in time $s$, in the moment-map framework
of Adams, Harnad and Previato, whose flows were integrated by Adams,
Harnad and Hurtubise
\cite{AdamsHarnadPreviato1988,AdamsHarnadHurtubise1990II}.
Their brackets are
$\{F_{rh},G_{jk}\}=-i\delta_{rj}\delta_{hk}$ for
$1\leq r,j\leq q$, $1\leq h,k\leq d$, and all other basic brackets
zero. The entries of $\mathsf G$ are independent partners of those
of $\mathsf F$. Set
\begin{equation}\label{eq:AHH-quartic-parent}
 \mathsf M=\mathsf G^{\trans}\mathsf F,\qquad
 \mathcal H_4=\tr(\mathsf D\mathsf M^2),\qquad
 \mathsf C=\mathsf D\mathsf M+\mathsf M\mathsf D.
\end{equation}
Here $\mathsf D$ is the diagonal matrix in
\cref{eq:canonical-pencil}; $\mathcal H_4$ is a selected pullback in
the moment-map framework of \cite{AdamsHarnadPreviato1988}, rather
than a separately distinguished model in that paper.
In scalar modes its full dependence is
\begin{equation}\label{eq:AHH-quartic-components}
 \mathcal H_4=
 \sum_{h=1}^{d}\sum_{k=1}^{d}\sum_{r=1}^{q}\sum_{j=1}^{q}
 D_{hh}G_{rh}F_{rk}G_{jk}F_{jh}.
\end{equation}
Each interaction therefore contains four mode factors, in contrast to
the three-mode products in the cubic Hamiltonian in
\cref{eq:canonical-H} that is the focus of this work.
Hamilton's equations and their closed bilinear projection are
\begin{equation}\label{eq:AHH-parent-rates}
 i\frac{d\mathsf F}{ds}=\mathsf F\mathsf C,\qquad
 i\frac{d\mathsf G}{ds}=-\mathsf G\mathsf C^{\trans},\qquad
 \widehat{\mathsf X}=-\mathsf G^{\trans}\mathsf F,
 \qquad i\frac{d\widehat{\mathsf X}}{ds}
       =[\mathsf D,\widehat{\mathsf X}^{,2}].
\end{equation}
The diagonal blocks of $\widehat{\mathsf X}$ are constant because
$\mathsf D$ is scalar on each of the three groups. Fix them as in
\cref{eq:canonical-X}. Define the constant block-diagonal phase-rate
matrix $\mathsf K_{\rm ph}$ and the transformed coefficient matrix by
\begin{equation}\label{eq:AHH-time-phase-map}
 \mathsf K_{\rm ph}=\diag(-3\mathsf B_L,3\mathsf B_M,0_N),\qquad
 \mathsf X(t)=e^{-it\mathsf K_{\rm ph}}
       \widehat{\mathsf X}(-t)e^{it\mathsf K_{\rm ph}}.
\end{equation}
The conjugation applies the required mode phase rotations, while
$\widehat{\mathsf X}(-t)$ reverses the quartic evolution time $s=-t$.
Then
\begin{equation}\label{eq:AHH-cubic-projection}
 i\dot{\mathsf X}=-[\mathsf D,\mathsf X^2]
                         +[\mathsf K_{\rm ph},\mathsf X],
\end{equation}
whose off-diagonal blocks are exactly
\cref{eq:canonical-matrix-rates} for every positive $L,M,N$.
For $q=d$ every initial $\mathsf X$ admits a factorization, for example
$\mathsf F(0)=\Id_d$, $\mathsf G(0)=-\mathsf X(0)^{\trans}$.
Thus the equation-level projection includes arbitrary canonical
initial modes of the cubic system. The adapted BA factorization is
given in Appendix~\ref{sec:AHH-dictionary}.

\subsubsection{The two matrix brackets}

The quartic mode system in $\mathsf F,\mathsf G$ and our cubic mode
system in $\boldsymbol u,\boldsymbol v,\boldsymbol w$ and their barred
partners both have constant canonical brackets. Their
relationship is seen by distinguishing the two brackets on
$\mathsf X$. For this bracket comparison write $[f,h]_a=i\{f,h\}_a$ to
suppress the common physical factor $-i$. The bilinear projection
from $\mathsf F,\mathsf G$ induces the linear Lie--Poisson bracket
\begin{equation}\label{eq:AHH-two-brackets}
\begin{aligned}
 [X_{hi},X_{jk}]_1&=\delta_{hk}X_{ji}-\delta_{ij}X_{hk},\\
 [X_{hi},X_{jk}]_0&=(D_{hh}-D_{ii})\delta_{hk}\delta_{ij},
 &&1\leq h,i,j,k\leq d.
\end{aligned}
\end{equation}
The second, constant bracket is exactly
\cref{eq:canonical-brackets} under the block scalings of
\cref{eq:canonical-X}. Its same-group blocks are Casimirs, so their
fixed values specify the canonical cubic mode space. With gradients
defined by $dh=\tr((\nabla h)d\mathsf X)$, the two brackets give
\begin{equation}\label{eq:AHH-Lenard}
 [\mathsf X,\tr(\mathsf D\mathsf X^2)]_1
 =[\mathsf X,\tr(\mathsf X^3)/3]_0
 =[\mathsf D,\mathsf X^2].
\end{equation}
They are compatible: translating $\mathsf X$ by $a\mathsf D$ in the
first bracket gives $P_1(\mathsf X+a\mathsf D)=P_1(\mathsf X)-aP_0$.
This is the argument-shift Poisson pencil already underlying
\cref{sec:canonical}; it explains the change from a quartic pullback
to a cubic canonical Hamiltonian. Our cubic Hamiltonian in
\cref{eq:canonical-H}, expressed through the common matrix
$\mathsf X$ and the power charges, is, up to an additive constant,
\begin{equation}\label{eq:AHH-cubic-H}
 H=-\frac13\tr(\mathsf X^3)
   +3\sum_{\ell=1}^{L}b_\ell^L Q_\ell^L
   +3\sum_{m=1}^{M}b_m^M Q_m^M.
\end{equation}
The last two terms generate precisely the phase conjugation in
\cref{eq:AHH-time-phase-map}. This compares vector fields on two
Poisson realizations: the bilinear map is many-to-one, and the
canonical spaces have $qd$ and $LM+MN+NL$ pairs, respectively.
Thus the two descriptions reproduce the same spectral matrix evolution,
after the stated time and phase changes, but the bilinear projection
is not a canonical transformation between their mode Hamiltonians.
The quartic and cubic systems use different independent modes and
induce different brackets on the common matrix variables.

\subsubsection{BA convolution and interaction factorization}

In the quartic-to-cubic correspondence, a weighted sum of products of
two intermediate BA evaluations collapses to a single endpoint BA
function. This
provides a step-down mechanism from quartic to cubic interaction
expressions. We use the Szeg\H{o}-kernel composition identity of
Ben-Zvi and Biswas \cite[Proposition~5.2.1]{BenZviBiswas2003},
expressed in the normalization appropriate to our BA modes.

First introduce the theta-ratio part of the BA pair mode,
written in scalar local half-differential frames as
\begin{equation}\label{eq:AHH-Szego}
 \mathscr S_z(P,Q)=
 \frac{\theta(z+A(P)-A(Q))}{\theta(z)E(P,Q)}.
\end{equation}
The fixed divisor factors and exponential BA normalization are
temporarily suppressed; restoring them gives the BA pair functions
defined in \cref{eq:general-kernel-normalization,eq:general-kernel-infinity}.
Let $\Lambda,W$ be meromorphic functions
with poles only at distinct points $P_1,\ldots,P_d$, with $\Lambda$
having a simple pole at each. Choose a complete regular fiber
$\Lambda^{-1}(\alpha)=Q_1+\cdots+Q_d$. In coordinates
$t_h=1/\Lambda$ at $P_h$ and $s_i$ at $Q_i$, write
\begin{equation}\label{eq:AHH-fiber-data}
 \frac{W}{\Lambda-\alpha}
       =y_h+c_ht_h+O(t_h^2),\qquad
 \gamma_i=\frac{W(Q_i)}{\partial_{s_i}\Lambda(Q_i)},
 \quad 1\leq h,i\leq d.
\end{equation}
Assume $W$ has at most simple poles at the $P_h$. Then
\begin{equation}\label{eq:AHH-fiber-convolution}
 \sum_{i=1}^{d}\gamma_i
      \mathscr S_z(P_h,Q_i)\mathscr S_z(Q_i,P_k)
 =\begin{cases}
 (y_h-y_k)\mathscr S_z(P_h,P_k),&h\ne k,\\
 c_h,&h=k.
 \end{cases}
\end{equation}
For distinct endpoints, the left-hand side contains products of two
intermediate evaluations, while the right-hand side contains one
endpoint evaluation multiplied by a constant.
The contracted half-differentials at $Q_i$ make the formula independent
of the coordinates $s_i$. To see it directly, take residues in $Q$
of $(W/(\Lambda-\alpha))\mathscr S_z(P_h,Q)
\mathscr S_z(Q,P_k)$. For $h\ne k$, the endpoint residues are
$-y_h\mathscr S_z(P_h,P_k)$ and
$y_k\mathscr S_z(P_h,P_k)$. For $h=k$, the kernel product has
expansion $(-t_h^{-2}+O(1))dt_h$, giving residue $-c_h$.
The remaining residues are exactly the displayed sum. This also
proves the cancellation of the extra theta-divisor pole in each
product sum.

For disjoint endpoint lists, assemble
$\mathsf A_{hi}=\mathscr S_z(P_h,Q_i)$,
$\mathsf B_{ik}=\mathscr S_z(Q_i,P_k)$,
$\mathsf K_{hk}=\mathscr S_z(P_h,P_k)$ and
$\mathsf\Gamma=\diag(\gamma_1,\ldots,\gamma_d)$, with
$\mathsf Y_{\rm left},\mathsf Y_{\rm right}$ the corresponding
endpoint diagonals of the $y$ values. The off-diagonal identity is
\begin{equation}\label{eq:AHH-Sylvester}
 \mathsf A\mathsf\Gamma\mathsf B
   =\mathsf Y_{\rm left}\mathsf K-\mathsf K\mathsf Y_{\rm right}.
\end{equation}
This Sylvester displacement relation makes the interaction change
explicit. For compatible vectors $\boldsymbol a,\boldsymbol b$,
\[
 \boldsymbol a^{\trans}
 (\mathsf Y_{\rm left}\mathsf K-\mathsf K\mathsf Y_{\rm right})
 \boldsymbol b
 =\boldsymbol a^{\trans}\mathsf A\mathsf\Gamma\mathsf B\boldsymbol b.
\]
Inserting the convolution sum between the two outer mode factors
turns the cubic endpoint expression on the left into the quartic
intermediate expression on the right; collapsing the sum performs
the reverse, step-down operation. The weights and the complete fiber
ensure that the summed products reduce to endpoint BA functions. In the
Adams--Harnad--Hurtubise realization this is the function-level counterpart of
\cref{eq:AHH-parent-rates}; changing the list of independent modes
also requires their brackets and evolution, supplied here by
\cref{eq:AHH-two-brackets,eq:AHH-cubic-projection}.

\subsection{The matrix constrained-KP correspondence}
\label{sec:BA-constrained-KP}

The weighted BA identities also have a constrained-KP operator
interpretation for arbitrary positive $L,M,N$. We give the operator
dictionary and its higher-time normalization, then relate those times
to the canonical flows. The one-point specialization in
\cref{sec:BA-one-point-specialization} recovers scalar operators.

\subsubsection{Matrix operators from the BA bases}

For this operator comparison put $x=-it$ and
$\partial=\partial_x=i\partial_t$. Assemble the
ordinary and dual BA bases into columns
\[
 \boldsymbol\psi(P)=(\psi_1(P),\ldots,\psi_N(P))^{\trans},\qquad
 \boldsymbol\psi^\vee(P)=(\psi_1^\vee(P),\ldots,\psi_N^\vee(P))^{\trans}.
\]
The kernel rate in \cref{eq:general-kernel-rate} specifies the paired
primitive
\begin{equation}\label{eq:KP-paired-primitive}
 \partial^{-1}\!\left(
 (\boldsymbol\psi^\vee(Q))^{\trans}\boldsymbol\psi(P)\right)
 =\Omega(P,Q),
\end{equation}
with its integration constant fixed by the kernel normalization.
Define the diagonal puncture-constant matrices
$\mathsf C_A=\diag(a_{A,1},\ldots,a_{A,N})$, $A\in\{L,M\}$,
using \cref{eq:general-h-normalization}. The weighted equations give
\begin{equation}\label{eq:BA-constrained-operators}
\begin{aligned}
 \mathcal A_L&=\Id_N\partial+\mathsf C_L+
  2\sum_{\ell=1}^{L}\kappa_\ell\boldsymbol\psi(Q_\ell)
     \partial^{-1}(\boldsymbol\psi^\vee(Q_\ell))^{\trans},\\
 \mathcal A_M&=\Id_N\partial+\mathsf C_M+
  2\sum_{j=L+1}^{K}\kappa_j\boldsymbol\psi(Q_j)
     \partial^{-1}(\boldsymbol\psi^\vee(Q_j))^{\trans},
 \qquad K=L+M.
\end{aligned}
\end{equation}
The matrix size $N$ counts punctures; the two sums contain $L$ and
$M$ source columns, respectively. Each integral term pairs an input
column with the dual row, takes the
scalar primitive and multiplies by the ordinary column. Thus
\begin{equation}\label{eq:KP-matrix-spectral}
 \mathcal A_A\boldsymbol\psi(P)=h_A(P)\boldsymbol\psi(P),\qquad
 \mathcal A_A^{\rm ad}\boldsymbol\psi^\vee(P)
        =h_A(P)\boldsymbol\psi^\vee(P),\qquad A\in\{L,M\}.
\end{equation}
Here $\mathrm{ad}$ is formal transpose-adjoint, with
$\partial^{\rm ad}=-\partial$ and no complex conjugation.
The normalized expansions at all $N$ punctures determine an
invertible matrix wave operator. Both $\mathcal A_L,\mathcal A_M$
dress time-independent diagonal Laurent operators, so
\begin{equation}\label{eq:KP-matrix-commutation}
 [\mathcal A_L,\mathcal A_M]=0.
\end{equation}
Appendix~\ref{app:KP-normalization} supplies this formal argument and
the higher-time details.

Finite-rank integral operators of this form belong to established
matrix constrained-KP theory
\cite[equations (7)--(14)]{ChvartatskyiSydorenko2013}.
The scalar vector-constrained case, its rational reductions and
Grassmannian description are treated in
\cite[Theorems~4.1 and~5.1]{HelminckVanDeLeur1998} and
\cite{KricheverRational1995}. Residue-based BA self-consistency is
developed in \cite[Section~1]{KricheverSelfConsistent1995}.
The explicit dictionary here uses the two separate interpolation
groups and their puncture constants.

\subsubsection{Normalized KP times and source evolution}

The puncture constants determine how this operator form is normalized.
Define the diagonal gauge matrix $\mathsf G(x)=e^{-x\mathsf C_L}$ and put
$\widetilde{\boldsymbol\psi}=\mathsf G^{-1}\boldsymbol\psi$,
$\widetilde{\boldsymbol\psi}^{\vee}=\mathsf G^{\trans}\boldsymbol\psi^\vee$.
Then the first operator becomes
\begin{equation}\label{eq:KP-normalized-L}
 \mathcal L_{\rm KP}=\mathsf G^{-1}\mathcal A_L\mathsf G
 =\Id_N\partial+
 2\sum_{\ell=1}^{L}\kappa_\ell\widetilde{\boldsymbol\psi}(Q_\ell)
 \partial^{-1}(\widetilde{\boldsymbol\psi}^{\vee}(Q_\ell))^{\trans}.
\end{equation}
The second, $\mathcal M_{\rm KP}=\mathsf G^{-1}\mathcal A_M\mathsf G$,
retains the constant term
\begin{equation}\label{eq:KP-residual-phase}
 \mathsf C_M-\mathsf C_L
 =2\diag(b_1^N,\ldots,b_N^N).
\end{equation}
Thus the residual matrix is precisely the third-group phase data of
the canonical Hamiltonian. A common gauge removes both constant
terms, after additive shifts of $h_L,h_M$, only when their difference
is scalar; this holds automatically for $N=1$.

Use $h_L(P)$ as the local spectral coordinate at every puncture.
The normalized KP times $x_n$, with $x_1=x$, prescribe leading
ordinary factors $\exp(\sum_{n=1}^{n_*}x_n h_L(P)^n)e_k$ at $R_k$;
the dual factors have the opposite exponent. Here $n_*$ is any finite
number of active times and $e_k$ is the $k$th coordinate vector.
In the original coordinates,
\[
 \operatorname{pp}_{R_k}h_L^2=s_k^{-2}+2a_{L,k}s_k^{-1}.
\]
The regular constant $\operatorname{reg}_{R_k}h_L^2$ also supplies a
component phase. These terms explain why the second KP time
need not coincide with an uncorrected double-pole BA time.

\begin{proposition}[Constrained-KP evolution on the BA realization]
\label{prop:KP-hierarchy}
Let $\mathcal B_n=(\mathcal L_{\rm KP}^{\,n})_+$, where $+$ retains
the nonnegative differential powers. On a regular BA time domain,
\begin{equation}\label{eq:KP-source-flows}
\begin{aligned}
 \partial_{x_n}\widetilde{\boldsymbol\psi}(P)
   &=\mathcal B_n\widetilde{\boldsymbol\psi}(P),\\
 \partial_{x_n}\widetilde{\boldsymbol\psi}^{\vee}(P)
   &=-\mathcal B_n^{\rm ad}\widetilde{\boldsymbol\psi}^{\vee}(P),
 \qquad n\geq1.
\end{aligned}
\end{equation}
Both operators retain their finite-rank representations and satisfy
\begin{equation}\label{eq:KP-operator-flows}
 \partial_{x_n}\mathcal L_{\rm KP}
       =[\mathcal B_n,\mathcal L_{\rm KP}],\qquad
 \partial_{x_n}\mathcal M_{\rm KP}
       =[\mathcal B_n,\mathcal M_{\rm KP}].
\end{equation}
In particular, evaluation of \cref{eq:KP-source-flows} at the fixed
$Q_a$ gives the ordinary and dual source evolution for both groups.
\end{proposition}
\begin{proof}
In the $h_L$ local coordinates, the normalized matrix wave operator
dresses $\partial$ to $\mathcal L_{\rm KP}$. Subtracting the indicated
differential action from each BA time derivative leaves a section
with zero leading values at every puncture; BA uniqueness gives
\cref{eq:KP-source-flows}. The finite-rank commutator identity then
gives \cref{eq:KP-operator-flows}. The normalization and adjoint
arguments are detailed in Appendix~\ref{app:KP-normalization}.
\end{proof}

\subsubsection{Relation to the canonical mode flows}

The sources in \cref{eq:BA-constrained-operators} are precisely the
ordinary and dual evaluations used for the $v,w$ modes in
\cref{eq:multipoint-mode-arrays}. Evaluating the $L$ spectral problem
at the $M$-group points and the $M$ problem at the $L$-group points
recovers their weighted equations; the kernel derivative supplies
the $u$ equations. Multiwave equations involving eigenfunctions,
adjoint eigenfunctions and pair potentials also occur in
\cite[equations (57)--(58)]{AratynGomesNissimovPacheva2001}.

On each birational BA level, \cref{thm:BA-higher-flows} supplies
spectral Hamiltonians spanning all Abel directions. Matching a KP
velocity is therefore a finite linear problem. The remaining endpoint
phases are generated by the power charges in
\cref{eq:canonical-charges}. Each normalized KP time consequently
has a canonical spectral-plus-gauge generator on that level, with
its coefficients held fixed after choosing the level.
Appendix~\ref{app:KP-normalization} gives the construction, and
\cref{sec:numeric-KP-time} checks the second time in genus two.

This identifies the operators and their BA evolution. The pointwise
source arrays contain $NL+NM$ canonical pairs, whereas the full mode
system also retains $LM$ pair-composite pairs through its kernel
primitives. Identifying a KP field Poisson bracket with the complete
finite canonical bracket requires a further constraint calculation.

\subsection{The one-point specialization}
\label{sec:BA-one-point-specialization}

The general construction becomes a vector--matrix--vector system
when $N=1$. We give its elementary-mode notation, specialize the
higher-flow phases and display the scalar conserved operator coefficients.
The curve remains arbitrary throughout this subsection.

\subsubsection{Elementary modes and their canonical equations}
\label{sec:BA-one-point}

For $N=1$ put $R_1=P_\infty$, $s_1=s$ and $e_1=1$.
The divisors have degree $g$ and there is a single ordinary/dual pair
$\psi,\psi^\vee$. Choose reference normalization
$\psi(P,0)=\psi^\vee(P,0)=1$, and subtract the constant term of
each $h$ at $P_\infty$. Thus $a_{L,1}=a_{M,1}=0$.
This member has two elementary vectors and a pair-mode matrix.

The combined index and number of canonical pairs are
\begin{equation}\label{eq:BA-mode-counts}
 K=L+M,\qquad p=K+LM.
\end{equation}
We use single indices for the $K$ elementary pairs and double indices
for the $LM$ cross-group pair composites. Their dictionary with the
$H_0$ modes of \cref{eq:canonical-phase-gauge} is
\begin{equation}\label{eq:BA-mode-dictionary}
 \xi_{\ell}=\tilde w_{1\ell},\qquad
 \xi_{L+m}=\tilde v_{m1},\qquad
 \xi_{\ell,L+m}=\tilde u_{\ell m},
 \quad1\leq\ell\leq L,\quad1\leq m\leq M,
\end{equation}
with the same rule for barred partners. Put
\begin{equation}\label{eq:BA-phase-dictionary}
 b_{\ell}=2b_{\ell}^L+b_1^N,\qquad b_{L+m}=2b_m^M+b_1^N,
 \quad1\leq\ell\leq L,\quad1\leq m\leq M.
\end{equation}
The inherited brackets satisfy $\{\xi_k,\bar\xi_r\}=-i\delta_{kr}$, with the corresponding rule for pair modes, and
\begin{equation}\label{eq:BA-H0-flat}
 H_0=\sum_{k=1}^{K}b_k\xi_k\bar{\xi}_k
 +\sqrt2\sum_{\ell=1}^{L}\sum_{j=L+1}^{K}
 (\xi_{\ell}\xi_j\xi_{\ell j}+\bar{\xi}_{\ell}\bar{\xi}_j\bar{\xi}_{\ell j}).
\end{equation}

The equivalent one-point theta and sigma formulas, with
$\Xi=s^{-1}+O(s)$ and
$\Xi_\sigma=\Xi-\boldsymbol U^{\trans}\varkappa A(P)$, are
\begin{equation}\label{eq:general-BA-theta}
\begin{aligned}
 \psi(P,t)
 &=e^{-it\Xi(P)}
 \tfrac{\theta(z(t)+A(P))\theta(z_0)}
 {\theta(z(t))\theta(z_0+A(P))}
 =e^{-it\Xi_\sigma(P)}
 \tfrac{\sigma(z(t)+A(P))\sigma(z_0)}
 {\sigma(z(t))\sigma(z_0+A(P))},\\[4pt]
 \psi^\vee(P,t)
 &=e^{it\Xi(P)}
 \tfrac{\theta(z(t)-A(P))\theta(z_0)}
 {\theta(z(t))\theta(z_0-A(P))}
 =e^{it\Xi_\sigma(P)}
 \tfrac{\sigma(z(t)-A(P))\sigma(z_0)}
 {\sigma(z(t))\sigma(z_0-A(P))}.
\end{aligned}
\end{equation}
In this notation the unphased BA modes are
\begin{equation}\label{eq:general-elementary-modes}
\begin{aligned}
 \xi_{\ell}&=c_{\ell} \psi^\vee(Q_{\ell}),&\bar{\xi}_{\ell}&=\kappa_{\ell} \psi(Q_{\ell})/c_{\ell},
 &&1\leq\ell\leq L,\\
 \xi_j&=c_j \psi(Q_j),&\bar{\xi}_j&=\kappa_j \psi^\vee(Q_j)/c_j,
 &&L+1\leq j\leq K,
\end{aligned}
\end{equation}
and
\begin{equation}\label{eq:general-pair-modes}
 \xi_{\ell j}=\frac{\sqrt2\kappa_{\ell}\kappa_j}{c_{\ell}c_j}\Omega(Q_{\ell},Q_j),
 \qquad \bar{\xi}_{\ell j}=-\sqrt2c_{\ell}c_j\Omega(Q_j,Q_{\ell}),
 \quad1\leq\ell\leq L<j\leq K.
\end{equation}

An \emph{elementary mode} has the shift of one curve-point lift
relative to $P_\infty$. A \emph{pair composite} has the difference
of the two lifts; in the hyperelliptic sheet convention this becomes
the sum of the elementary shifts. The terms specify the shift data,
not a factorization at a common $z$. For $N>1$, the basis sections
instead have shifts $A(P)-A(R_k)$; all sections share the same curve
and kernel, but generally no single base point makes both arrays
one-point elementary vectors.

The raw phase values and equations reduce to
\begin{equation}\label{eq:one-point-phase-values}
 b_{\ell}=-h_M(Q_{\ell}),\quad1\leq\ell\leq L,\qquad
 b_j=h_L(Q_j),\quad L+1\leq j\leq K,
\end{equation}
\begin{equation}\label{eq:general-network-rates}
\begin{aligned}
 i\dot\xi_{\ell}&=b_{\ell}\xi_{\ell}+\sqrt2\sum_{j=L+1}^{K}\bar{\xi}_j\bar{\xi}_{\ell j},&
 i\dot{\bar{\xi}}_{\ell}&=-b_{\ell}\bar{\xi}_{\ell}-\sqrt2\sum_{j=L+1}^{K}\xi_j\xi_{\ell j},
 &&1\leq\ell\leq L,\\
 i\dot\xi_j&=b_j\xi_j+\sqrt2\sum_{\ell=1}^{L}\bar{\xi}_{\ell}\bar{\xi}_{\ell j},&
 i\dot{\bar{\xi}}_j&=-b_j\bar{\xi}_j-\sqrt2\sum_{\ell=1}^{L}\xi_{\ell}\xi_{\ell j},
 &&L+1\leq j\leq K,\\
 i\dot\xi_{\ell j}&=\sqrt2\bar{\xi}_{\ell}\bar{\xi}_j,&
 i\dot{\bar{\xi}}_{\ell j}&=-\sqrt2\xi_{\ell}\xi_j,
 &&1\leq\ell\leq L<j\leq K.
\end{aligned}
\end{equation}
These are Hamilton's equations for \cref{eq:BA-H0-flat}.
For this phase gauge, write $\mathsf A_L=\diag(b_1,\ldots,b_L)$,
$\mathsf A_M=\diag(b_{L+1},\ldots,b_K)$.
An optional scalar spectral origin $\rho$ gives
\begin{equation}\label{eq:BA-H0-pencil}
\begin{aligned}
 \mathsf L_0(\nu)&=\begin{pmatrix}
 \rho\Id_L-\mathsf A_L+\nu\Id_L&\sqrt2\boldsymbol\xi_{LM}&\bar{\boldsymbol\xi}_L\\
 \sqrt2\bar{\boldsymbol\xi}_{LM}^{\trans}&\rho\Id_M-\mathsf A_M-\nu\Id_M&\boldsymbol\xi_M\\
 -\boldsymbol\xi_L^{\trans}&-\bar{\boldsymbol\xi}_M^{\trans}&\rho
 \end{pmatrix},\\
 \mathsf T_0(\nu)&=\begin{pmatrix}
 0&0&\bar{\boldsymbol\xi}_L\\0&0&-\boldsymbol\xi_M\\
 -\boldsymbol\xi_L^{\trans}&\bar{\boldsymbol\xi}_M^{\trans}&-\nu
 \end{pmatrix}.
\end{aligned}
\end{equation}
Here $\boldsymbol\xi_L,\boldsymbol\xi_M$ are the column vectors of
the two elementary groups and $\boldsymbol\xi_{LM}$ is their pair
matrix. This is the one-point form of the canonical pencil after
its phase gauge; $\rho$ only shifts the spectral ordinate.

\subsubsection{Higher-flow phases in the one-point notation}
\label{sec:BA-one-point-higher-flows}

We specialize \cref{thm:BA-higher-flows} using the modes and pencil
just defined. This gives a modal-power expression for the finite
phase constants of a higher Hamiltonian.

For $N=1$, choose the ordinary gap complement at $P_\infty$.
Then $p_1=\sum_{\alpha=1}^{g}v_\alpha s^{-n_\alpha}$ for its
Weierstrass gaps $n_\alpha$, and $\varphi_1^N=0$.
The elementary phase rates are
$b_{\ell}^{(r,\nu_0)}=-\varphi_{\ell}^L$,
$b_{L+j}^{(r,\nu_0)}=\varphi_j^M$; pair phases are the opposite
sums of these rates. For example, in the one-point $H_0$ pencil
\cref{eq:BA-H0-pencil}, with optional spectral origin $\rho$, the
fourth trace has $q=\lambda^3/\nu$. Put $\epsilon_a=\rho-b_a$,
$\chi_a=1$ in the first ordinary group and $\chi_a=-1$ in the second.
Write $I_a=\xi_a\bar\xi_a$, $1\leq a\leq K$, and
$I_{\ell j}=\xi_{\ell j}\bar\xi_{\ell j}$, $1\leq\ell\leq L<j\leq K$.
For the conserved power combinations
$C_{\ell}=I_{\ell}-\sum_{j=L+1}^{K}I_{\ell j}$ and
$C_j=I_j-\sum_{\ell=1}^{L}I_{\ell j}$, perturbation at infinity gives
\[
 \lambda=\chi_a\nu+\epsilon_a-\chi_aC_a/\nu+O(\nu^{-2}),\qquad
 \operatorname{reg}_{Q_a}q=3\chi_a(\epsilon_a^2-C_a).
\]
The phase is this constant plus the finite part of $\mathfrak f$,
with the elementary sign just specified. Thus the one-point
modal-power interpretation is retained inside the multipoint theorem.

\subsubsection{Scalar constrained operators and conserved coefficients}

For $N=1$, the additive constants of $h_L,h_M$ may be chosen so that
$\mathsf C_L=\mathsf C_M=0$. The matrix operators in
\cref{eq:BA-constrained-operators} are then scalar, with
$\boldsymbol\psi=\psi$ and $\boldsymbol\psi^\vee=\psi^\vee$.
This specialization also exposes two conserved quantities.
Write $I_k=\xi_k\bar\xi_k$, $1\leq k\leq K$. Expanding $\partial^{-1}f=f\partial^{-1}-(\partial f)\partial^{-2}
+O(\partial^{-3})$ and using \cref{eq:general-network-rates} gives
\begin{equation}\label{eq:BA-constrained-H-coefficient}
 \mathcal A_L-\mathcal A_M
 =2\left(\sum_{\ell=1}^{L}I_\ell-\sum_{j=L+1}^{K}I_j\right)\partial^{-1}
      -2H_0\partial^{-2}+O(\partial^{-3}).
\end{equation}
The first two negative-order coefficients of the commutator are
the derivatives of these coefficient differences, up to sign.
Their vanishing recovers conservation of the group power difference
and $H_0$ on the BA realization.

\subsection{Hyperelliptic BA modes and polynomial interpolation}
\label{sec:BA-realization}

We specialize to $N=1$ with its puncture at infinity on an odd-degree
hyperelliptic curve. The general meromorphic residue criterion applies
for arbitrary $L,M,N$; the compact polynomial formulas derived here
use this one-point marking. Sheet reflection identifies ordinary and dual
functions, the kernel becomes a two-function Wronskian, and the residue
conditions become polynomial interpolation moments. We give the sigma
normalization and the conversion to \cref{thm:general-BA-realization},
then retain the explicit weights needed for the Mumford correspondence.
The mode counts and Hamiltonian remain
\cref{eq:BA-mode-counts,eq:BA-H0-flat}.

\subsubsection{Curve and function conventions}

Fix a smooth odd-degree curve, its Abel map and ordered differentials:
\begin{equation}\label{eq:BA-curve}
 y^2=f(x)=4x^{2g+1}+\sum_{r=0}^{2g}\lambda_r x^r,\qquad
 A(P)=\int_\infty^P du,\qquad
 du_j=\frac{x^{j-1}\,dx}{y},\quad1\leq j\leq g.
\end{equation}
The sheet involution is $P=(a,y)\mapsto\iota(P)=(a,-y)$;
compatible lifts give $A(\iota(P))=-A(P)$.
This is the Baker basis used in hyperelliptic sigma-function theory
\cite{BEL1997,Onishi2005}.
The vector $z$ in this hyperelliptic description uses these Baker Abel coordinates,
related by a fixed linear change to the normalized theta coordinates
of \cref{eq:general-BA-theta}. The scalar time follows the same first BA direction, with first BA
parameter $-it$.
In converting
from a monic equation $y_{\mathrm{mon}}^2=f_{\mathrm{mon}}(x)$ with
differentials $x^{j-1}dx/(2y_{\mathrm{mon}})$, set
$y=2y_{\mathrm{mon}}$ and $f=4f_{\mathrm{mon}}$; the Abel coordinates
then agree.

Use the canonical hyperelliptic Baker sigma function, with
$\wp_{ij}=-\partial_{z_i}\partial_{z_j}\log\sigma$ and
\begin{equation}\label{eq:BA-sharp}
 \sigma_\sharp=\sigma_{2,4,6,\ldots},
\end{equation}
where indices stop at $g$ and an empty list means $\sigma$ itself
\cite{BEL1997,Onishi2005}. To fix the multiplicative sigma constant,
its lowest ordinary-degree term is
$\det(z_{i+j-1})_{1\leq i,j\leq\lfloor(g+1)/2\rfloor}$.

The compatible second-kind vector is
\begin{equation}\label{eq:BA-second-kind}
 \boldsymbol R(P)=\operatorname{FP}\int_\infty^P dr,\qquad
 dr_j=\frac1{4y}\sum_{k=j}^{2g-j}
          (k+1-j)\lambda_{k+1+j}x^k\,dx,
 \quad1\leq j\leq g,\qquad \lambda_{2g+1}=4.
\end{equation}
The finite part subtracts the polar primitive at infinity and fixes
its constant term to zero in the local parameter $x^{-1/2}$.
The first- and second-kind lifts use the same paths. With this
choice, second-kind half-periods satisfy
$2\eta=-\oint_a dr$, and the exponential in the BA function has
the minus sign
\begin{equation}\label{eq:BA-generapy}
 \Psi(P,z)=\frac{\sigma(A(P)-z)}
 {\sigma_\sharp(A(P))\sigma(z)}
 \exp\bigl(-\boldsymbol R(P)^{\trans}z\bigr).
\end{equation}
The second-kind and Schr\"odinger conventions follow
\cite{Christiansen2000}; see also the normalized genus-two formula in
\cite[equation (3.3)]{Braden2005}. These conventions agree with
Generapy \cite{Generapy}.

The physical time follows the last Abel direction:
\begin{equation}\label{eq:BA-time-polynomial}
 z(t)=z_0-it e_g,\qquad
 \mathcal P(z,x)=x^g-\sum_{j=1}^{g}\wp_{gj}(z)x^{j-1},
 \qquad \mathcal P_{z_g}=\partial_{z_g}\mathcal P=i\dot{\mathcal P}.
\end{equation}
Here $z_0$ is unrestricted and $e_g$ is the last coordinate vector.
The polynomial $f(x)$ fixes the curve and has degree $2g+1$;
$\mathcal P(z,x)$ has degree $g$ and describes the moving divisor.

The standard one-point identities, in these conventions, are
\begin{equation}\label{eq:BA-identities}
\begin{aligned}
 \Psi(P,z)\Psi(\iota(P),z)&=(-1)^g\mathcal P(z,a),\\
 \partial_{z_g}\log\Psi(P,z)
 &=\frac{\mathcal P_{z_g}(z,a)-y}{2\mathcal P(z,a)},\\
 \partial_{z_g}^2\Psi(P,z)
 &=\left(a+2\wp_{gg}(z)+\frac{\lambda_{2g}}4\right)\Psi(P,z).
\end{aligned}
\end{equation}
They hold for a free $z$ and a single curve-point shift $A(P)$
\cite{BEL1997,Eilbeck2000}. In particular, raw sheet partners have
the sign $(-1)^g$ in their product. We absorb that constant sign
into the mode normalization so that their power is a weighted
evaluation of the monic $\mathcal P$.

\subsubsection{Sigma modes and the general normalization}

Use the mode indices and canonical partners of
\cref{eq:BA-mode-dictionary}. For the hyperelliptic modes the constants
$\kappa_k$ multiply the raw sigma products in
\cref{eq:BA-powers-rates}; their conversion to the normalized
residue weights of \cref{eq:general-residue-weights} is given in
\cref{eq:BA-general-dictionary}. The involution $\iota$ denotes sheet reflection; bars label independent Hamiltonian partners.

Choose $K$ points $Q_k=(a_k,y_k)$ on \cref{eq:BA-curve} with
finite distinct abscissae, and nonzero constants $\kappa_k,c_k$,
$1\leq k\leq K$. Define all elementary modes by
\begin{equation}\label{eq:BA-elementary}
 \xi_k=c_k\Psi(Q_k,z),\qquad
 \bar{\xi}_k=\frac{(-1)^g\kappa_k}{c_k}\Psi(\iota(Q_k),z),
 \qquad1\leq k\leq K.
\end{equation}
Their powers and logarithmic rates are, for $1\leq k\leq K$,
\begin{equation}\label{eq:BA-powers-rates}
\begin{aligned}
 \xi_k\bar{\xi}_k&=\kappa_k\mathcal P(z,a_k),\\
 \frac{i\dot\xi_k}{\xi_k}
 &=\frac{\mathcal P_{z_g}(z,a_k)-y_k}{2\mathcal P(z,a_k)},&
 \frac{i\dot{\bar{\xi}}_k}{\bar{\xi}_k}
 &=\frac{\mathcal P_{z_g}(z,a_k)+y_k}{2\mathcal P(z,a_k)}.
\end{aligned}
\end{equation}
Define the Wronskian in the Abel direction by
$W(f_1,f_2)=f_1\partial_{z_g}f_2-(\partial_{z_g}f_1)f_2$
for arbitrary functions $f_1,f_2$ of $z$. Along the physical flow,
$W(f_1,f_2)=i(f_1\dot f_2-\dot f_1f_2)$.
Each elementary pair has the constant Abel-direction Wronskian
$W(\xi_k,\bar\xi_k)=\kappa_k y_k$, $1\leq k\leq K$.
All elementary modes share the potential
$2\wp_{gg}+\lambda_{2g}/4$ in \cref{eq:BA-identities}.

For every cross-group pair set
\begin{equation}\label{eq:BA-pair-composites}
\begin{aligned}
 \xi_{\ell j}
 &=\frac{\sqrt2}{a_j-a_{\ell}}
        W(\bar{\xi}_{\ell},\bar{\xi}_j),\\
 \bar{\xi}_{\ell j}
 &=-\frac{\sqrt2}{a_j-a_{\ell}}
        W(\xi_{\ell},\xi_j),
 \qquad 1\leq\ell\leq L,\quad L+1\leq j\leq K.
\end{aligned}
\end{equation}
The kernel specialization in \cref{eq:BA-hyperelliptic-kernel} identifies these Wronskians with
sigma ratios having shifts $A(Q_{\ell})+A(Q_j)$ and its negative,
with fixed normalizing factors
\cite[Theorem~6.2, $n=2$]{Nakayashiki2026}. Their differential equations have
a particularly direct explanation: the common potential cancels,
leaving
\begin{equation}\label{eq:BA-pair-rates}
 i\dot\xi_{\ell j}=\sqrt2\bar{\xi}_{\ell}\bar{\xi}_j,
 \qquad
 i\dot{\bar{\xi}}_{\ell j}=-\sqrt2\xi_{\ell}\xi_j,
 \qquad1\leq\ell\leq L,\quad L+1\leq j\leq K.
\end{equation}
These are exactly the pair-mode equations of $H_0$. The elementary equations are supplied by the general realization criterion.
Primitives of products of ordinary and dual BA functions also occur in
the Cauchy--BA kernel construction
\cite[Section~2.4, equation~(2.13)]{GrinevichOrlov1989} and in
squared-eigenfunction-potential theory
\cite[Corollary~3.1]{AratynNissimovPacheva1997}.
The residue criterion takes the explicit polynomial form developed in
\cref{sec:BA-interpolation}.

Here is the explicit dictionary. Put $\epsilon=(-1)^g$ and
$P_0(x)=\mathcal P(z_0,x)$. The ordinary and dual functions of
\cref{eq:general-BA-asymptotics} are
\begin{equation}\label{eq:BA-normalized-sigma}
 \psi(P,t)=\frac{\epsilon\Psi(\iota(P),z(t))}{\epsilon\Psi(\iota(P),z_0)},
 \qquad
 \psi^\vee(P,t)=\frac{\Psi(P,z(t))}{\Psi(P,z_0)},\qquad
 d\Omega_0=P_0(x)\frac{dx}{y}.
\end{equation}
The raw product identity gives $\psi\psi^\vee=\mathcal P/P_0$. For a regular
initial chart $P_0(a_k)\ne0$, take the general construction's points,
weights and gauges to be
\begin{equation}\label{eq:BA-general-dictionary}
\begin{aligned}
 \widehat Q_{\ell}&=Q_{\ell},&&1\leq\ell\leq L,&
 \widehat Q_j&=\iota(Q_j),&&L+1\leq j\leq K,\\
 \widehat\kappa_k&=\kappa_kP_0(a_k),&&1\leq k\leq K,&
 \widehat c_k&=c_k\Psi(Q_k,z_0),&&1\leq k\leq K.
\end{aligned}
\end{equation}
The hats identify the normalized constants used in
\cref{eq:general-elementary-modes,eq:general-pair-modes}.
The kernel is
\begin{equation}\label{eq:BA-hyperelliptic-kernel}
 \Omega(P,Q)=\frac{\psi^\vee(Q)\partial_{z_g}\psi(P)-\partial_{z_g}\psi^\vee(Q)\psi(P)}{x(P)-x(Q)}.
\end{equation}
Its Abel-direction derivative is $\psi(P)\psi^\vee(Q)$ by the common Schr\"odinger equation.
Substituting \cref{eq:BA-general-dictionary} into
\cref{eq:general-pair-modes} gives exactly the Wronskians in
\cref{eq:BA-pair-composites}, including their signs and $\sqrt2$
factors. Thus the elementary and pair functions are the same modes,
expressed in normalized BA and raw sigma conventions.

\subsubsection{Explicit interpolation weights and phases}
\label{sec:BA-interpolation}

In the local parameter $s=x^{-1/2}$ at infinity,
$du_\alpha=-s^{2g-2\alpha}(1+O(s^2))ds$ for
$1\leq\alpha\leq g$. At the finite nodes, use $x-a_k$.
The general residue conditions therefore become
\begin{equation}\label{eq:BA-moments-final}
 \sum_{\ell=1}^{L}\kappa_{\ell} a_{\ell}^r
 =\sum_{j=L+1}^{K}\kappa_j a_j^r
 =\begin{cases}
 0,&0\leq r\leq g-2,\\
 \tfrac12,&r=g-1.
 \end{cases}
\end{equation}
Reflection changes both the principal coefficient of the interpolation
function and the differential coefficient, leaving identical moments
for the two groups.

To construct all weights with these moments, collect their simple
fractions over the node polynomials
\begin{equation}\label{eq:BA-node-polynomials}
 R_L(x)=\prod_{\ell=1}^{L}(x-a_{\ell}),\qquad
 R_M(x)=\prod_{j=L+1}^{K}(x-a_j).
\end{equation}
Define their numerator polynomials $q_L,q_M$ by
\begin{equation}\label{eq:BA-weight-generating}
 \frac{q_L(x)}{R_L(x)}
 =2\sum_{\ell=1}^{L}\frac{\kappa_{\ell}}{x-a_{\ell}},\qquad
 \frac{q_M(x)}{R_M(x)}
 =2\sum_{j=L+1}^{K}\frac{\kappa_j}{x-a_j}.
\end{equation}
Before imposing the moments, these numerators have degrees at most
$L-1,M-1$. The moments say precisely that each rational function has expansion
$x^{-g}+O(x^{-g-1})$ at infinity. Hence $L,M\geq g$, and the
numerators must be monic of degrees $L-g,M-g$. Taking residues gives
\begin{equation}\label{eq:BA-residue-weights}
 \kappa_{\ell}=\frac{q_L(a_{\ell})}{2R_L'(a_{\ell})},
 \quad1\leq\ell\leq L,\qquad
 \kappa_j=\frac{q_M(a_j)}{2R_M'(a_j)},
 \quad L+1\leq j\leq K.
\end{equation}
Primes denote differentiation in $x$. Conversely, every pair of
monic numerators of these degrees gives the required moments;
their values at the respective nodes must be nonzero to keep the
mode weights nonzero. Thus $q_L,q_M$
parametrize the available interpolation weights: they equal $1$
when the corresponding group has exactly $g$ points, while larger
groups give free polynomial coefficients.

\begin{corollary}\label{cor:BA-realization}
Let the elementary modes and pair composites in
\cref{eq:BA-elementary,eq:BA-pair-composites} range over a nonempty
open subset of the Jacobian of the fixed curve
\cref{eq:BA-curve}. They satisfy all Hamilton equations of
\cref{eq:BA-H0-flat} if and only if their weights obey the interpolation
moments \cref{eq:BA-moments-final} and their phase coefficients satisfy
\begin{equation}\label{eq:BA-phase-final}
\begin{aligned}
 b_{\ell}
 &=-\sum_{j=L+1}^{K}\kappa_j
       \frac{y_j-y_{\ell}}{a_j-a_{\ell}},&&1\leq\ell\leq L,\\
 b_j
 &=-\sum_{\ell=1}^{L}\kappa_{\ell}
       \frac{y_{\ell}-y_j}{a_{\ell}-a_j},&&L+1\leq j\leq K.
\end{aligned}
\end{equation}
\end{corollary}
\begin{proof}
In the unnormalized hyperelliptic weights the two interpolation
functions are
\begin{equation}\label{eq:BA-hyperelliptic-h}
 h_L(P)=\sum_{\ell=1}^{L}\kappa_{\ell}\frac{y(P)+y_{\ell}}{x(P)-a_{\ell}},\qquad
 h_M(P)=\sum_{j=L+1}^{K}\kappa_j\frac{y(P)-y_j}{x(P)-a_j}.
\end{equation}
Their finite poles are at $Q_{\ell}$ and $\iota(Q_j)$, respectively.
The moments make their puncture expansion $s^{-1}+O(s)$.
With \cref{eq:BA-normalized-sigma}, half the residue of
$h\,d\Omega_0$ is $\kappa_kP_0(a_k)$, precisely the normalized
weight in \cref{eq:BA-general-dictionary}.
Their opposite-group values give \cref{eq:BA-phase-final}.
The equivalence now follows from \cref{thm:general-BA-realization}.
At a branch point use its regular ordinate coordinate for the residue;
the formula extends without division by its zero ordinate.
\end{proof}

The phase conditions \cref{eq:BA-phase-final} explain the role of the
independent constant diagonal blocks. For $N=1$, the printed
Jiang--Rauch-Wojciechowski specialization gives
$b_k=b_1^N$ for every $1\leq k\leq K$, whereas these conditions
generically require different rates at different marked points.
The outer diagonal parameters in \cref{eq:BA-phase-dictionary}
supply exactly that freedom. The constants $c_k$ specify initial mode
normalizations and leave these logarithmic rates unchanged.

For example, take $g=L=M=2$ and let $Y(x)$ be the polynomial of degree
at most three with $Y(a_k)=y_k$, $1\leq k\leq4$. The moment weights
\cref{eq:BA-residue-weights} give
\begin{equation}\label{eq:BA-phase-differences}
 b_1-b_2=-\tfrac12[x^3]Y(x)(a_1-a_2),\qquad
 b_3-b_4=-\tfrac12[x^3]Y(x)(a_3-a_4).
\end{equation}
Thus a nonzero cubic interpolation coefficient requires distinct
phase rates within both groups. This is the generic minimal genus-two
realization. \Cref{thm:canonical-independence} applies both to these
distinct rates and to their coincident specializations.

\Cref{prop:BA-prescribed-spectral-curve} identifies this hyperelliptic
family as a spectral level of genus $g$ with total defect $LM-g$.

\subsection{Half-period shifts, al functions and elliptic examples}
\label{sec:classical-BA-specializations}

We specialize the explicit formulas to half-period shifts and genus
one, connecting the construction to classical functions and providing
examples with different mode counts.

\subsubsection{Half-period shifts and al functions}
\label{sec:BA-al-functions}

Branch-point evaluations give a classical specialization of these
canonical modes. Let $B_k=(a_k,0)$ be a finite branch point and
$\delta_k=A(B_k)$. Since
$\operatorname{div}(x-a_k)=2B_k-2\infty$, Abel's theorem gives
$2\delta_k\in\Lambda$, where
$\Lambda=2\omega\mathbb Z^g+2\omega'\mathbb Z^g$.
The two sheet-partner functions are then proportional, and their
Wronskian vanishes. After a constant normalization they are
Weierstrass's hyperelliptic al functions
\cite{Weierstrass1856,MatsutaniPreviato2015}, satisfying
\begin{equation}\label{eq:BA-al-power}
 \operatorname{al}_k(z)=d_k\Psi(B_k,z),\qquad
 \operatorname{al}_k(z)^2=\mathcal P(z,a_k),\qquad d_k^2=(-1)^g.
\end{equation}
Each al function is meromorphic on a double cover of the Jacobian;
its square is Abelian. These functions generalize the Jacobi elliptic
functions \cite[Appendix]{MatsutaniPreviato2015}.

For an arbitrary-genus example containing all $2g+1$ finite branch
points, partition them into groups of sizes $L=g$, $M=g+1$.
Choose $q_L=1$ and $q_M(x)=x-\alpha$, with $\alpha$ distinct from
the second group's abscissae, and use the weights in
\cref{eq:BA-residue-weights}. They satisfy
\cref{eq:BA-moments-final}. All ordinates are zero, so
\cref{eq:BA-phase-final} gives $b_k=0$, $1\leq k\leq2g+1$.
Thus \cref{cor:BA-realization} realizes the entire elementary al
family in the purely cubic Hamiltonian \cref{eq:BA-H0-flat}.
The cross-group pair modes are their Wronskians in
\cref{eq:BA-pair-composites}, with half-period sums as shifts.
The canonical partners remain independent in the ambient system;
they become proportional on this al solution family.

Since all phases vanish, this family also solves the original
Jiang--Rauch-Wojciechowski system of
Appendix~\ref{sec:JRW-dictionary}; its complete integral count is
provided by the repeated-phase part of
\cref{thm:canonical-independence}, with moment integrals from blocks
of sizes $g$ and $g+1$.
In the plane characteristic curve $\Gamma$ of
\cref{prop:BA-prescribed-spectral-curve}, these groups meet the boundary
at two points of multiplicities $g$ and $g+1$.
Within each group the coincident intercepts give pairwise branch
contact order two. A point of multiplicity $m$ therefore has total
defect $m(m-1)$, of which $\binom m2$ is excess over the ordinary
multiple-point defect. The two excesses are $g(g-1)/2$ and
$g(g+1)/2$. Their sum $g^2=LM-g$ accounts for the entire additional
defect; the total defect is $2g^2$. Consequently the affine curve
is smooth. In the product-surface model the ordinary infinity
contributions have been separated, leaving total boundary defect $g^2$.

For genus at least two, the elementary half-period mechanism is
specific to hyperelliptic curves: $2A(P)\in\Lambda$ for $P\ne P_\infty$ makes
$2P-2P_\infty$ principal and supplies a degree-two map.
Other covering structures give higher-torsion analogues. On a cyclic
trigonal curve $y^3=\prod_{r=1}^{4}(x-b_r)$, for example,
$\operatorname{div}(x-b_r)=3B_r-3\infty$ gives order-three shifts;
trigonal al functions are developed in
\cite[Section~8]{MatsutaniPreviato2015}.

\subsubsection{Genus one and many elementary modes}

In Weierstrass coordinates take
$y^2=4x^3-g_2x-g_3$. Then $A(P)=\mu$,
$a=\wp(\mu)$, $y=\wp'(\mu)$ and
$R(P)=-\zeta(\mu)$, so the elementary function is
\begin{equation}\label{eq:BA-elliptic}
 \Psi(P,z)=\frac{\sigma(\mu-z)}{\sigma(\mu)\sigma(z)}
                  e^{\zeta(\mu)z},\qquad z=z_0-it.
\end{equation}
Its normalized mode powers are
$\xi_k\bar{\xi}_k=\kappa_k(a_k-\wp(z))$, $1\leq k\leq K$.
The only moment conditions are
\[
 \sum_{\ell=1}^{L}\kappa_{\ell}
 =\sum_{j=L+1}^{K}\kappa_j=\tfrac12.
\]
Thus arbitrarily large groups of elementary elliptic modes fit the
same cubic-interaction construction, with their $LM$ pair composites.
Consequently, for arbitrary positive $L,M$, the Hamiltonian
\cref{eq:BA-H0-flat} admits a genus-one BA realization with
$L+M+LM$ canonical mode pairs. This gives a family of cubic
multiwave Hamiltonians with elliptic solutions extending beyond the
three-wave case; its explicit Weierstrass formulation merits further study.

For $L=M=1$, so $K=2$, write
$Y=(y_2-y_1)/(2(a_2-a_1))$.
The two weights are $1/2$, and phase matching gives
$b_1=b_2=-Y$. The resulting three-pair Hamiltonian is
\begin{equation}\label{eq:BA-three-wave}
 H_0=-Y(\xi_1\bar{\xi}_1+\xi_2\bar{\xi}_2)
 +\sqrt2(\xi_1\xi_2\xi_{12}
             +\bar{\xi}_1\bar{\xi}_2\bar{\xi}_{12}).
\end{equation}
This recovers the familiar detuned three-wave form
\cite{MenyukChenLee1983Integrable}. Increasing $L,M$ retains elliptic
elementary solutions while adding pair-composite interactions; it
also increases the genus of the regular ambient spectral curve.

\subsection{Mode powers and the Mumford subsystem}
\label{sec:mumford-correspondence}

Throughout this subsection $N=1$, with the hyperelliptic data of
\cref{sec:BA-realization}.

The elementary mode powers in \cref{cor:BA-realization} are
evaluations of the moving divisor polynomial $\mathcal P$. Together
with their derivatives, they determine a closed projected subsystem
in divisor coordinates, which we identify with the odd Mumford flow.
We call this the Mumford subsystem. We reconstruct its polynomials
from the modes and describe the mode information retained by the
BA formulas. The curve, marked points and weights remain fixed throughout this subsection. We write $\mathcal P_{z_g}=\partial_{z_g}\mathcal P=i\dot{\mathcal P}$ and
$\mathcal P_{z_gz_g}=\partial_{z_g}^2\mathcal P=-\ddot{\mathcal P}$.

\subsubsection{Reconstructing the divisor polynomials}

Define the elementary powers $I_k=\xi_k\bar{\xi}_k$, $1\leq k\leq K$.
Their derivatives follow directly from the canonical Hamiltonian
\cref{eq:BA-H0-flat}: the linear phase terms cancel, leaving
\begin{equation}\label{eq:mumford-power-rates}
\begin{aligned}
 i\dot I_{\ell}&=\sqrt2\sum_{j=L+1}^{K}
 (\bar{\xi}_{\ell}\bar{\xi}_j\bar{\xi}_{\ell j}-\xi_{\ell}\xi_j\xi_{\ell j}),&&1\leq\ell\leq L,\\
 i\dot I_j&=\sqrt2\sum_{\ell=1}^{L}
 (\bar{\xi}_{\ell}\bar{\xi}_j\bar{\xi}_{\ell j}-\xi_{\ell}\xi_j\xi_{\ell j}),&&L+1\leq j\leq K.
\end{aligned}
\end{equation}
Thus $\dot I_k=\{I_k,H_0\}$ is a polynomial in the modes, rather
than additional derivative data. On the BA family,
\begin{equation}\label{eq:mumford-evaluations}
 \frac{I_k}{\kappa_k}=\mathcal P(z,a_k),\qquad
 \frac{i\dot I_k}{\kappa_k}=\mathcal P_{z_g}(z,a_k),
 \qquad1\leq k\leq K.
\end{equation}

Since $L\geq g$, use the first $g$ marked abscissae and their
Lagrange basis polynomials
\begin{equation}\label{eq:mumford-lagrange}
 \ell_k(x)=\prod_{\substack{r=1\\r\ne k}}^{g}
                  \frac{x-a_r}{a_k-a_r},\qquad1\leq k\leq g.
\end{equation}
The leading coefficient of $\mathcal P$ is one, so interpolation
of its remaining coefficients and of $\mathcal P_{z_g}$ gives
\begin{equation}\label{eq:mumford-reconstruction}
\begin{aligned}
 \mathcal P(z,x)
 &=x^g+\sum_{k=1}^{g}
       \left(\frac{I_k}{\kappa_k}-a_k^g\right)\ell_k(x),\\
 \mathcal P_{z_g}(z,x)
 &=\sum_{k=1}^{g}\frac{i\dot I_k}{\kappa_k}\ell_k(x).
\end{aligned}
\end{equation}
Any $g$ distinct marked abscissae give the same reconstruction on
this family. The additional evaluations supply the remaining mode
powers and their linear interpolation relations.

Jacobi inversion gives $\mathcal P\mid f-\mathcal P_{z_g}^{\,2}$:
at each moving divisor point $(x_r,y_r)$ one has
$\mathcal P_{z_g}(z,x_r)=y_r$ and hence $f(x_r)=\mathcal P_{z_g}(z,x_r)^2$.
Consequently define the polynomial
\begin{equation}\label{eq:mumford-W}
 \mathcal W(z,x)=\frac{f(x)-\mathcal P_{z_g}(z,x)^2}
                         {4\mathcal P(z,x)},\qquad
 f=\mathcal P_{z_g}^{\,2}+4\mathcal P\mathcal W.
\end{equation}
Here $\mathcal P$ and $\mathcal W$ are monic of degrees $g$ and
$g+1$, while $\deg_x\mathcal P_{z_g}\leq g-1$. The factor four
makes $\mathcal W$ monic for the curve normalization
\cref{eq:BA-curve}. Write Mumford's three polynomials as
$(\mathcal U,\mathcal V,\mathcal W)$, with
\begin{equation}\label{eq:mumford-dictionary}
 \mathcal U=\mathcal P,\qquad
 \mathcal V=\mathcal P_{z_g}/2,\qquad
 \mathcal U\mathcal W+\mathcal V^2=f/4.
\end{equation}
These are the degree constraints and spectral relation of the
ordinary odd Mumford system
\cite[Chapter~IIIa]{Mumford1984}; see also
\cite[\S3.1]{FernandesVanhaecke2001}. Its monic spectral curve is
obtained from ours by dividing the ordinate by two.

\subsubsection{The projected equations of motion}

Write the common scalar potential as
\begin{equation}\label{eq:mumford-potential}
 \mathcal Q(z)=2\wp_{gg}(z)+\frac{\lambda_{2g}}4
 =\frac{\lambda_{2g}}4-2[x^{g-1}]\mathcal P(z,x).
\end{equation}
The shared second-order BA equation and constant sheet-pair
Wronskian in \cref{eq:BA-identities,eq:BA-powers-rates} give
\begin{equation}\label{eq:mumford-product-identity}
 2\mathcal P\mathcal P_{z_gz_g}-\mathcal P_{z_g}^{\,2}
       -4(x+\mathcal Q)\mathcal P^2=-f(x).
\end{equation}
For example, differentiate the product of the BA functions at an
arbitrary curve point with abscissa $x$, use their common scalar
equation, and eliminate the difference of their derivatives with
their Wronskian. Since the point is arbitrary, this gives a
polynomial identity in $x$.

Substituting \cref{eq:mumford-W} into
\cref{eq:mumford-product-identity}, then differentiating the
spectral relation, yields the closed polynomial flow
\begin{equation}\label{eq:mumford-flow}
 \mathcal P_{z_gz_g}=2\bigl((x+\mathcal Q)\mathcal P-\mathcal W\bigr),
 \qquad
 i\dot{\mathcal W}=-(x+\mathcal Q)\mathcal P_{z_g}.
\end{equation}
Together with $\mathcal P_{z_g}$ as the second polynomial
coordinate, these are the odd Mumford infinity flow. The time
normalization is fixed by $i\dot{\mathcal U}=2\mathcal V$
and the BA direction $z(t)=z_0-it e_g$.

The correspondence can also be checked in a two-by-two Lax form.
Define
\begin{equation}\label{eq:mumford-lax-matrices}
 \mathsf A(x)=\begin{pmatrix}
 \mathcal V&\mathcal U\\
 \mathcal W&-\mathcal V
 \end{pmatrix},\qquad
 \mathsf B(x)=\begin{pmatrix}0&1\\x+\mathcal Q&0\end{pmatrix}.
\end{equation}
Then \cref{eq:mumford-flow} is equivalent to
\begin{equation}\label{eq:mumford-lax}
 i\dot{\mathsf A}=[\mathsf A,\mathsf B],\qquad
 \det\left(\frac y2\Id_2-\mathsf A(x)\right)
                  =\frac{y^2-f(x)}4.
\end{equation}
Thus the mode-power projection preserves both the divisor dynamics
and the prescribed hyperelliptic curve.

\subsubsection{Recovering the modes from divisor data}

At a generic state of the fixed-curve Mumford fiber, set
$\mathcal P_k=\mathcal P(a_k)$ and
$\mathcal P_{z_g,k}=\mathcal P_{z_g}(a_k)$,
$1\leq k\leq K$. Choose nonzero elementary values $\xi_k$ and put
\begin{equation}\label{eq:mumford-mode-inverse}
 \bar{\xi}_k=\frac{\kappa_k\mathcal P_k}{\xi_k},
 \qquad1\leq k\leq K.
\end{equation}
Substitution of the BA logarithmic rates into the Wronskians
\cref{eq:BA-pair-composites} determines every composite algebraically:
\begin{equation}\label{eq:mumford-composite-inverse}
\begin{aligned}
 \xi_{\ell j}
 &=\frac{\bar{\xi}_{\ell}\bar{\xi}_j}{\sqrt2(a_j-a_{\ell})}
       \left(\frac{\mathcal P_{z_g,j}+y_j}{\mathcal P_j}
             -\frac{\mathcal P_{z_g,i}+y_{\ell}}{\mathcal P_{\ell}}\right),\\
 \bar{\xi}_{\ell j}
 &=-\frac{\xi_{\ell}\xi_j}{\sqrt2(a_j-a_{\ell})}
       \left(\frac{\mathcal P_{z_g,j}-y_j}{\mathcal P_j}
             -\frac{\mathcal P_{z_g,i}-y_{\ell}}{\mathcal P_{\ell}}\right),
 \qquad1\leq\ell\leq L<j\leq K.
\end{aligned}
\end{equation}
These formulas apply on the chart $\mathcal P_k\ne0$,
$1\leq k\leq K$. Jacobi inversion supplies the corresponding
$z$, and the constants $c_k$ in \cref{eq:BA-elementary} match the
chosen elementary values. Under the conditions of
\cref{cor:BA-realization}, the lifted modes then follow the
canonical mode equations.

The polynomial triple determines the divisor state; the $K$
elementary normalizations specify its mode lift. In particular,
the phase-dependent individual BA functions are retained by the
mode description, and the composites are fixed once those
normalizations have been chosen. After forgetting this gauge
freedom, the fixed-curve BA family gives an open part of the usual
Mumford fiber, a Jacobian with its theta divisor removed
\cite[Proposition~3.1]{FernandesVanhaecke2001}. This identifies
the fixed-curve solutions and their vector fields. A correspondence
of Poisson structures would additionally require a reduction of
the canonical mode bracket.

\subsubsection{Mode gauges and the divisor lift}
\label{sec:mumford-gauge-lift}

We specialize the general symmetry interpretation of
\cref{sec:gauge-Abel-symmetries} to identify the mode phases omitted
by the divisor projection.

The $K=L+M$ independent one-point charges generate the reciprocal
mode gauges
\begin{equation}\label{eq:BA-gauge-action}
\begin{aligned}
 \xi_k&\longmapsto e^{\theta_k}\xi_k,&
 \bar{\xi}_k&\longmapsto e^{-\theta_k}\bar{\xi}_k,
 &&1\leq k\leq K,\\
 \xi_{\ell j}&\longmapsto e^{-(\theta_{\ell}+\theta_j)}\xi_{\ell j},&
 \bar{\xi}_{\ell j}&\longmapsto e^{\theta_{\ell}+\theta_j}\bar{\xi}_{\ell j},
 &&1\leq\ell\leq L<j\leq K.
\end{aligned}
\end{equation}
Here the $\theta_k$ are constant complex parameters; purely imaginary
values preserve the conjugate real form. Every term of $H_0$ is
invariant. These symmetries follow from the polynomial Hamiltonian
itself; the BA realization identifies them with changes of the
normalization constants $c_k$.

For the prescribed hyperelliptic BA family, $g$ divisor coordinates and
$K$ mode normalizations specify the lift described in
\cref{eq:mumford-mode-inverse,eq:mumford-composite-inverse}.
Its spectral level is identified in
\cref{prop:BA-prescribed-spectral-curve}. The realized $H_0$
flow translates $z$ in the $-ie_g$ direction.
\Cref{thm:BA-higher-flows} identifies the further spectral flows
with translations of this same BA family and the constant mode phases
in \cref{eq:BA-higher-time-phase}. Their Abel directions span its
Jacobian, while the power charges generate the reciprocal gauges.

\subsubsection{Mechanical coordinate realizations}

An explicit connection with Garnier coordinates follows by taking
$L=g$. Then $q_L=1$ and $\kappa_{\ell}=1/(2R_L'(a_{\ell}))$,
$1\leq\ell\leq g$. Introduce local coordinates $r_{\ell}$ by
$r_{\ell}^2=4I_{\ell}$. Lagrange interpolation gives
\begin{equation}\label{eq:mumford-garnier-potential}
 \sum_{\ell=1}^{g}r_{\ell}^2
 =2\left([x^{g-1}]\mathcal P+\sum_{\ell=1}^{g}a_{\ell}\right).
\end{equation}
Using \cref{eq:mumford-product-identity} to differentiate these
coordinates gives
\begin{equation}\label{eq:mumford-garnier}
 -\ddot r_{\ell}
 =\left(\alpha_{\ell}-\sum_{j=1}^{g}r_j^2\right)r_{\ell}
   -\frac{f(a_{\ell})}{R_L'(a_{\ell})^2r_{\ell}^3},\qquad
 \alpha_{\ell}=a_{\ell}+\frac{\lambda_{2g}}4+2\sum_{j=1}^{g}a_j,
 \quad1\leq\ell\leq g.
\end{equation}
In the geometric Abel time $-it$, this is the Garnier form with centrifugal terms
\cite{Kostov1999,Vanhaecke2001}. At marked branch points the
corresponding inverse-cube terms vanish. The elementary BA modes
provide a description retaining the sheet-pair phases as well as
these mechanical coordinates.

For the Neumann realization of the same divisor geometry, see
\cite[Chapter~IIIa, \S4]{Mumford1984}. Vanhaecke discusses its
connections with Garnier and integrable H\'enon--Heiles systems
\cite{Vanhaecke2001}; the genus-two case~(ii) H\'enon--Heiles
system and its BA description are treated explicitly in
\cite[\S4]{Braden2005}.

\paragraph{Higher-rank Mumford systems.}
Pedroni and Vanhaecke extend the Mumford construction to polynomial
matrices in $\mathfrak{sl}_r$ \cite{PedroniVanhaecke1998}.
Our preliminary calculations connect a stationary $(3,4)$ trigonal
BA family to their reduced $\mathfrak{sl}_3$ flow, suggesting a
higher-rank counterpart of the hyperelliptic mode-power correspondence.
We leave its systematic development to future work.

\section{Numerical validation}\label{sec:numerical-validation}

We test the forward realization, higher-flow normalization and inverse
reconstruction separately, comparing BA formulas with independent
Hamiltonian evolution.

\subsection{Prescribed hyperelliptic and trigonal BA solutions}
\label{sec:numeric-prescribed-BA}

We compare the one-point BA construction ($N=1$) with independent integration of
Hamilton's equations on a genus-two hyperelliptic curve and a smooth
genus-three trigonal curve. Analytic BA values and derivatives use
Generapy's reference-normalized BA jets; directional sigma jets supply
the pair-mode potentials \cite{Generapy}. Periods and first- and
second-kind lifts are computed once per example, with fixed reference
normalizations. The canonical vector field is
then evaluated separately from the polynomial Hamiltonian
\cref{eq:BA-H0-flat}.

\subsubsection{Curves and prescribed BA data}

The hyperelliptic example uses
\begin{equation}\label{eq:numeric-hyperelliptic-curve}
 y^2=4x(x^2-1)(x^2-4),\qquad L=M=2,
\end{equation}
and hence has eight canonical mode pairs. The ordered point
abscissae are $0.1,2.3,0.5,3.0$, with positive real ordinates in the
first group and negative ordinates in the second. We use the Baker
differentials of \cref{eq:BA-curve,eq:BA-second-kind} and reference
vector $z_0=\omega e_2+\tfrac{4i}{5}(1,1)^{\trans}$, where
$\omega$ is the first-kind half-period matrix. These are the normalized
BA points $\widehat Q_k$ of \cref{eq:BA-general-dictionary}.
This choice gives real phase coefficients, positive normalized weights
and the all-positive real form
$\bar\xi_k=\xi_k^*$, $1\leq k\leq K$, and
$\bar\xi_{ij}=\xi_{ij}^*$, $1\leq i\leq L<j\leq K$.
Thus every mode power is a squared modulus on this explosive-type solution.

The trigonal example uses
\begin{equation}\label{eq:numeric-trigonal-curve}
 y^3=x^4+2x^3+3x^2+5x+7,\qquad L=M=3,
\end{equation}
and hence has fifteen canonical mode pairs. The compatible
first- and second-kind bases are
\begin{equation}\label{eq:numeric-trigonal-bases}
 du=\frac{(1,x,y)^{\trans}}{3y^2}\,dx,\qquad
 dr=\frac{((5x^2+6x+3)y,\,2xy,\,x^2)^{\trans}}{3y^2}\,dx.
\end{equation}
At the unique point at infinity write $x=s^{-3}$ and
$y=s^{-4}w(s)$, where
$w(s)^3=1+2s^3+3s^6+5s^9+7s^{12}$ and $w(0)=1$.
Let $\epsilon_3=e^{2\pi i/3}$. The points have $x=r^3$ and
$y=\epsilon_3^j f(r^3)^{1/3}$, where $f(x)$ is the quartic in
\cref{eq:numeric-trigonal-curve}, its displayed cube root is positive,
and the ordered $(r,j)$ lists are
\[
 (0.6,0),\ (0.9,1),\ (1.4,2)
 \quad\text{and}\quad
 (0.7,0),\ (1.1,1),\ (1.8,2).
\]
The lifts follow the corresponding infinity-chart paths. The finite
part of $\int dr_3$ has zero constant term in $s$; explicitly its
principal-sheet value is
\[
 R_3(s)=s^{-1}+\int_0^s\frac{1-w(q)^{-2}}{q^2}\,dq.
\]
The reference vector is
$z_0=(0.6+0.3i,\,0.2+0.15i,\,0.1+0.2i)$.

In both cases we use reference-normalized sigma ratios
\begin{equation}\label{eq:numeric-reference-sigma}
 \psi_\pm(P,z;z_0)=
 \frac{\sigma(A(P)\pm z)\sigma(z_0)}
 {\sigma(A(P)\pm z_0)\sigma(z)}
 \exp\bigl(\pm\boldsymbol R(P)^{\trans}(z-z_0)\bigr).
\end{equation}
These equal one at $z_0$. The residue criterion in
\cref{thm:general-BA-realization} determines the normalized weights
$\kappa_k$ and phase coefficients $b_k$, $1\leq k\leq K$.
The initial elementary gauges are $c_k=\sqrt{\kappa_k}$, using the
principal complex square root. Thus
$\xi_k(0)=\bar{\xi}_k(0)=\sqrt{\kappa_k}$; the pair-mode initial data
are fixed by \cref{eq:general-pair-modes}.
For the trigonal example the resulting constants, rounded to six
decimal places, are
\begin{center}
\begin{tabular}{ccc}
\hline
$k$ & $\kappa_k$ & $b_k$\\
\hline
1 & $0.278555-0.080913i$ & $6.227141+15.533670i$\\
2 & $-0.105763-0.895590i$ & $-5.060995+6.549498i$\\
3 & $-0.084898+0.199820i$ & $-3.591234+0.946300i$\\
4 & $0.238558-0.001427i$ & $13.329291+16.279427i$\\
5 & $0.005090-0.994442i$ & $-3.184404+7.790703i$\\
6 & $-0.044599+0.514974i$ & $-2.564030-1.368484i$\\
\hline
\end{tabular}
\end{center}

The numerical integrations use the full stored values.

To evaluate the trigonal pair-composite modes, we use an explicit
bilinear formula for the kernel in terms of the BA functions and
their derivatives. Differentiating it also gives analytic mode rates
without substituting Hamilton's equations.
Write $\partial=\partial_{z_3}$,
$f_i=\psi_+(Q_i,z;z_0)$ and $g_j=\psi_-(Q_j,z;z_0)$.
The scalar BA operator and its formal adjoint have the form
\[
 \mathscr L_3=\partial^3+p_3\partial+q_3,\qquad
 \mathscr L_3^{\rm ad}=-\partial^3-p_3\partial+q_3-\partial p_3,
 \qquad p_3=3\partial^2\log\sigma(z),
\]
with $\mathscr L_3 f_i=x_i f_i$ and
$\mathscr L_3^{\rm ad}g_j=x_jg_j$. Such scalar operators are supplied
by the BA construction \cite{Krichever1977}. Their Green identity gives
\begin{equation}\label{eq:numeric-trigonal-kernel}
 \Omega(Q_i,Q_j)=
 \frac{g_j\partial^2f_i-(\partial g_j)(\partial f_i)
              +(\partial^2g_j)f_i+p_3g_jf_i}{x_i-x_j}.
\end{equation}
Indeed the derivative of the numerator is
$g_j\mathscr L_3 f_i-f_i\mathscr L_3^{\rm ad}g_j
=(x_i-x_j)f_ig_j$; the coefficient $q_3$ cancels.
The kernel normalization fixes the primitive, and
$i\dot\Omega(Q_i,Q_j)=f_ig_j$ along $z=z_0-it e_3$.

\subsubsection{Independent evolution and accuracy}

Both examples use real time $t$, with $z=z_0-it e_g$,
$0\leq t\leq6$, in the stated Abel coordinates. The independent
integrator solves Hamilton's equations for $H_0$ with brackets $-i$,
exactly as in \cref{eq:general-network-rates}. The trigonal example
uses genuinely complex weights $\kappa_k$ and phase coefficients $b_k$,
$1\leq k\leq K$, and retains independent Hamiltonian partners.

Sigma jets are evaluated at 24 decimal digits. Mode assembly and the
independent fourth-order Runge--Kutta integration use double-precision
complex arithmetic. We evaluate the BA formulas at 41 points and
plot the numerical trajectory at 601 points. Integrations with
6,000 and 12,000 steps test convergence. The kernel is evaluated
from BA jets through the two-function Wronskian in the hyperelliptic
case and the bilinear formula \cref{eq:numeric-trigonal-kernel} in the
trigonal case.
Its derivative is computed by differentiating that pairing, without
substituting Hamilton's equations.

For each mode, the scaled discrepancy between values $a,b$ is
$|a-b|/(1+\max(|a|,|b|))$. Maxima are taken over all modes and all
41 comparison points, including modes not displayed in the figures.
The fine-step results are
\begin{center}
\begin{tabular}{lcc}
\hline
Check & Hyperelliptic & Trigonal\\
\hline
Mode discrepancy & $1.7\times10^{-11}$ & $1.8\times10^{-11}$\\
Mode-power discrepancy & $1.6\times10^{-11}$ & $1.2\times10^{-11}$\\
Canonical vector-field residual & $1.2\times10^{-15}$ & $1.2\times10^{-12}$\\
BA conserved-quantity drift & $1.0\times10^{-15}$ & $1.6\times10^{-12}$\\
Runge--Kutta conserved-quantity drift & $2.1\times10^{-14}$ & $8.6\times10^{-13}$\\
35-digit RK4 refinement ratio & $16.0$ & $16.0$\\
Extrapolated mode discrepancy & $2.8\times10^{-12}$ & $3.1\times10^{-12}$\\
\hline
\end{tabular}
\end{center}

\begin{figure}[p]
 \centering
 \includegraphics[width=\textwidth,height=0.82\textheight,keepaspectratio]
 {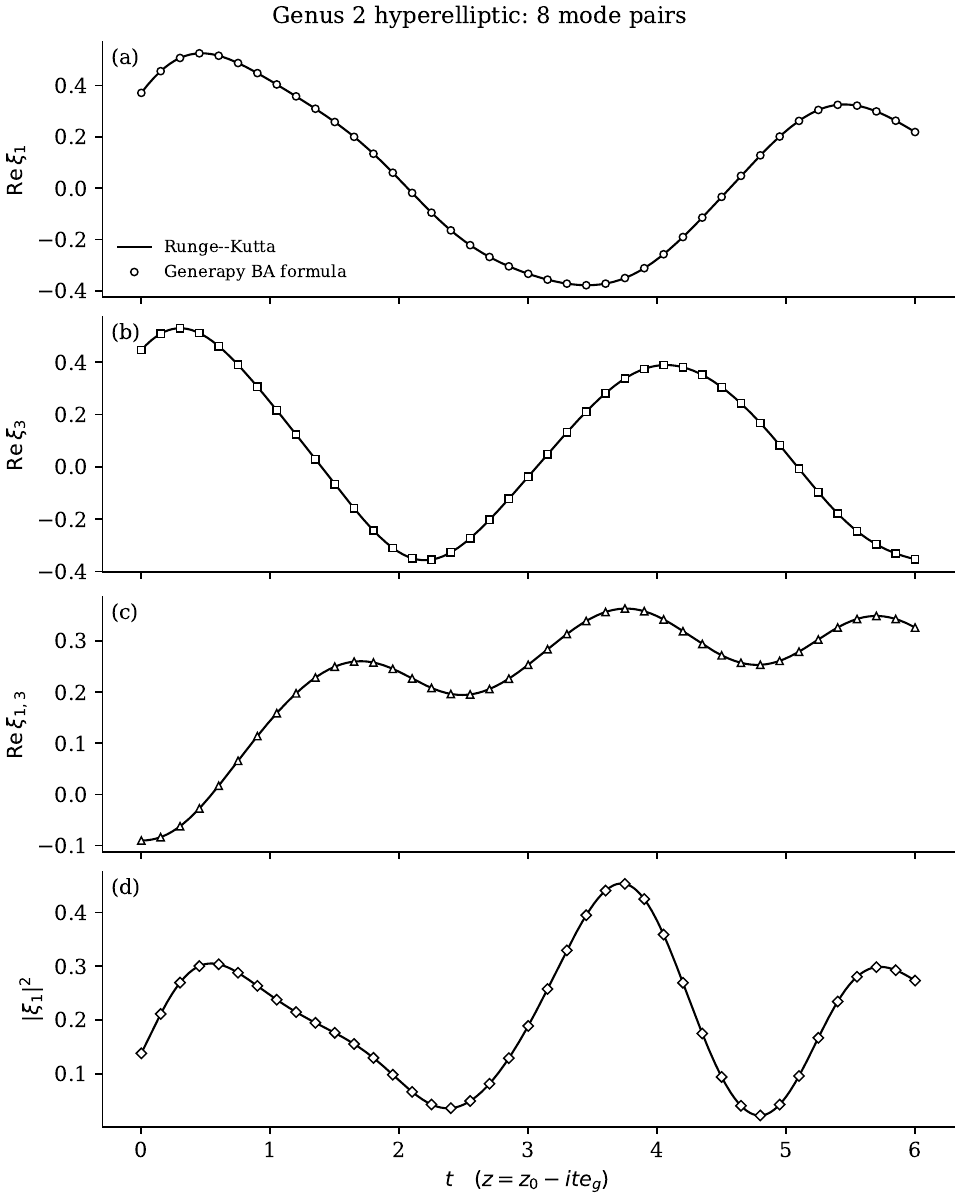}
 \caption{Genus-two hyperelliptic realization on
 \cref{eq:numeric-hyperelliptic-curve}. The panels show
 $\operatorname{Re}\xi_1$, $\operatorname{Re}\xi_3$,
 $\operatorname{Re}\xi_{1,3}$ and
 $|\xi_1|^2$.
 Open markers are direct Generapy BA evaluations; solid black lines
 are independent Runge--Kutta solutions of the canonical equations
 generated by $H_0$ with brackets $-i$, and $z=z_0-it e_2$.}
 \label{fig:numeric-hyperelliptic}
\end{figure}

\begin{figure}[p]
 \centering
 \includegraphics[width=\textwidth,height=0.82\textheight,keepaspectratio]
 {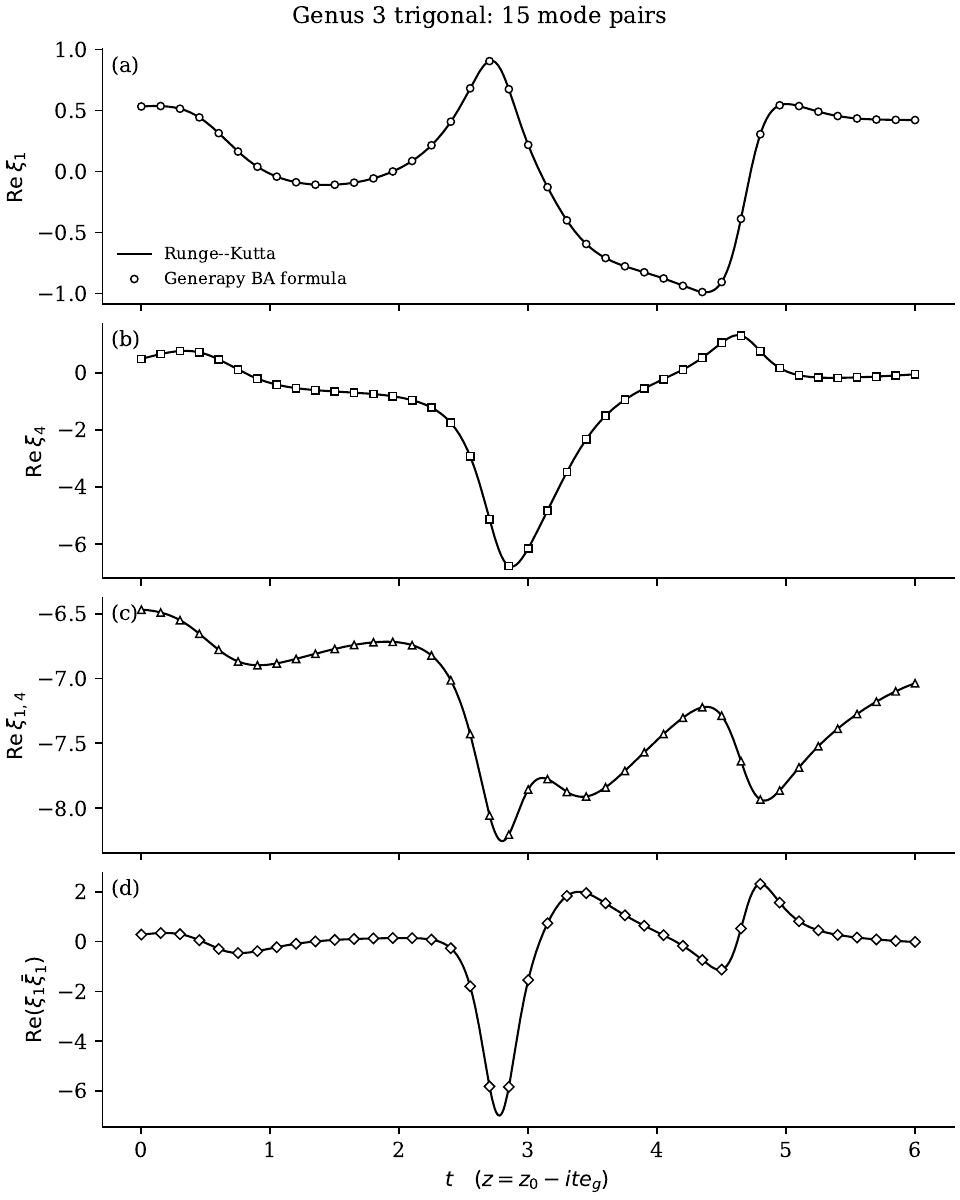}
 \caption{Genus-three trigonal realization on
 \cref{eq:numeric-trigonal-curve}. The panels show
 $\operatorname{Re}\xi_1$, $\operatorname{Re}\xi_4$,
 $\operatorname{Re}\xi_{1,4}$ and
 $\operatorname{Re}(\xi_1\bar{\xi}_1)$.
 Open markers are direct Generapy BA evaluations; solid black lines
 are independent Runge--Kutta solutions of the canonical equations
 generated by $H_0$ with brackets $-i$, and $z=z_0-it e_3$.}
 \label{fig:numeric-trigonal}
\end{figure}

We also check the spectral map
\cref{eq:BA-prescribed-spectral-map} at twelve curve points at each
of three BA states. The maximum ratio of the smallest to largest
singular value of
$\lambda(P)\Id-\mathsf L_0(\nu(P))$ is
$2.2\times10^{-16}$ for the hyperelliptic example and
$6.8\times10^{-17}$ for the trigonal example. The corresponding
total excess singularity defects in
\cref{eq:BA-prescribed-genus-defect} are two and six.
The vector-field residual directly tests the BA functions against
Hamilton's equations; trajectory discrepancies also include numerical
integration and stored-data errors. To separate these effects, a
35-digit Runge--Kutta calculation promotes the same stored coefficients
and initial values and uses $6{,}000$, $12{,}000$ and $24{,}000$ steps.
For both examples the ratio of successive trajectory differences is
$16.0$, confirming fourth-order convergence. The table also gives the
Richardson-extrapolated discrepancies; these retain the precision limits
of the stored parameters and BA values. The original double-precision
BA-comparison ratios therefore do not measure the integrator's order.
For the hyperelliptic run, the maximum scaled discrepancy between
barred partners and actual complex conjugates is $3.0\times10^{-16}$
for the BA evaluations and $1.8\times10^{-12}$ for Runge--Kutta.
The conservation checks include $H_0$ and every row and column power charge
\[
 I_i-\sum_{j=L+1}^{K}I_{ij},\quad1\leq i\leq L,
 \qquad
 I_j-\sum_{i=1}^{L}I_{ij},\quad L+1\leq j\leq K,
 \qquad I_{ij}=\xi_{ij}\bar{\xi}_{ij}.
\]
\Cref{fig:numeric-hyperelliptic,fig:numeric-trigonal} show representative
elementary modes, a pair composite and an elementary mode power.
The scripts \path{code/ba_numeric_examples.py} and
\path{code/reference_sigma_ba.py}, together with the accompanying
parameter records, are available in the
accompanying project repository at
\url{https://github.com/HeskethGD/ba-canonical-ham}
and reproduce the comparisons.

\subsection{Multipuncture solutions and independent Abel directions}
\label{sec:numeric-multipuncture-flows}

We test the full three-array construction and its higher-flow phases,
including two independent Abel directions on one genus-two curve.

\subsubsection{A multipuncture elliptic example}
\label{sec:numeric-multipoint}

To test all three nonsingleton arrays, take $L=M=N=2$ on
$\C/(\mathbb Z+\tau\mathbb Z)$ with $\tau=0.23+1.17i$ and punctures
$R_1=0.13+0.14i$, $R_2=0.41+0.29i$.
Write $\Theta(P)=\theta[\frac12,\frac12](P\mid\tau)$ and
$E_1=\partial_P\log\Theta$. Choose
\[
 d\Omega_0=\beta(P)dP,\qquad
 \beta(P)=E_1'(P-R_1)+E_1'(P-R_2)+0.23-0.19i.
\]
Its four zeros split into two ordinary and two dual divisor points;
the ordinary points, rounded here, are
\[
 D_1=1.13522461+0.80548828i,\qquad D_2=0.16709062+0.35074039i.
\]

The reference is $z_0=R_1+R_2-D_1-D_2$, with these exact lifts,
and $z(t)=z_0+2it$. For
$\mathfrak a(P)=\prod_{k=1}^{2}\Theta(P-R_k)/
\prod_{a=1}^{2}\Theta(P-D_a)$ and
$\Xi(P)=\sum_{k=1}^{2}E_1(P-R_k)$, the kernel is
\[
 \Omega(P,Q)=\frac{\mathfrak a(P)}{\beta(Q)\mathfrak a(Q)}
 e^{-it(\Xi(P)-\Xi(Q))}
 \frac{\Theta'(0)\Theta(z(t)+P-Q)}{\Theta(z(t))\Theta(P-Q)}.
\]
Its puncture limits give the two normalized bases. Use the four
ordinary evaluation points
\[
 (Q_1,Q_2,Q_3,Q_4)
 =(0.07+0.03i,\ 0.57+0.08i,\ 0.78+0.21i,\ 0.26+0.63i)
\]
and principal coefficients $(r_1,r_2,r_3,r_4)=(-0.7,-1.3,-0.9,-1.1)$.
Each group sum is $-2$, the required genus-one Abel velocity.
Thus $h_L=\Xi+\sum_{i=1}^{2}r_iE_1(P-Q_i)$ and
$h_M=\Xi+\sum_{j=3}^{4}r_jE_1(P-Q_j)$ give the interpolation
functions, weights and phases. This genus-one realization has
excess spectral defect six, since the system's generic genus is seven.

Generapy's existing public theta jets evaluate the kernel and its
analytic derivatives at $55$ digits. The multipuncture basis and
fixed divisor factors are assembled from these data. The sigma versions in
\cref{sec:multipoint-BA-data} are likewise computable from its sigma
jets once the puncture and second-kind data have been supplied.
At $t=0,0.031,0.087$ we compare all $24$ analytic mode rates with
the separately assembled canonical field, and check both spectral
eigenvectors and the reference-normalized bilinear kernel.
For the shifted trace orders $r=3,4$, with $\nu_0=0.17+0.09i$,
meromorphic finite parts determine the higher BA time and all vertex
phases. Their analytic rates are compared with the independent matrix
field $i\dot{\mathsf X}=[\mathsf D,(\mathsf X+\nu_0\mathsf D)^{r-1}]$:
\begin{center}
\begin{tabular}{lr}
\hline
Check & Maximum scaled residual\\
\hline
Canonical mode rates & $2.9\times10^{-55}$\\
Rank-two kernel rate & $5.1\times10^{-56}$\\
Left and right spectral eigenvectors & $1.9\times10^{-55}$\\
Reference-normalized kernel & $2.6\times10^{-56}$\\
Shifted trace, order 3 & $5.0\times10^{-55}$\\
Shifted trace, order 4 & $6.7\times10^{-55}$\\
\hline
\end{tabular}
\end{center}

The scripts \path{code/multipoint_BA_elliptic_check.py} and
\path{code/multipoint_higher_flow_check.py} record the full data and
these residuals. Exact rational-curve checks and conserved block
moments provide separate symbolic checks of the signs and the
repeated-phase basis construction.

\subsubsection{Two independent genus-two multipuncture flows}
\label{sec:numeric-multipoint-genus2}

For a higher-genus test of the full arrays, retain $L=M=N=2$ and
use the smooth genus-two curve of
\cref{eq:numeric-hyperelliptic-curve}. Write
$P(x)=(x,2\sqrt{x(x^2-1)(x^2-4)})$, with the principal square root
at the specified points. Choose
\[
 R_1=P(0.25),\quad R_2=P(2.4),\qquad
 (Q_1,Q_2,Q_3,Q_4)=(P(0.4),P(2.6),P(0.8),P(3.3)),
\]
with the first two $Q$ points in the $L$ group. The ordinary divisor
has points $D_a=P(d_a)$, $1\leq a\leq3$, where
\[
 (d_1,d_2,d_3)=(0.1+0.15i,\ 0.7+0.25i,\ 1.4+0.2i).
\]
Here the local parameters at the punctures are $s_k=x-x(R_k)$,
$1\leq k\leq2$. Reference sections are constructed as rational functions with poles
at the three $D_a$ and unit leading vectors at the two punctures.
The duality differential is determined by its zeros at $D_a$ and
leading parts $-ds_k/s_k^2$; its remaining three zeros give the dual
divisor. Rational interpolation gives $h_L,h_M$, the weights and
phase constants. Generapy Abel maps and theta jets then evaluate
\cref{eq:multipoint-BA-theta,eq:multipoint-mode-arrays}. The full
construction is in \path{code/multipoint_BA_genus2_check.py}.

We compare the cubic Hamiltonian $H$ with the commuting fourth-trace
Hamiltonian $H_{4,\nu_0}=\tr(\mathsf X+\nu_0\mathsf D)^4/4$,
$\nu_0=0.17$. Meromorphic time normalization uses the two simple
puncture principal parts as representatives of the Abel directions.
In normalized Abel coordinates their trajectories are
$z=z_0-it\boldsymbol U$ and $z=z_0-it_4\boldsymbol U_4$, with
\[
 \boldsymbol U\simeq
 \begin{pmatrix}0.05678315\\0.49752720\end{pmatrix},\qquad
 \boldsymbol U_4\simeq
 \begin{pmatrix}-0.07762562\\-3.74854497\end{pmatrix},\qquad
 \det(\boldsymbol U,\boldsymbol U_4)\simeq-0.17423333.
\]
Thus the two flows exercise independent Jacobian directions on the
same curve with the same initial modes. At $32$ digits, analytic
mode rates and the left and right pencil eigenvectors have scaled
residuals below $2\times10^{-29}$; the fourth-trace mode rates have
residual below $3\times10^{-26}$. The distinct third-group phases
also satisfy the birationality criterion of
\cref{prop:BA-prescribed-spectral-curve}. This is a genus-two level
of the generically genus-seven system, with spectral defect five.

\begin{figure}[p]
 \centering
 \includegraphics[width=\textwidth,height=0.82\textheight,keepaspectratio]
 {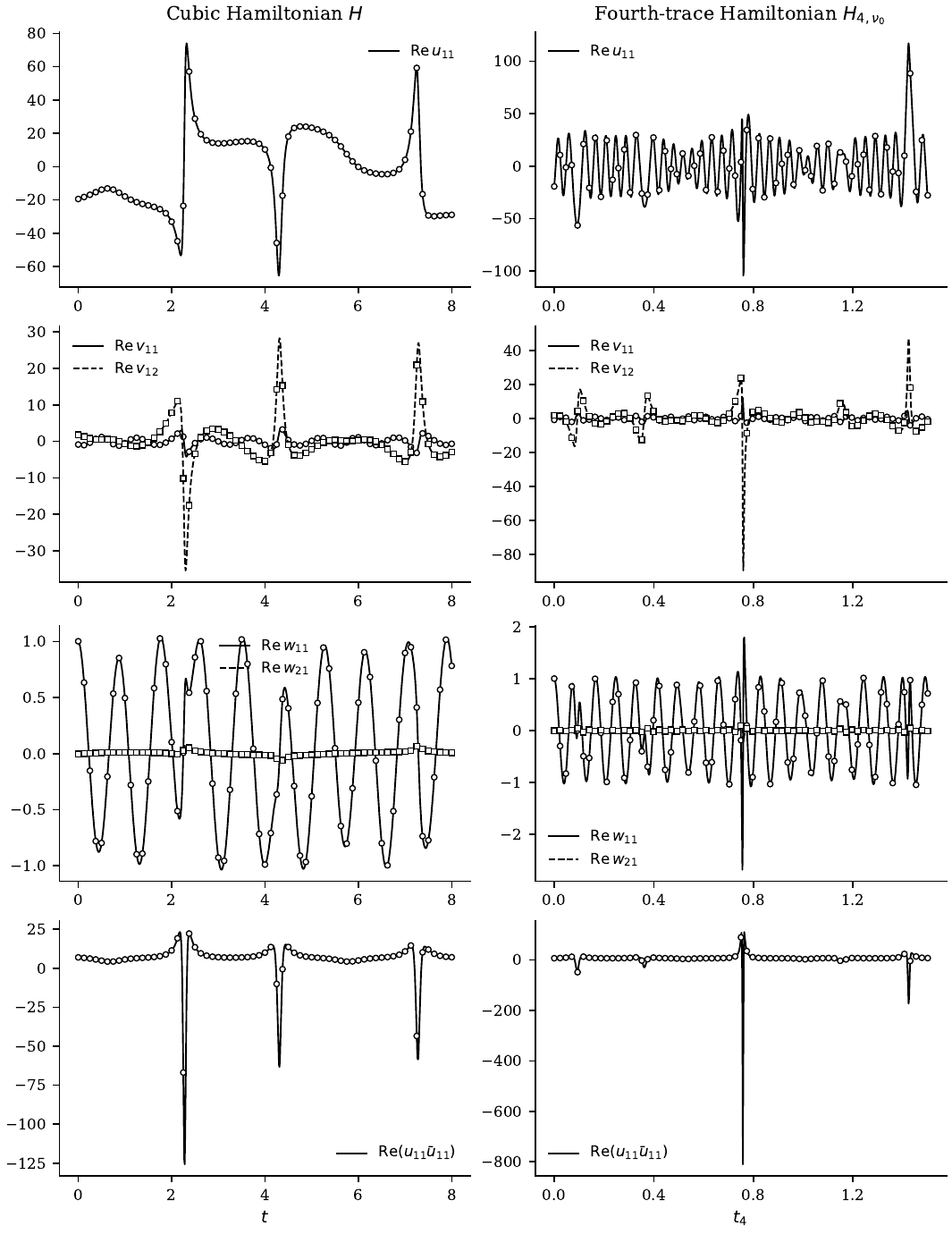}
 \caption{Genus-two realization with $L=M=N=2$. Left: the cubic
 Hamiltonian $H$; right: the commuting fourth-trace Hamiltonian
 $H_{4,\nu_0}$, with $\nu_0=0.17$. Each column uses the same initial modes. The panels
 show $\operatorname{Re}u_{11}$, the two puncture-indexed modes
 $\operatorname{Re}v_{11}$ and $\operatorname{Re}v_{12}$,
 $\operatorname{Re}w_{11}$ and $\operatorname{Re}w_{21}$, and
 $\operatorname{Re}(u_{11}\bar u_{11})$.
 Open circles and squares are direct Generapy-based BA evaluations;
 solid and dashed black lines are independent RK4 trajectories.
 The barred modes are independent Hamiltonian partners.}
 \label{fig:numeric-multipoint-genus2}
\end{figure}

\Cref{fig:numeric-multipoint-genus2} compares $65$ direct BA samples
per flow with dense independent RK4 solutions. The cubic flow is
integrated to $t=8$ with $8000$ and $16000$ steps; the fourth-trace
flow is integrated to $t_4=1.5$ with $32000$ and $64000$ steps.
The table reports maximum scaled errors over all $24$ modes and
integral drifts for the Hamiltonian, power charges and shifted
spectral traces:
\begin{center}
\begin{tabular}{lrr}
\hline
Check & Cubic flow & Fourth-trace flow\\
\hline
Direct BA versus fine RK4 & $9.1\times10^{-9}$ & $5.5\times10^{-8}$\\
BA integral drift & $6.3\times10^{-15}$ & $8.8\times10^{-15}$\\
RK4 integral drift & $3.0\times10^{-10}$ & $3.5\times10^{-9}$\\
\hline
\end{tabular}
\end{center}

At $40$ digits the analytic residuals decrease further; both precision
records and the trajectories accompany the code.

\subsubsection{The second normalized matrix-KP time}
\label{sec:numeric-KP-time}

Using the same genus-two $L=M=N=2$ data, we test
\cref{prop:KP-hierarchy} with the local spectral coordinate $h_L$.
The principal parts of $h_L^2$ determine the second KP time. A rational
correction reduces those double poles to the simple-pole time
representatives, retaining their endpoint and puncture phases.
Complete theta jets through order two give the BA time derivatives
independently of the KP differential equations.
For $\mathcal L_{\rm KP}=\partial+V\partial^{-1}+O(\partial^{-2})$,
the second differential generator is $\mathcal B_2=\partial^2+2V$.
We compare its ordinary and dual source equations, its bilinear
kernel derivative, and all $24$ mode rates at three states:
\begin{center}
\begin{tabular}{lr}
\hline
Check & Maximum scaled residual\\
\hline
Ordinary/source second-time equations & $3.8\times10^{-31}$\\
Dual/source second-time equations & $2.3\times10^{-31}$\\
Second-time kernel identity & $6.5\times10^{-31}$\\
All canonical rates: spectral-plus-gauge match & $2.3\times10^{-30}$\\
\hline
\end{tabular}
\end{center}

The canonical comparison solves the Abel-direction matching problem
once and fixes the gauge rates at the initial state, then checks
those same coefficients at two further states. The script
\path{code/matrix_kp_hierarchy_check.py} records the coefficients,
principal-part correction and $192$ exact ambient source-bracket
checks. Its stored results generate the displayed table.

\subsection{BA reconstruction from canonical initial modes}
\label{sec:numeric-BA-converse}

We start from canonical initial modes to test the inverse construction
independently of examples initialized from prescribed BA data.

To test the reverse construction of \cref{thm:BA-spectral-converse},
start with $L=M=1$, $N=2$ and independent canonical launch data,
\[
 b_1^L=\tfrac17,\quad b_1^M=-\tfrac29,\quad
 (b_1^N,b_2^N)=(\tfrac13,-\tfrac15),
\]
\[
 \mathsf X(0)=\begin{pmatrix}
 -2/7&(7+2i)/10&(9+2i)/10&(4-i)/10\\
 (4-3i)/10&4/9&(6+i)/10&(-3+5i)/10\\
 -(5-4i)/10&-(8-2i)/10&1/3&0\\
 -(-2+3i)/10&-(2+4i)/10&0&-1/5
 \end{pmatrix}.
\]
The off-diagonal entries specify all five canonical mode pairs by
\cref{eq:canonical-X}; no BA data are used to choose them.
Exact symbolic computation gives
$F(\lambda,\nu)=A(\lambda)\nu^2+B(\lambda)\nu+C(\lambda)$.
Its discriminant $B^2-4AC$ is a square-free polynomial of degree six,
so the smooth compactification of $y^2=B^2-4AC$ has genus two.
Here $g=G=2$:
this tests the generic spectral genus, with no excess defect.

The adjugate normalization of \cref{eq:BA-converse-functions}, with
$a=(0,0,1,1)^{\trans}$, determines the ordinary and dual divisors
from the zeros of
$a^{\trans}\operatorname{adj}(\lambda\Id-\mathsf L)a$.
The two types of zero are distinguished by which of the left and
right anchor contractions vanishes. Both divisors have degree three.
A branch-point change $x=(\lambda-e)^{-1}$, $Y=y x^3$ gives an
odd-degree model for Generapy's period and Abel-map evaluation.
The branch point sent to infinity fixes the numerical Abel base
and its Riemann characteristic. The punctures and divisors are transported with this change.
The BA parameters remain $s_k=-1/\nu$, so their principal
coefficients and finite parts are recomputed in the $x$ chart.
Their reference sections and duality differential recover every
initial mode algebraically before any theta evaluation. We then
evolve the reconstructed theta formulas and compare all ten analytic
rates with the canonical field:
\begin{center}
\begin{tabular}{lr}
\hline
Check & Maximum scaled error\\
\hline
Algebraic recovery of initial modes & $2.6\times10^{-41}$\\
Recovery of residue weights & $3.3\times10^{-41}$\\
Reference eigenline and kernel recovery & $5.8\times10^{-41}$\\
All ten analytic mode rates & $3.9\times10^{-40}$\\
Direct BA versus fine RK4 & $4.1\times10^{-13}$\\
\hline
\end{tabular}
\end{center}

The RK4 comparison uses $65$ direct BA samples over $0\leq t\leq2$,
with $1600$ and $3200$ steps. The error ratio is $16.01$.
The script \path{code/ba_genus2_converse_check.py} and its stored
records contain the exact characteristic polynomial, both recovered
divisors and the complete reconstruction data.

Generapy's tests against mathematical identities and its reproductions
of published results are documented in its repository and
documentation \cite{Generapy}.

\clearpage

\section{Discussion}\label{sec:discussion}

We bring together the functional and Hamiltonian correspondences to
explain the common BA setting and identify directions for further work.

The canonical multiwave system provides a canonical
model for scalar BA dynamics whose interaction structure is independent
of the curve type. The two-way correspondence for arbitrary positive $L,M,N$ and the
higher-flow result give this interpretation its substance: prescribed
BA evaluations find a home in the cubic equations, generic canonical
initial modes recover their BA data, and the spectral Hamiltonians
supply every Abel direction on birational spectral levels together
with the mode phases.
This correspondence motivates further study of the system's
Hamiltonian hierarchy and reductions across different curve families.
The invariant real forms in \cref{sec:canonical-modes} apply to real
initial data; explicit curve-and-divisor reality criteria, particularly
for decay-form BA realizations, remain to be developed.

The realization criterion turns prescribed curve data into canonical
multiwave dynamics: residue equations determine compatible weights,
and opposite-group interpolation values determine the linear phase
rates. The kernel lemma gives the differential identities behind this
construction. The matrix constrained-KP correspondence in
\cref{sec:BA-constrained-KP} packages these identities into two
commuting finite-rank operators for arbitrary $N$. Choosing the local
spectral coordinate fixes their KP times and source evolution;
spectral Hamiltonians and power charges give their canonical
generators on each birational BA level. The one-point member recovers
scalar operators and their conserved coefficients.
\Cref{prop:BA-prescribed-spectral-curve,thm:BA-spectral-converse}
identify the realized curve and reconstruct BA data from generic
canonical initial modes of the full three-array system.

The comparison with Adams, Harnad and Hurtubise gives two compatible
views of the same matrix evolution. Their canonical moment-map lift
has quartic interactions, while the cubic modes use the constant
member of the matrix Poisson pencil. Complete-fiber BA convolution
turns the bilinear lift into endpoint BA functions and expresses this
change of interaction order as a Sylvester displacement relation.
This connection complements the prescribed-point realization: it
explains how the direct cubic modes sit within the established
spectral construction while retaining a simple canonical bracket.

For the full three-matrix family, the spectral and principal-moment
integrals give the complete Liouville count for every fixed phase
profile, with the detailed proof in Appendix~\ref{sec:integrability-proof}. The symmetry interpretation in
\cref{sec:gauge-Abel-symmetries} distinguishes mode gauges, equal-phase
changes of basis and flows that become translations in spectral
divisor coordinates. These different symmetries explain why power
charges alone need not complete the integral count.

In the one-point hyperelliptic specialization, the Mumford projection
has a useful optical analogy in the passage from
complex field modes to Stokes variables. In elliptically solvable optical
systems, Stokes variables describe closed projected dynamics and determine
the logarithmic rates needed to reconstruct the individual modes.
The Mumford variables play a similar role here: the powers and their
derivatives determine the moving divisor polynomials, whose evolution
closes and whose values determine the BA logarithmic rates in
\cref{eq:BA-powers-rates}. The BA formulas give the mode reconstruction
explicitly, with the remaining gauge choices carried by the constant
normalizations $c_k$. Thus both descriptions connect projected dynamics
to the evolution of the underlying complex modes. The Stokes projection
omits one common optical phase, while the Mumford projection here omits
the $K$ independent reciprocal mode gauges.

The broader spectral family offers further function-theoretic problems.
Its marked plane models describe different covering degrees and branches
at infinity; special levels and involutions produce lower-genus
normalizations and Prym descriptions. The BA construction applies across these curve types: one puncture
gives the vector--matrix--vector member, while several punctures
produce the full arrays through a common scalar BA basis and a
rank-$N$ kernel derivative. The puncture count is independent of
the genus or the number of infinities in another plane presentation.
Singular spectral levels and nonsemisimple moment
blocks also require limiting constructions beyond the generic converse.

The commuting spectral Hamiltonians give a multi-time view of the
interactions. \Cref{thm:BA-higher-flows} computes their puncture times and constant
mode phases by finite meromorphic interpolation. On a birational
level, finitely many spectral coefficients generate every Abel direction.
A scalar evolution selects one such direction, and higher Hamiltonians
introduce higher-order mixing terms on the same mode space. This makes
the translation and gauge symmetries explicit in the BA representation.
The interaction orders and combinations associated with particular
directions provide a further problem for constructing polynomial
Hamiltonians from BA identities.

\section*{AI assistance and validation}

The author conceived the research project, formulated its mathematical
objectives and directed its development. OpenAI GPT~6.1 Sol, with High
reasoning, assisted with mathematical exploration and proof development,
manuscript preparation, symbolic and numerical verification code, and
plot generation. Anthropic Fable~5.2, with High reasoning, provided
iterative manuscript reviews. The author critically assessed the proposed
arguments and results, selected the material presented, and takes
responsibility for the final manuscript.

The work was checked through iterative mathematical review, exact
symbolic calculations of selected identities and compatibility conditions,
and numerical comparisons between direct BA function evaluations using
Generapy and separate Runge--Kutta integration of Hamilton's equations.
Hyperelliptic and trigonal examples were tested, including refinement
calculations to assess numerical convergence, as described in
\cref{sec:numerical-validation}. The verification code, parameters and
recorded results accompany the manuscript.

Generapy was also developed by the author with AI assistance. Its
automated validation includes unit tests against known mathematical
identities, integration tests connecting its curve, Abelian-integral,
theta-function and Kleinian-function components, and numerical
reproductions of results from the published literature. The tests and
literature-based examples are available in the Generapy repository and
documentation \cite{Generapy}.

\appendix
\addtocontents{toc}{\protect\setcounter{tocdepth}{1}}
\section{Comparison with Menyuk, Chen and Lee}
\label{sec:MCL-dictionary}

The restricted shared-mode Hamiltonian and the broader wave construction
in \cite{MenyukChenLee1983Integrable} give two distinct comparisons
with the canonical equations used here.

\subsection{The restricted shared-mode family}

Write the shared mode as $p$ and the remaining pairs as $q_j,r_j$,
$1\leq j\leq R$. Complexify the conjugates as independent partners.
With brackets $\{p,\bar p\}=-i$ and the same convention for the
other pairs, the action--angle Hamiltonian in
\cite[equation (1.1)]{MenyukChenLee1983Integrable} is
\begin{equation}\label{eq:MCL-star-H}
 H_{\rm MCL}=\sum_{j=1}^{R}\left[
 \frac{\Delta_j}{2}(q_j\bar q_j-r_j\bar r_j)
 -\frac{\epsilon_j}{2}
 (p\bar q_jr_j+\bar p q_j\bar r_j)\right].
\end{equation}
Here $\Delta_j$ is the frequency mismatch and $\epsilon_j$ the
coupling. For the equal-coupling family $\epsilon_j=\epsilon$, set
\[
 u=p,\quad\bar u=\bar p,\qquad
 v_j=i\bar q_j,\quad\bar v_j=iq_j,\qquad
 w_j=r_j,\quad\bar w_j=\bar r_j,
 \qquad1\leq j\leq R.
\]
This is a complex canonical change: for example,
$\{i\bar q_j,iq_j\}=-i$. It gives
\begin{equation}\label{eq:MCL-star-map}
 H_{\rm MCL}=-\frac12\sum_{j=1}^{R}\Delta_j
 (v_j\bar v_j+w_j\bar w_j)
 +\frac{i\epsilon}{2}\sum_{j=1}^{R}
 (uv_jw_j+\bar u\bar v_j\bar w_j).
\end{equation}
Thus the restricted model has our interaction arrays with
$(L,M,N)=(1,1,R)$, up to the constant coupling and time normalization.
Its original physical real form becomes $\bar v_j=-v_j^*$,
$\bar u=u^*$ and $\bar w_j=w_j^*$ under this complex dictionary;
the transformed coupling is imaginary. This is a different convention
from imposing actual conjugation on all partner arrays.

Menyuk, Chen and Lee give a complete set of $2R+1$ independent
commuting integrals for this restricted family, including coincident
mismatches \cite[Section~III]{MenyukChenLee1983Integrable}.
Their Section~V assumes equal mismatches as well as equal couplings:
the shared power is elliptic, its phase is an elliptic integral of
the third kind, and the remaining modes are recovered by quadratures.

\subsection{The broader matrix wave construction}

Their equation (2.13), attributed in its reality-reduced form to
Ablowitz and Haberman, is
\begin{equation}\label{eq:MCL-matrix-wave}
 \alpha_{hi}\partial_x N_{hi}
 =\partial_t N_{hi}
 +\sum_{\substack{j=1\\j\ne h,i}}^{d}
 N_{hj}N_{ji}(\alpha_{ji}-\alpha_{hj}),
 \qquad1\leq h,i\leq d,\quad h\ne i.
\end{equation}
For unequal source spectral constants,
$\alpha_{hi}=(c_i-c_h)/(i(a_i-a_h))$; coincident $a_h=a_i$,
$c_h=c_i$ permit separate choices. Section~IV takes
$N_{hi}(x,t)=N_{hi}(t)e^{\Delta_{hi}x}$ with
$\Delta_{hi}=\Delta_{hj}+\Delta_{ji}$, producing linear rates and
quadratic mode interactions.

Here is the reduction to our arbitrary-size canonical system.
Use the $H_0$ modes in \cref{eq:canonical-X} and write
\[
 \widetilde{\mathsf X}=E+Z,\qquad
 E=\diag(-2\mathsf B_L,-2\mathsf B_M,\mathsf B_N).
\]
The matrix $Z$ has zero diagonal blocks. On the three groups of
indices choose symmetric coefficients
\[
 \alpha_{LM}=0,\qquad\alpha_{LN}=1,\qquad
 \alpha_{MN}=-1,
\]
and zero coefficients within each group. These follow from the
complexified source parameters $a=i,-i,0$ and $c=0,0,1$ on the
respective groups. Define $(T_0)_{hi}=\alpha_{hi}Z_{hi}$.
Our $H_0$ equations give $i\dot Z=[E+Z,T_0]$. Substituting
\begin{equation}\label{eq:MCL-matrix-map}
 N(x,t)=i e^{-ixE}Z(t)e^{ixE}
\end{equation}
into \cref{eq:MCL-matrix-wave} gives exactly this commutator:
at $x=0$, $\partial_xN_{hi}=(E_{hh}-E_{ii})Z_{hi}$ and the
quadratic sum is $-[Z,T_0]_{hi}$. The exponential rates
$\Delta_{hi}=-i(E_{hh}-E_{ii})$ obey the required consistency
condition. Equal coefficients to the third group preserve the zero
diagonal blocks. This comparison applies to all positive $L,M,N$;
\cref{eq:canonical-phase-gauge} gives the symmetric phase convention.

The substitution in \cref{eq:MCL-matrix-map} identifies our
phase-dependent equations within the matrix-wave framework of
Menyuk, Chen and Lee. The independent phase parameters are a
constant-diagonal extension relative to the printed 1991 Hamiltonian,
as detailed in Appendix~\ref{sec:JRW-dictionary}. Their determination
from prescribed BA data is the role they play in the present paper.
Their explicit complete-integral calculation concerns
the restricted shared-mode family; the all-size canonical count used
here is stated in \cref{thm:canonical-independence} and proved in
Appendix~\ref{sec:integrability-proof}.

\section{Comparison with Jiang and Rauch-Wojciechowski}
\label{sec:JRW-dictionary}

We give the source dictionary to distinguish the printed Hamiltonian
from its constant-diagonal phase extension and fix the conventions
used by the BA realization.

In \cite[equations (2.7)--(2.12)]{JRW1991}, take the row count to
be $L+M$ and the column count to be $N$. Choose the signature matrices
$\mathsf S_{\rm row}=\Id_{L+M}$, $\mathsf S_{\rm col}=\Id_N$ and
$\mathsf S_{\pm}=\diag(\Id_L,-\Id_M)$. Write the original variables as
\begin{equation}\label{eq:JRW-variables}
 \tau=\begin{pmatrix}a\\b\end{pmatrix},\qquad
 \varpi=\begin{pmatrix}0&P\\Q&0\end{pmatrix},\qquad
 E=\diag(\epsilon_1,\ldots,\epsilon_N),
\end{equation}
where $a$ is $L\times N$, $b$ is $M\times N$ and $P$ is
$L\times M$. The original Hermitian relations are complexified by
treating their partner entries as independent. The mode change of variables is
\begin{equation}\label{eq:JRW-mode-map}
\begin{aligned}
 \tilde w_{n\ell}&=a_{\ell n},&
 \bar{\tilde w}_{n\ell}&=\bar{a}_{\ell n},&&1\leq n\leq N,\quad1\leq\ell\leq L,\\
 \tilde v_{mn}&=\bar{b}_{mn},&
 \bar{\tilde v}_{mn}&=b_{mn},&&1\leq m\leq M,\quad1\leq n\leq N,\\
 \tilde u_{\ell m}&=\frac{\mathrm i}{\sqrt2}Q_{m\ell},&
 \bar{\tilde u}_{\ell m}&=\frac{\mathrm i}{\sqrt2}P_{\ell m},
 &&1\leq\ell\leq L,\quad1\leq m\leq M.
\end{aligned}
\end{equation}
Indeed, the original brackets on $a,b$ have signs $+1,-1$,
respectively, while $\{P_{\ell m},Q_{ji}\}=2\delta_{\ell i}\delta_{mj}$.
Multiplying the original Poisson bracket by $-i$ gives our time
convention, equivalently $t_{\rm J}=-it$ for its original time.
The brackets then become \cref{eq:canonical-brackets} under
\cref{eq:JRW-mode-map}. The letter $b$ in \cref{eq:JRW-variables}
denotes the original mode array, distinct from the fixed coefficients
$b_\ell^L,b_m^M$.

The Hamiltonian printed by Jiang and Rauch-Wojciechowski becomes
\begin{equation}\label{eq:JRW-H}
 H_{\mathrm{source}}=\mathrm i\,\tr
 \left(\varpi\tau\bar{\tau}^{\trans}+\tau E\bar{\tau}^{\trans}\right).
\end{equation}
This is exactly \cref{eq:canonical-H0} with
$b_\ell^L=b_m^M=0$ and $b_n^N=\mathrm i\epsilon_n$ for all indicated
indices. In particular, its three rectangular interaction arrays are
already present for general column count $N$, together with the
column phase parameters $\epsilon_n$.

To obtain the phase parameters in \cref{eq:canonical-H}, retain a
constant diagonal upper block
\begin{equation}\label{eq:JRW-Z}
 Z=\diag(\delta_1^L,\ldots,\delta_L^L,
         \delta_1^M,\ldots,\delta_M^M)
\end{equation}
in the original pencil and Hamiltonian:
\begin{equation}\label{eq:JRW-extension}
\begin{aligned}
 X_{{\rm J},Z}(\mu)
 &=\begin{pmatrix}Z+\varpi&\tau\\\bar{\tau}^{\trans}&E\end{pmatrix}
   +\mu\diag(-\mathsf S_{\pm},0_N),\\
 H_Z&=\mathrm i\,\tr\left[
       (\varpi-Z)\tau\bar{\tau}^{\trans}+\tau E\bar{\tau}^{\trans}\right].
\end{aligned}
\end{equation}
With the same time convention, the original time matrix gives $i\dot X_{{\rm J},Z}=[X_{{\rm J},Z},T_{\rm J}]$. Under
\cref{eq:JRW-mode-map}, this is $H_0$ with
\begin{equation}\label{eq:JRW-phase-map}
 b_\ell^L=-\frac{\mathrm i}{2}\delta_\ell^L,\quad1\leq\ell\leq L,
 \qquad b_m^M=-\frac{\mathrm i}{2}\delta_m^M,\quad1\leq m\leq M,
 \qquad b_n^N=\mathrm i\epsilon_n,\quad1\leq n\leq N.
\end{equation}
Thus the constant-diagonal extension is explicit: $Z=0$ recovers the
printed 1991 Hamiltonian, and arbitrary $Z$ supplies independent
outer phase coefficients. Applying \cref{eq:canonical-phase-gauge}
then gives the symmetric Hamiltonian \cref{eq:canonical-H}.
For $N=1$, $Z=0$ forces all elementary phase coefficients to
coincide; arbitrary $Z$ accommodates \cref{eq:general-phase-values}.

The original pencil and the pencil of \cref{sec:canonical-lax} are related by
\begin{equation}\label{eq:JRW-pencil-map}
 \mathrm i X_{{\rm J},Z}(\mu)
 =\mathsf G\,\widetilde{\mathsf L}(-\mathrm i\mu)^{\trans}\mathsf G^{-1},
 \qquad \mathsf G=\diag(\Id_L,\Id_M,\mathrm i\Id_N).
\end{equation}
Here $\trans$ is ordinary transpose. The matrix $\widetilde{\mathsf L}$
is built from \cref{eq:canonical-X,eq:canonical-pencil} using the
solutions $\tilde{\boldsymbol u},\tilde{\boldsymbol v},\tilde{\boldsymbol w}$
of $H_0$ defined in \cref{sec:canonical-phase-gauges}, with the same
constant diagonal matrices. The symmetric-phase solutions of $H$ and
the zero-$u$-phase solutions of $H_0$ give pencils related by
\begin{equation}\label{eq:canonical-pencil-gauge}
 \mathsf L(\nu,t)=\mathsf S(t)\widetilde{\mathsf L}(\nu,t)
                         \mathsf S(t)^{-1},\qquad
 \mathsf S(t)=\diag(e^{-it\mathsf B_L},e^{it\mathsf B_M},\Id_N).
\end{equation}
Consequently $H$ and $H_0$ have identical spectral polynomials along
the solutions related by \cref{eq:canonical-phase-gauge}.

The completeness statement of \cref{thm:canonical-independence} also
applies at $Z=0$, for arbitrary row count $L+M$ and column count $N$, with repeated or distinct entries of $E$.
Jiang and Rauch-Wojciechowski prove the two- and three-row cases and
recover extra integrals at coincident column parameters in the
two-row case by perturbation \cite[Section II.C, equation (2.23)]{JRW1991}.
The equal-phase moment matrices and insertion lemma provide a single
independence argument for all these sizes and coincidences.

\section{Comparison with Adams, Harnad and Hurtubise}
\label{sec:AHH-dictionary}

We record the source and BA evaluation dictionaries to make the
quartic--cubic correspondence of \cref{sec:AHH-comparison} explicit.

We use the polynomial-flow convention of
\cite[equation (2.1)]{AdamsHarnadHurtubise1990II}. Their matrix size
$r$ is our $d=L+M+N$, and we take polynomial degree one. After
translating a single source pole to the origin, write
\begin{equation}\label{eq:AHH-source-flow}
 \widehat{\mathsf L}(\nu)=\widehat{\mathsf X}+\nu\mathsf D,
 \qquad
 \frac{d\widehat{\mathsf L}}{d\upsilon}
   =[\widehat{\mathsf B},\widehat{\mathsf L}],\qquad
 \widehat{\mathsf B}
   =(\widehat{\mathsf L}^{,2}/\nu)_+
   =\nu\mathsf D^2+\mathsf D\widehat{\mathsf X}
                           +\widehat{\mathsf X}\mathsf D.
\end{equation}
The subscript $+$ means the polynomial part in $\nu$.
The source complex time is $\upsilon=it$, followed by conjugation
by $e^{-it\mathsf K_{\rm ph}}$. Equivalently, in the physical
quartic convention \cref{eq:AHH-parent-rates}, set $s=-t$.
The time matrix of our pencil is
\begin{equation}\label{eq:AHH-source-time}
 \mathsf T=\mathsf B-\mathsf K_{\rm ph}-\nu\Id_d.
\end{equation}
The scalar last term does not affect the commutator. The blocks of
$\mathsf X$ are exactly those of \cref{eq:canonical-X}; explicitly,
\[
 u_{\ell m}=X_{\ell,L+m}/\sqrt2,\quad
 \bar u_{\ell m}=X_{L+m,\ell}/\sqrt2,\quad
 v_{mn}=X_{L+m,L+M+n},\quad
 \bar v_{mn}=-X_{L+M+n,L+m},
\]
\[
 w_{n\ell}=-X_{L+M+n,\ell},\qquad
 \bar w_{n\ell}=X_{\ell,L+M+n},
 \qquad1\leq\ell\leq L,\ 1\leq m\leq M,\ 1\leq n\leq N.
\]
This gives the dictionary for unequal sizes and for $N>1$ without
changing the matrix evolution. In the moment-map convention
$\mathsf X=-\mathsf G^{\trans}\mathsf F$, reversing the sign of the
source bilinear matrix reverses the corresponding spectral ordinate.

\subsection*{The BA evaluation map}

Sections 2c--2g and 4 of \cite{AdamsHarnadHurtubise1990II} reconstruct
the polynomial matrix and its canonical lift from a line bundle on
the normalized curve. The normalized sections in equations
(4.17), (4.21)--(4.22) and (4.25)--(4.26), evaluated by
(2.24)--(2.25), yield the following concrete comparison. Use a single
source pole with a complete regular fiber of $d$ points and an
adapted row basis for its full-rank moment level. In the plus
convention $\mathsf B=\mathsf G^{\trans}\mathsf F$, take
\[
 \mathsf Y=-\mathsf D,\qquad
 \mathsf L_{\rm src}=(\Lambda-\alpha)\mathsf Y+\mathsf B,
 \qquad \det(W\Id_d-\mathsf L_{\rm src})=0.
\]
Let $R(P)$ denote the common fixed meromorphic section appearing in
the source theta formulas; it has simple zeros at the infinity
points $P_h$. Put $T(P)=(\Lambda(P)-\alpha)R(P)$, whose zeros
are the fiber points $Q_i$. In compatible fixed local frames,
with $R'_h=\partial_{t_h}R(P_h)$ and
$\Lambda'_i=\partial_{s_i}\Lambda(Q_i)$, their evaluation formulas
become
\begin{equation}\label{eq:AHH-evaluation-dictionary}
 F_{ih}=\frac{R(Q_i)}{R'_h}\mathscr S_z(Q_i,P_h),\qquad
 G_{ih}=\frac{W(Q_i)R'_h}{\Lambda'_iR(Q_i)}
                 \mathscr S_z(P_h,Q_i),\quad1\leq i,h\leq d.
\end{equation}
Here fixed exponential BA gauges have been suppressed. They cancel
at the intermediate point and leave the endpoint difference rate
in each product. The source row-moment eigenvalues are $W(Q_i)$.
Since $\partial_{s_i}T(Q_i)=\Lambda'_iR(Q_i)$ and $T(P_h)=R'_h$,
\cref{eq:AHH-fiber-convolution} gives
\begin{equation}\label{eq:AHH-BA-projection}
 B_{hk}=\frac{R'_h}{R'_k}(y_h-y_k)\mathscr S_z(P_h,P_k)
 \quad(h\ne k),\qquad B_{hh}=c_h.
\end{equation}
Thus a fixed diagonal similarity removes $R'_h/R'_k$ and the
bilinear sums become single endpoint BA ratios. Here $\mathsf B=-\widehat{\mathsf X}$ and
$\mathsf L_{\rm src}=-\widehat{\mathsf L}$, with
$\nu=\Lambda-\alpha$; hence its ordinate is the negative of
that for $\widehat{\mathsf L}$. This fixes both the leading-matrix
and ordinate signs in converting the plus moment map to our minus
convention. Within a repeated-slope group
$y_h=y_k$, so its off-diagonal entries vanish in this adapted basis;
the constants $c_h$ supply its diagonal entries. Normalization
separates the infinity branches even when leading slopes repeat.

Choose $\alpha$ away from the BA essential fiber so that these are
regular evaluations. Translating the source spectral coordinate
changes the polynomial generator only by a constant diagonal term:
\[
 (\mathsf L_{\rm src}^{,2}/(\Lambda-\alpha))_+
 -(\mathsf L_{\rm src}^{,2}/\Lambda)_+=\alpha\mathsf Y^2=\alpha\mathsf D^2.
\]
Its effect is another fixed diagonal phase conjugation. These
formulas apply on the regular spectral locus with an unramified
chosen fiber and nonspecial evaluation data. They identify the
quartic lift and the cubic realization at the equation and BA
function levels; the finite prescribed-point interpolation and
completion criterion of \cref{thm:general-BA-realization,cor:BA-finite-list}
are developed directly in this paper.

\section{The complete Liouville count for arbitrary phases}
\label{sec:integrability-proof}

We prove \cref{thm:canonical-independence} using the commuting algebra
$\mathcal A$ defined in \cref{sec:canonical-independence}.
The argument inserts one index at a time. At each step the new
canonical pairs split into directions detected by spectral invariants
and directions detected by equal-phase moment invariants.
\Cref{lem:canonical-insertion} constructs complementary independent derivatives for these two
sets; its final cyclic-vector test shows that they occur simultaneously.

The lemma constructs an independence certificate in the mode
coordinates. A vector $v$ is \emph{cyclic} for a matrix $\mathsf Z$ of size
$a$ if $v,\mathsf Z v,\ldots,\mathsf Z^{a-1}v$ span $\C^a$.

\begin{lemma}[Increasing a matrix dimension]\label{lem:canonical-insertion}
Increase one of $L,M,N$ by one. Suppose the new phase already
occurs $r$ times in that group and $c$ other indices there have
different phases. Let the other group sizes be $m_1,m_2$, and put
$n=m_1+m_2$. Assume either $r\geq n$ or
\begin{equation}\label{eq:canonical-insertion-condition}
 |m_1-m_2|\leq r+c+1.
\end{equation}
If the old algebra $\mathcal A$ supplies the complete Liouville count,
the enlarged algebra does too. The insertion supplies
$n-\min(r,n)$ new independent spectral functions and $\min(r,n)$
new independent functions of the principal moments.
\end{lemma}
\begin{proof}
Write the new outgoing entries as a row $\boldsymbol a\in\C^{1\times n}$ and
the incoming entries as a column $\boldsymbol b\in\C^{n\times1}$.
We test the Jacobian at $\boldsymbol a=0$. The enlarged determinant then equals
$(\lambda-e-s\nu)F(\lambda,\nu)$, where $e,s$ are the new intercept
and slope. More generally the bordered determinant is this product
minus a term bilinear in the new outgoing and incoming entries.
Multiplication by $\lambda-e-s\nu$ is an injective linear map on
polynomial coefficients. A fixed left inverse therefore lifts each
old spectral coefficient to a function of the enlarged coefficients.
At $\boldsymbol a=0$ its old-coordinate derivative is unchanged,
and its derivative in the new incoming column is zero. The old
principal moments have the same properties. We need $n$ further
functions with zero old-coordinate
derivatives and independent derivatives in $\boldsymbol a$.

The certificate will have the block form
\[
 \frac{\partial(\text{lifted old functions},\text{new functions})}
      {\partial(\text{selected old coordinates},\boldsymbol a)}
 =\begin{pmatrix}J_{\rm old}&*\\0&J_{\rm new}\end{pmatrix}.
\]
Here $J_{\rm old}$ is already nonsingular. We construct $J_{\rm new}$
from moment derivatives on one summand of $\C^n$ and spectral
derivatives on its complement. Establishing this at one point is
enough, since nonvanishing of the minor is a Zariski-open condition.

Order the old indices as the $r$ equal-phase indices, the $c$
different-phase indices of the same original group, and the $n$
outside indices:
\begin{equation}\label{eq:canonical-insertion-blocks}
 \mathsf X=
 \begin{pmatrix}e\Id_r&0&\mathsf F\\0&E_C&\mathsf F_C\\\mathsf G&\mathsf G_C&X_{\rm out}\end{pmatrix},
 \qquad
 \mathsf W=\diag\left(\frac1{s-s_h}\right)_{h\ {\rm outside}}.
\end{equation}
Here $e\Id_c-E_C$ is invertible. The growing moment block and its
enlargement are
\[
 \mathsf J=\mathsf F\mathsf W \mathsf G,\qquad
 \mathsf J_+=\begin{pmatrix}\mathsf J&\mathsf F\mathsf W \boldsymbol b\\\boldsymbol a\mathsf W \mathsf G&\boldsymbol a\mathsf W \boldsymbol b\end{pmatrix}.
\]
The principal-moment differences
$\Delta_k=(\tr \mathsf J_+^k-\tr \mathsf J^k)/k$ vanish at $\boldsymbol a=0$, and
\begin{equation}\label{eq:canonical-insertion-moment}
 \left.\mathrm d\Delta_k\right|_{\delta \boldsymbol a}
   =\delta \boldsymbol a\,\mathsf W \mathsf G \mathsf J^{k-2}\mathsf F\mathsf W \boldsymbol b,
 \qquad 2\leq k\leq\min(r,n)+1.
\end{equation}
Newton's identities express the old trace of order $r+1$ in its
first $r$ traces, so every $\Delta_k$ belongs to the enlarged algebra.
The first difference, $\Delta_1$, has derivative
$\delta\boldsymbol a\mathsf W\boldsymbol b$. We do not need it:
the displayed higher differences will already span the moment
directions, and the spectral functions will supply their complement.

First suppose $0<r<n$. Generically $\mathsf J$ is invertible. Set
\begin{equation}\label{eq:canonical-insertion-projection}
 \Pi=\Id_n-\mathsf W \mathsf G \mathsf J^{-1}\mathsf F,\qquad
 \C^n=\ker \mathsf F\oplus\operatorname{im}(\mathsf W \mathsf G).
\end{equation}
The $r$ vectors in \cref{eq:canonical-insertion-moment} span the
second summand if $\mathsf F\mathsf W \boldsymbol b$ is cyclic for $\mathsf J$.
For the other summand use the coefficients of
$R(\nu)=F_+(e+s\nu,\nu)$, where $F_+$ is the enlarged determinant.
They vanish at $\boldsymbol a=0$. With
$\mathsf K(\nu)=(e+s\nu)\Id-\mathsf X-\nu\mathsf D_{\rm old}$,
the Schur complement gives
\[
 R(\nu)=-\boldsymbol a\,
       \operatorname{adj}(\mathsf K(\nu))_{\rm out,out}\boldsymbol b,
 \qquad
 \left.\mathrm dR(\nu)\right|_{\delta \boldsymbol a}
 =-\delta \boldsymbol a\,\det \mathsf K(\nu)\,
            (\mathsf K(\nu)^{-1})_{\rm out,out}\boldsymbol b.
\]
Eliminating the $c$ different-phase indices leaves an outside block
$\nu\mathsf W^{-1}+\mathsf C$, with
\[
 \mathsf C=e\Id_n-X_{\rm out}-\mathsf G_C(e\Id_c-E_C)^{-1}\mathsf F_C,\qquad
 \mathsf R_0=\left.\Pi\mathsf W \mathsf C\right|_{\ker \mathsf F},\quad \boldsymbol c=\Pi\mathsf W \boldsymbol b.
\]
The remaining linear equations give
$(\mathsf K^{-1})_{\rm out,out}\boldsymbol b=(\nu\Id+\mathsf R_0)^{-1}\boldsymbol c$.
To compute its determinant, first eliminate the $c$ different-phase
indices, multiply the outside rows by $\mathsf W$, and use the
decomposition in \cref{eq:canonical-insertion-projection}. The
equal-phase rows pair with $\operatorname{im}(\mathsf W\mathsf G)$
through $\mathsf J$, leaving $\nu\Id+\mathsf R_0$ on $\ker\mathsf F$.
Consequently
\[
 \det\mathsf K(\nu)
 =(-1)^r\det(e\Id_c-E_C)\det(\mathsf W^{-1})\det\mathsf J\,
       \det(\nu\Id+\mathsf R_0).
\]
The prefactor is nonzero and independent of $\nu$.
Hence the coefficient vectors of the new spectral
derivatives span $\ker \mathsf F$ if $\boldsymbol c$ is cyclic for $\mathsf R_0$.
These coefficients are triangular combinations of
$\boldsymbol c,\mathsf R_0\boldsymbol c,\ldots,\mathsf R_0^{n-r-1}\boldsymbol c$.
There are $n-r$ such coefficients, since $\deg R\leq n-r-1$.
When $r=0$, use $\Pi=\Id_n$ and omit the moment differences.

We verify that both cyclicity conditions occur. Choose $\mathsf F$ to select
$r$ outside coordinate directions and $\mathsf G$ so that $\mathsf J$ is diagonal
with distinct nonzero entries. Then $\Pi$ selects the remaining
directions. Condition~\eqref{eq:canonical-insertion-condition} lets
these directions have outside group sizes differing by at most $c+1$.
Pair them across the two groups, leaving at most $c+1$ unpaired
indices. At this test point $\mathsf R_0$ is a direct sum of blocks
\[
 \begin{pmatrix}d_h&a_h\\b_h&d'_h\end{pmatrix}
 \quad\text{and scalar blocks}\quad d_j+\delta_j.
\]
The $d_h,d'_h,d_j$ are fixed by the outside phases; the products
$a_hb_h$ are free. Each paired block can therefore have two distinct
eigenvalues avoiding all eigenvalues already chosen. For all but at most one unpaired
index, use a different-phase index to adjust its diagonal entry
through the term $-\mathsf G_C(e\Id_c-E_C)^{-1}\mathsf F_C$;
these are the independent shifts $\delta_j$. Thus at most one scalar
eigenvalue remains fixed, and the full spectrum can be made simple.
This explains the size hypothesis: even when all unpaired outside
phases coincide, the $c$ available shifts separate all but one of
the at most $c+1$ unpaired directions.
Finally $\boldsymbol b\mapsto(\mathsf F\mathsf W \boldsymbol b,\Pi\mathsf W \boldsymbol b)$ is invertible onto
$\C^r\oplus\ker \mathsf F$, so the two cyclic vectors can be chosen
independently. The required conditions are nonempty Zariski-open
conditions.

If $r\geq n$, choose $\mathsf F,\mathsf G$ so that $\mathsf R=\mathsf W \mathsf G \mathsf F$ has distinct
nonzero eigenvalues. The $n$ vectors in
\cref{eq:canonical-insertion-moment} are now
$\mathsf R^{k-1}\mathsf W \boldsymbol b$, $2\leq k\leq n+1$, and span $\C^n$ for a
cyclic $\mathsf W \boldsymbol b$. No new spectral functions are needed.

In either case the open conditions intersect the old nonzero-Jacobian
locus. The old-coordinate and new-$\boldsymbol a$ derivatives form a block-triangular
Jacobian with nonzero diagonal minors. This proves the insertion step.
\end{proof}

\subsection{Proof of the complete count}

\begin{proof}[Proof of \cref{thm:canonical-independence}]
For $(1,1,1)$ two quadratic spectral coefficients give independent
power combinations. A third contains the sum of the oriented triangle
products, which varies at fixed powers. Their rank is three for every
phase triple.

A positive size triple satisfying the triangle inequalities can be
reduced to $(1,1,1)$ by decreasing a largest entry; the preceding triple
still satisfies those inequalities. Reverse this process, appending
indices last within their equal-phase blocks. At each step
$|m_1-m_2|\leq r+c$, which is stronger than the sufficient
bound $r+c+1$ in \cref{lem:canonical-insertion}.

If one group exceeds the sum of the others, first build the balanced
triple where its size equals that sum, using initial segments of its
prescribed phase blocks. Append the remaining indices. The growing
group has size $r+c\geq m_1+m_2$, so the lemma applies at every step.
Each insertion adds $n$ canonical pairs and $n$ independent commuting
functions. This proves the full count. Since $H$ commutes with them,
it is locally a function of a complete set and can replace a suitable
member wherever $\mathrm dH\ne0$.
\end{proof}

\subsection{The spectral deficit at repeated phases}

The proof also counts the spectral deficit explicitly. For each
equal-phase block $B$, let $r_B=|B|$, let $o_B$ be the size outside its
original group, and set
\begin{equation}\label{eq:canonical-spectral-deficit}
 k_B=\min(r_B,o_B),\qquad
 q_B=r_Bk_B-\frac{k_B(k_B+1)}2.
\end{equation}
The generic spectral rank is $p-\sum_{B\in\mathcal B}q_B$; the proof
supplies $\sum_{B\in\mathcal B}q_B$ additional independent
principal-moment functions, where $\mathcal B$ is the set of all
equal-phase blocks.
For the lower bound, each insertion increases its block's $q_B$ by
$\min(r,n)$ and leaves the other blocks' $q_B$ unchanged along the
chosen construction. The spectral derivatives supply the complementary
$n-\min(r,n)$ directions. For the upper bound, $\mathsf J_B$ has
$k_B$ distinct nonzero eigenvalues and a semisimple zero eigenvalue of
multiplicity $r_B-k_B$, so its conjugacy orbit has dimension $2q_B$.
The moment-generated tangent space therefore has symplectic rank
$2\sum_{B\in\mathcal B}q_B$. Commuting spectral fields lie in its
symplectic orthogonal, whose maximal isotropic dimension is
$p-\sum_{B\in\mathcal B}q_B$.

For example, with $L=M=2$, $N=1$ and constant phases within each
outer group, there are two size-two equal-phase blocks, each with
$q_B=1$. The spectral rank is six; two additional principal-moment
functions complete the eight-integral count.

\section{Explicit spectral integrals for distinct phase coefficients}
\label{sec:distinct-phase-integrals}

For distinct phase coefficients within each group, the complete integral
set can be chosen directly from spectral values and derivatives. This
provides an explicit specialization of \cref{thm:canonical-independence}.
The selection points are determined by
the fixed diagonal entries of the pencil. In the $(\nu,\lambda)$
plane, the diagonal factor $\lambda-X_{hh}-\nu D_{hh}$ vanishes
on the straight line $\lambda=X_{hh}+\nu D_{hh}$,
$1\leq h\leq d$. From \cref{eq:canonical-X,eq:canonical-pencil},
these are the three families
\begin{equation}\label{eq:canonical-diagonal-lines}
\begin{aligned}
 \lambda&=\nu-2b_\ell^L,&&1\leq\ell\leq L,\\
 \lambda&=-\nu-2b_m^M,&&1\leq m\leq M,\\
 \lambda&=b_n^N,&&1\leq n\leq N.
\end{aligned}
\end{equation}
Thus a family is simply a list of parallel straight lines: there are
$L$ lines of slope $1$, $M$ of slope $-1$, and $N$ horizontal lines.
The $\lambda$-intercept is the value at $\nu=0$, namely $X_{hh}$;
the three lists of intercepts are $-2b_\ell^L,-2b_m^M,b_n^N$.

An intersection $q=(\lambda_q,\nu_q)$ is fixed by these phase
coefficients. Evaluating the spectral determinant there gives the
function $F(q)=F(\lambda_q,\nu_q)$ of the evolving modes.
These values are conserved and commute by
\cref{prop:canonical-commutation}. Distinct intercepts within each
family ensure that at most one line from each family passes through
any point, so intersections involve either two or three lines.

\begin{theorem}\label{thm:distinct-phase-evaluations}
Suppose each of the three lists of phase coefficients
\[
 b_1^L,\ldots,b_L^L;\qquad
 b_1^M,\ldots,b_M^M;\qquad
 b_1^N,\ldots,b_N^N
\]
is pairwise distinct.
At every two-line intersection $q$, take $F(q)$; at every triple
intersection, take $F(q),\partial_\lambda F(q),\partial_\nu F(q)$.
These $p=LM+MN+NL$ functions are generically independent commuting
first integrals. Hence \cref{eq:canonical-H} is Liouville integrable
for every positive $L,M,N$. A complete local set can include $H$
wherever $\mathrm dH\ne0$.
\end{theorem}
\begin{proof}
There are $LM+MN+NL$ pairs of lines from different families.
Ordinarily each pair gives one evaluation. If three lines meet,
their three pairwise intersections coincide; the value and its two
first spectral derivatives provide three functions at that point.
Thus the total count remains $p$. The derivatives also commute,
by differentiating \cref{prop:canonical-commutation} with respect
to the fixed spectral arguments.

To prove independence, it suffices to find one nonzero Jacobian minor.
Use the $p$ upper off-diagonal mode entries $x_{hi}=X_{hi}$ as
column coordinates, with $1\leq h<i\leq d$ in different blocks;
use the proposed integrals as rows. Evaluate this Jacobian at the
test states $X_{hi}=0$, $X_{ih}=\epsilon$, keeping all phase
coefficients fixed. Here $\epsilon$ is a parameter selecting the
test state. We find the leading determinant as $\epsilon\to0$.

Write $l_h=\lambda-X_{hh}-\nu D_{hh}$, $1\leq h\leq d$,
and let $R_q$ be the product of the nonzero $l_h(q)$ over
$1\leq h\leq d$. At a two-line intersection, two diagonal
factors vanish. A determinant term linear in $x_{hi}$ can remove
both zeros by pairing it with its reverse entry $\epsilon$;
this gives $-\epsilon R_q$ only for the edge joining those two
lines. All other columns vanish to higher order. Dividing this
row by $\epsilon$ therefore leaves the single nonzero leading
entry $-R_q$.

At a triple intersection, three diagonal factors vanish. The Jacobian
row for $F(q)$ first detects a three-edge cycle, of order $\epsilon^2$.
A spectral derivative removes one vanishing factor, so
$F_\lambda(q)$ and $F_\nu(q)$ detect two-edge cycles of order
$\epsilon$. Order the three relevant columns as $L$--$M$,
$L$--$N$, $M$--$N$. After dividing these three rows by
$\epsilon^2,\epsilon,\epsilon$, respectively, determinant
differentiation gives the leading block
\begin{equation}\label{eq:canonical-confluent-block}
 R_q\begin{pmatrix}0&-1&0\\-1&-1&-1\\0&-1&1\end{pmatrix},
 \qquad \det=-R_q^3\ne0.
\end{equation}
Leading entries on edges outside the corresponding intersection
vanish, because a vanishing diagonal factor remains. Consequently
the full normalized Jacobian is block diagonal: one nonzero scalar
block for each two-line intersection and one invertible three-by-three
block for each triple. Its determinant is nonzero. The unscaled
minor is therefore a nonzero polynomial in $\epsilon$, proving
independence on a dense open set of the mode phase space. Since $H$
commutes with this complete set, it is locally a function of these
integrals and can replace a suitable member wherever $\mathrm dH\ne0$.
\end{proof}

The $d-1$ independent power charges can be retained in a complete set,
leaving
\[
 G=p-d+1=LM+MN+NL-L-M-N+1.
\]
spectral integrals. Away from triple intersections, there is a simple
group-symmetric selection: regard the modes as edges of the complete
tripartite graph, choose any spanning tree, and retain the evaluations
on its $G$ complementary edges. Indeed, at the proof's test state the
charge differentials form a weighted incidence matrix, whose tree
minor is nonzero; the complementary spectral rows give the remaining
nonzero diagonal block. At triple intersections, differential row
selection from the confluent set gives the required complement.
For $N=1$, the count becomes $L+M$ power charges and $LM$ spectral
integrals.

\section{Details of the marked spectral-curve classification}
\label{sec:geometry-details}

We prove the marked-curve statements of \cref{sec:spectral-geometry},
including existence and reconstruction, to justify the fixed-size
classification used in the spectral discussion.

\subsection{Existence of regular spectral curves}

The regular locus is nonempty for every positive $L,M,N$, with
phases allowed to vary. To see this, choose intercepts $e_h$ distinct
within each group so that the lines
$\lambda-e_h-s_h\nu=0$, $1\leq h\leq d$, have no triple affine
intersection. Set $X=\diag(e_h)+\varepsilon A$, with zero diagonal
blocks in $A$ and $A_{hi}A_{ih}\ne0$ between different groups.
At $\varepsilon=0$ the curve is the union of these lines.
Near an intersection of lines $h,i$ of different slopes, use their
equations as local coordinates $x,y$. A Schur complement gives,
up to a nonvanishing factor,
\[
 xy-\varepsilon^2 A_{hi}A_{ih}
   +O\bigl(\varepsilon^2x,\varepsilon^2y,\varepsilon^3\bigr).
\]
Its critical point has $x,y=O(\varepsilon^2)$, where the displayed
nonzero quadratic term prevents a singularity for small nonzero
$\varepsilon$. Blowing up the three infinity points separates their
distinct tangent branches. They meet the exceptional curves
transversely, so remain smooth there uniformly for small $\varepsilon$.
Compactness then
excludes further singularities. On this blown-up surface the central
fiber is a connected reduced union of lines, so nearby smooth fibers
are connected, hence irreducible. Thus small generic couplings give
regular curves. Smoothness and irreducibility are open conditions,
so they hold on a nonempty Zariski-open subset of the parameter space.

\subsection{Proof of the marked-curve characterization}

\begin{proof}[Proof of \cref{thm:geometry-classification}]
The implication from~1 to~2 follows from
\cref{prop:geometry-genus}. The expansions
\cref{eq:geometry-branches} give~3: subtracting or adding $\nu$
cancels the pole on the indicated branch family. The two affine
coordinates generate the function field because the plane model
is birational to $C$.

Conversely, the pole conditions in~3 imply that $\nu$ has exactly
$d$ simple poles, one at each marked point. The functions
$1,\lambda,\nu$ define a map to the plane with no base points:
at each marked point at least one of $\lambda,\nu$ has a simple
pole, and away from them the constant section is nonzero.
The map is birational because $\lambda,\nu$ generate $\C(C)$.
The image has degree $d$, since a generic linear combination of
these three functions has $d$ poles. The three pole cancellations
put the marked branches at $[1:1:0]$, $[-1:1:0]$, $[0:1:0]$.
Their simple poles make the branches smooth and transverse to infinity.
The genus assumption forces their total singularity defect to be
exactly the sum in \cref{prop:geometry-genus}. Hence their tangents
are distinct and there are no additional singularities, proving~2.

Finally, a plane model in~2 can be reconstructed as a pencil
$\mathsf X+\nu\mathsf D$. The spectral line-bundle construction
\cite{Beauville1990} represents multiplication by $\lambda$ as an
affine matrix function of $\nu$. Its leading matrix has eigenvalues
$1,-1,0$, with multiplicities $L,M,N$. The distinct branch intercepts
allow the three constant diagonal blocks to be diagonalized; the
remaining six blocks are precisely the unrestricted mode blocks
of \cref{eq:canonical-X}. \Cref{sec:geometry-converse}
gives the construction, proving~1.
\end{proof}

\subsection{Reconstructing a pencil from its spectral curve}
\label{sec:geometry-converse}

We prove the converse used in \cref{thm:geometry-classification}.
The construction follows the spectral line-bundle method
\cite[\S1]{Beauville1990}, with the prescribed infinity branches
determining the repeated eigenvalues of the leading matrix.

Let $C$ be the normalization of the plane model in statement~2 of
\cref{thm:geometry-classification}. Choose projective coordinates so
that the common line is $\tau=0$ and its three marked plane points
are $[1:1:0]$, $[-1:1:0]$, $[0:1:0]$.
Transversality makes $\nu$ have $d$ simple poles. Thus
\[
 \pi=\nu:C\longrightarrow\mathbb P^1,\qquad
 \pi^{-1}(\infty)=\mathcal D_\infty
 =\mathcal D_L+\mathcal D_M+\mathcal D_N
\]
is a finite degree-$d$ map, and the poles of $\lambda$ are bounded
by $\mathcal D_\infty$.

Choose a line bundle $\mathcal E$ of degree $G-1$ with
$H^0(C,\mathcal E)=0$, and put
$\mathcal L=\mathcal E(\mathcal D_\infty)$. Such $\mathcal E$
exist outside the theta divisor. The pushforward $\pi_*\mathcal L$
is a rank-$d$ bundle on $\mathbb P^1$ of degree zero, by
Riemann--Roch. Moreover, the projection formula gives
\[
 H^0\bigl(\mathbb P^1,(\pi_*\mathcal L)(-1)\bigr)
 =H^0(C,\mathcal E)=0.
\]
Every vector bundle on $\mathbb P^1$ splits into line bundles.
The vanishing makes all their degrees nonpositive, while their sum
is zero. Therefore $\pi_*\mathcal L\simeq\mathcal O^{d}$.

Multiplication by $\lambda$ now defines
\[
 \mathcal O^{d}\longrightarrow\mathcal O^{d}(1).
\]
In a trivialization it is an affine pencil $\mathsf X+\nu\mathsf D$.
At $\nu=\infty$ its leading operator is multiplication by
$\lambda/\nu$ on the $d$ separate fibers over infinity. Its values
are $1,-1,0$, with multiplicities $L,M,N$; hence a constant basis
change makes $\mathsf D=\diag(\Id_L,-\Id_M,0_N)$.
The characteristic polynomial is the plane equation: over a generic
$\nu$, the operator has the $d$ values of $\lambda$ as eigenvalues.

The eigenvalues of the three diagonal blocks of $\mathsf X$ are
the branch intercepts. Ordinary multiplicity gives distinct
intercepts within each block, so a block-diagonal basis change
diagonalizes all three blocks while preserving $\mathsf D$.
Write their entries as $-2b_\ell^L$, $-2b_m^M$, $b_n^N$,
with $1\leq\ell\leq L$, $1\leq m\leq M$, $1\leq n\leq N$.
The six remaining blocks are unrestricted. Assigning them the
factors and signs of \cref{eq:canonical-X} gives the three mode
matrices and their independent partners, and hence initial data
for \cref{eq:canonical-H}. This proves the required realization.

\subsection{Proof of the hyperelliptic criterion}

\begin{proof}[Proof of \cref{prop:geometry-hyperelliptic}]
Two singleton groups give a degree-two map in
\cref{eq:geometry-degrees}, proving sufficiency.
For necessity, sort the three sizes as $a\leq b\leq c$ and
suppose $b\geq2$. Two of the generating functions have degrees
$a+b$ and $a+c$, and
\[
 G-(a+c-1)=(a+b-2)c+(a-1)(b-2)>0.
\]
Both degrees are therefore at most $G$.
On a hyperelliptic curve of genus $G$, every meromorphic function
of degree at most $G$ factors through the degree-two hyperelliptic
map. Indeed, otherwise that map and the function would generate
the function field, and the Castelnuovo--Severi inequality would give
$G\leq\deg(f)-1$; see \cite[Theorem~14]{KhawajaSiksek2024}.
Both generating functions would thus belong to the rational
function field of the hyperelliptic map, a contradiction.
\end{proof}

\section{Generic BA reconstruction with repeated phases}
\label{app:BA-active-converse}

This appendix supplies the regularity step in
\cref{thm:BA-spectral-converse} for arbitrary positive $L,M,N$ and
every fixed phase profile. Bases may change only within equal-phase
blocks, preserving the canonical bracket and Hamiltonian.

\subsection{Constant canonical bases and inactive directions}

For constant invertible matrices $\mathsf S_L,\mathsf S_M,\mathsf S_N$
respecting the phase blocks, the canonical transformations are
\begin{equation}\label{eq:BA-block-bases}
\begin{aligned}
 \boldsymbol u&\mapsto\mathsf S_L\boldsymbol u\mathsf S_M^{-1},&
 \bar{\boldsymbol u}&\mapsto\mathsf S_L^{-\trans}\bar{\boldsymbol u}\mathsf S_M^{\trans},\\
 \boldsymbol v&\mapsto\mathsf S_M\boldsymbol v\mathsf S_N^{-1},&
 \bar{\boldsymbol v}&\mapsto\mathsf S_M^{-\trans}\bar{\boldsymbol v}\mathsf S_N^{\trans},\\
 \boldsymbol w&\mapsto\mathsf S_N\boldsymbol w\mathsf S_L^{-1},&
 \bar{\boldsymbol w}&\mapsto\mathsf S_N^{-\trans}\bar{\boldsymbol w}\mathsf S_L^{\trans}.
\end{aligned}
\end{equation}
They conjugate the pencil by the corresponding three-block matrix.
For an equal-phase block $B$ in group $A$, the conserved moment is
\[
 \mathsf Q_B=\sum_{\substack{h=1\\h\ne A}}^{3}
 \frac{\mathsf X_{B,h}\mathsf X_{h,B}}{d_A-d_h},\qquad
 (d_1,d_2,d_3)=(1,-1,0).
\]
Here $\mathsf X_{B,h}$ has rows in $B$ and columns in the entire
$h$th group. Conservation follows from the moment map of the block
conjugation symmetry; these moments also give the second-order
spectral corrections at infinity. Their diagonalizing bases are
therefore constant in time.

For a block of size $r$, let $n$ be the number of outside directions:
$n=M+N$, $L+N$ or $L+M$ in the three groups respectively. Absorbing the
nonzero slope denominators into the incoming couplings writes its
moment as $\mathsf F\mathsf G$, with $\mathsf F$ of size $r\times n$ and $\mathsf G$ of size
$n\times r$. If $r>n$, then generically $\mathsf G\mathsf F$ is invertible and
\begin{equation}\label{eq:BA-inactive-splitting}
 \mathbb C^r=\operatorname{im}\mathsf F\oplus\ker\mathsf G
             =\operatorname{im}(\mathsf F\mathsf G)\oplus\ker(\mathsf F\mathsf G).
\end{equation}
The two subspaces are constant. In an adapted basis the $r-n$
directions in the second summand have zero incoming and outgoing
couplings. They stay zero and contribute only the spectral factor
$(\lambda-e_B-s_B\nu)^{r-n}$, where $e_B,s_B$ are their intercept
and slope. Delete them and repeat. The process terminates with
positive active dimensions $L_a,M_a,N_a$ such that every remaining block
has size at most its number of outside directions. Inverse basis
changes restore all original coordinates.

The reduction is generic: arbitrary maximal-rank coupling pairs can
be written as an embedded active pair, with zero extra directions,
followed by a block basis change. Thus the active coupling data are
not forced onto an additional exceptional algebraic subset.

\subsection{Regularity of the active spectral curve}

\begin{proposition}\label{prop:BA-active-regularity}
For every fixed phase profile, generic active initial modes have an
irreducible spectral curve, smooth in the affine plane, whose
normalization has distinct resolved infinity branches and nonspecial
scalar BA divisors. Each active moment block has simple nonzero
spectrum. The normalization genus is
\begin{equation}\label{eq:BA-active-genus}
 G_a=L_aM_a+M_aN_a+N_aL_a-L_a-M_a-N_a+1
 -\sum_{h=1}^{3}\sum_{j=1}^{m_h}
       \frac{|B_{h,j}|(|B_{h,j}|-1)}2,
\end{equation}
where $B_{h,1},\ldots,B_{h,m_h}$ are the equal-phase blocks in active
group $h$. The ordinary and dual multipoint divisors have degree
$G_a+N_a-1$ and give normalized bases of $N_a$ sections.
\end{proposition}

\begin{proof}
We check the affine curve, its connectivity, and its infinity data
separately. Generic smoothness and the spectral line-bundle
correspondence are standard \cite{Hartshorne1977,Beauville1990,AdamsHarnadHurtubise1990II}.
Explicit coupling and insertion arguments verify their
regularity hypotheses uniformly for the fixed repeated-phase profiles
of this mode space. For a block of size $r\leq n$, the product $\mathsf F\mathsf G$ can have
arbitrary distinct nonzero eigenvalues: choose $\mathsf F$ to select $r$
outside directions and $\mathsf G$ to give a prescribed diagonal product.
Thus maximal coupling ranks and simple nonzero moment spectra are
nonempty open conditions. There are finitely many blocks, and their
conditions hold simultaneously on a nonempty open set of the
irreducible active mode space.

For affine smoothness put
$\mathsf M=\lambda\Id-\mathsf X-\nu\mathsf D$.
A right kernel vector supported in one original group would, by
the diagonal equations, be supported in one equal-phase block.
The incoming coupling map on that block is injective, so the vector
is zero. The analogous statement holds for left kernels. At a point
of corank $q$, put bases for the right and left kernels in the
columns of a $d\times q$ matrix $\mathsf R$ and the rows of a
$q\times d$ matrix $\mathsf S$, respectively. The derivative
transverse to the matrix rank stratum is
\[
 \delta\mathsf X\longmapsto \mathsf S\,\delta\mathsf X\,\mathsf R.
\]
This map is onto: a nonzero $q\times q$ annihilator $\mathsf Z$
would produce $\mathsf R\mathsf Z\mathsf S$
orthogonal to all off-diagonal-block variations, hence block diagonal.
A nonzero column of that block-diagonal matrix would be a forbidden
group-supported right kernel vector. The off-diagonal parameters
therefore meet every rank stratum transversely. Corank at least two
has codimension at least four, while $(\lambda,\nu)$ provide only two
variables. Such points are absent for generic parameters. The
corank-one determinant hypersurface is smooth, and generic smoothness
of its projection to parameter space gives smooth affine fibers
\cite[Chapter~III]{Hartshorne1977}.

For irreducibility, insert one index at a time while a block's
size remains at most its outside dimension. Such an order exists.
Two blocked groups would have current sizes satisfying, after
relabeling, $L_{\rm cur}\geq M_{\rm cur}+N_{\rm cur}$ and
$M_{\rm cur}\geq L_{\rm cur}+N_{\rm cur}$, impossible because the
third group is positive. If only one group has pending indices,
its final active bound permits the next insertion.
The $(1,1,1)$ base has an irreducible specialization
\[
 (\lambda-e_N)(\lambda-\nu-e_L)(\lambda+\nu-e_M)-\delta_0=0,
 \qquad \delta_0\ne0.
\]
As a quadratic in $\nu$ its discriminant has an odd pole at
$\lambda=e_N$, so is not a square.

For an insertion, at zero new outgoing coupling the characteristic
polynomial is $\ell F_{\rm old}$, with
$\ell=\lambda-e-s\nu$. Suppose $r$ old indices have the same phase
as the new index and $n$ directions lie outside its group. Eliminating
the other indices in its group gives
\[
 \deg_\nu F_{\rm old}(e+s\nu,\nu)=n-r.
\]
For $r>0$ the leading coefficient is a nonzero constant times the
old moment-block determinant; for $r=0$ it is the nonzero outside
leading determinant. The allowed insertion has $r+1\leq n$, so
the old curve meets the new line at a finite point. At that point
the smooth old pencil has corank one with nonzero outside components
in both kernels. Choose new coupling vectors $\boldsymbol a,\boldsymbol b$,
supported on the outside indices, so that
$\boldsymbol a\operatorname{adj}(\mathsf M_{\rm old})\boldsymbol b$ does not vanish there.
The perturbed determinant is
\[
 F_\varepsilon=\ell F_{\rm old}
       -\varepsilon \boldsymbol a\operatorname{adj}(\mathsf M_{\rm old})\boldsymbol b.
\]
A factorization deforming the coprime monic factors
$\ell,F_{\rm old}$ would force the first perturbation into
$(\ell,F_{\rm old})$, hence to vanish at every intersection.
This obstruction also excludes a factorization after finite algebraic
parameter extension. Monicity and Gauss's lemma keep the factor coefficients integral
at $\varepsilon=0$. After a finite ramified parameter change they
admit power-series expansions. Put $\varepsilon=\tau^k$ and suppose
monic factors specialize to $\ell$ and $F_{\rm old}$. These are the
only nontrivial special factors, and monicity preserves their positive
$\lambda$-degrees under specialization. Write them as
\[
 G=\ell+\sum_{m\geq1}\tau^m a_m,\qquad
 J=F_{\rm old}+\sum_{m\geq1}\tau^m b_m,
 \qquad \deg_\lambda a_m<1,\quad
        \deg_\lambda b_m<\deg_\lambda F_{\rm old}.
\]
For $m<k$, induction on $m$ gives
$a_mF_{\rm old}+b_m\ell=0$. Coprimeness forces $\ell$ to divide
$a_m$, hence $a_m=b_m=0$ by the degree bounds. At order $k$, the
same coefficient equation would put
$-\boldsymbol a\operatorname{adj}(\mathsf M_{\rm old})\boldsymbol b$
in the ideal $(\ell,F_{\rm old})$, contrary to its nonzero value at
the selected intersection. Thus fractional parameter powers cannot
remove the obstruction. Induction proves generic geometric
irreducibility for every active phase profile.

Finally diagonalize the conserved moment blocks. Their simple
eigenvalues $\vartheta_i$ give the infinity expansions
\[
 \lambda=s_B\nu+e_B+\vartheta_i/\nu+O(\nu^{-2}),
 \qquad1\leq i\leq |B|.
\]
Branches in different phase blocks of one group have contact order
one, and those in the same block have contact order two. They are
distinct smooth points on the normalization. Counting these contacts
relative to the distinct-phase genus $L_aM_a+M_aN_a+N_aL_a-L_a-M_a-N_a+1$ gives
\cref{eq:BA-active-genus}. The same calculation includes the resolved branches in the third group.

For the eigenline step set $t=1/\nu$ and consider the analytic
matrix $\mathsf D+t\mathsf X$. Slopes split at order zero, distinct
intercepts at order $t$, and moment eigenvalues at order $t^2$ within
an equal-phase block. In the fixed moment basis the resolved
analytic eigenvectors form a matrix $\mathsf V(t)$ with invertible
$\mathsf V(0)$. Thus the branch modules at infinity are identified
with the original free matrix module by an invertible analytic
matrix. At finite points the smooth affine curve has corank one,
so its spectral module is already a line bundle. The normalized
component line bundle consequently has pushforward
$\mathcal O^{L_a+M_a+N_a}$.

Subtract the complete infinity fiber to obtain a bundle $\mathscr E$
with pushforward $\mathcal O(-1)^{L_a+M_a+N_a}$; both its zeroth
and first cohomology vanish. Adding the $N_a$ punctures gives the
exact leading-value isomorphism
$H^0(\mathscr E(\sum_{k=1}^{N_a}R_k))\simeq\C^{N_a}$.
Its degree is $G_a+N_a-1$, and the transpose module gives the dual
statement. These are precisely the nonspecial multipoint bases and
kernel required in \cref{thm:BA-spectral-converse}.
\end{proof}

The original coordinates are restored by the constant canonical
bases \cref{eq:BA-block-bases}; individual entries may therefore be
linear combinations of the active BA evaluations. This handles
generic initial modes for every fixed phase profile. Singular levels
and nonsemisimple moment specializations need additional limiting data.

\section{The Prym condition for the symmetric reduction}
\label{sec:geometry-prym-proof}

We establish the spectral involution and the Prym constraint for the
symmetric reduction, providing the geometric justification for
\cref{sec:special-spectral-levels}.

\subsection{The symmetric reduction}
\label{sec:geometry-prym}

Extra symmetry of the modes can confine the divisor motion to a
smaller Abelian variety inside the spectral Jacobian. For
$L=M=s$, $N=1$, impose
\begin{equation}\label{eq:geometry-fold}
 \mathsf B_L=\mathsf B_M,\qquad
 \boldsymbol u=\boldsymbol u^{\trans},\qquad
 \bar{\boldsymbol u}=(\bar{\boldsymbol u})^{\trans},\qquad
 \boldsymbol w=\boldsymbol v^{\trans},\qquad
 \bar{\boldsymbol w}=(\bar{\boldsymbol v})^{\trans}.
\end{equation}
Here bars on matrices label the independent partner arrays;
$\trans$ is ordinary transpose. The equations in \cref{eq:canonical-matrix-rates}
preserve these constraints. The restricted symplectic form is
nondegenerate, with $s$ vector-mode pairs and $s(s+1)/2$ symmetric
matrix-mode pairs. Constant rescalings put it in canonical form.

The resulting pencil satisfies a transpose symmetry giving
\begin{equation}\label{eq:geometry-involution}
 F(\lambda,\nu)=F(\lambda,-\nu),\qquad
 \iota:C\longrightarrow C,\quad
 \iota(\lambda,\nu)=(\lambda,-\nu).
\end{equation}
Assume the curve is regular and $F(\lambda,0)$ has distinct roots.
Its genus is $s^2$. The involution has $2s+1$ fixed points over
$\nu=0$ and one at the central infinity branch; it exchanges the
other two infinity groups. Riemann--Hurwitz therefore gives
\begin{equation}\label{eq:geometry-prym-dimensions}
 g(C/\iota)=\frac{s(s-1)}2,\qquad
 \dim\operatorname{Prym}(C/(C/\iota))=\frac{s(s+1)}2.
\end{equation}
The Prym is the identity component of the kernel of the norm map
$\operatorname{Jac}(C)\to\operatorname{Jac}(C/\iota)$; it describes
the divisor classes reversed by $\iota$
\cite[\S3.1]{BradlowSchaposnik2016}.
For this reduction, the moving eigenvector-divisor class lies in a
translate of the Prym. The eigenvector-divisor argument in this appendix proves this
assertion from the pencil's transpose symmetry.

For $s=2$, there are five canonical pairs: the spectral curve has
genus four, its quotient is elliptic, and the reduced divisor motion
has three Prym coordinates. Two mode-gauge directions complete the
count of five. More generally, the count is
$s(s+3)/2=s+\dim\operatorname{Prym}(C/(C/\iota))$.
This family illustrates how a spectral involution reduces the
Abelian variety carrying the motion, even when the spectral curve
itself is non-hyperelliptic.

\subsection{The eigenvector-divisor proof}

Under \cref{eq:geometry-fold}, define the constant matrix
\begin{equation}\label{eq:prym-transpose}
 \mathsf S=\begin{pmatrix}
 0&\Id_s&0\\ \Id_s&0&0\\ 0&0&-1
 \end{pmatrix}.
\end{equation}
The matrices in \cref{eq:canonical-X,eq:canonical-lax} obey
\begin{equation}\label{eq:prym-pencil-duality}
 \mathsf S^2=\Id_{2s+1},\qquad
 \mathsf X=\mathsf S\mathsf X^{\trans}\mathsf S,\qquad
 \mathsf S\mathsf D\mathsf S=-\mathsf D,\qquad
 \mathsf T(\nu)=-\mathsf S\mathsf T(-\nu)^{\trans}\mathsf S.
\end{equation}
Consequently
$\mathsf L(\nu)=\mathsf S\mathsf L(-\nu)^{\trans}\mathsf S$,
which proves \cref{eq:geometry-involution}. It also identifies a
right eigenvector at $\iota P$ with a left eigenvector at $P$.

Normalize the last component of the right eigenvector to one, and
let $\mathcal A(t)$ be its finite pole divisor on a generic smooth
spectral chart. Consider the last diagonal cofactor
\begin{equation}\label{eq:prym-cofactor}
 A(\lambda,\nu)=
 \det\bigl(\lambda\Id_{2s}-\mathsf L_{\mathrm{upper}}(\nu)\bigr),
\end{equation}
where $\mathsf L_{\mathrm{upper}}$ is the first $2s$ rows and
columns of the pencil. This cofactor is even in $\nu$.
At a smooth finite point of $C$, the adjugate of
$\lambda\Id_{2s+1}-\mathsf L$ has rank one. In local nonvanishing
eigenvector frames it is a nonzero scalar times a right vector and
a left vector. The zeros of its last diagonal entry thus split
between zeros of their last components. The transpose relation
identifies these two zero divisors by $\iota$, giving
\begin{equation}\label{eq:prym-divisor}
 (A)_0=\mathcal A(t)+\iota\mathcal A(t).
\end{equation}
This is stated on the open chart where these zeros are simple,
disjoint and finite.

Generically, the poles of $A$ have order $s-1$ on each of the $2s$
outer infinity branches and order $2s$ on the central branch.
These orders follow by substituting \cref{eq:geometry-branches}
into \cref{eq:prym-cofactor}. Thus its pole divisor is the fixed divisor
\[
 \mathcal A_\infty
 =(s-1)\left(\sum_{\ell=1}^{s}P_\ell^L
                  +\sum_{m=1}^{s}P_m^M\right)+2s P_1^N,
 \qquad \deg\mathcal A_\infty=2s^2.
\]
It follows that $\deg\mathcal A(t)=s^2$ and
$\mathcal A(t)+\iota\mathcal A(t)\sim\mathcal A_\infty$.
For a fixed reference divisor $\mathcal A(0)$, the degree-zero class
\[
 \xi(t)=[\mathcal A(t)-\mathcal A(0)]\in\operatorname{Jac}(C)
\]
therefore satisfies $\xi(t)+\iota(\xi(t))=0$.
Since $\xi(0)=0$, connected evolution puts $\xi(t)$ in the identity
component, namely the Prym. The standard isospectral
eigenvector-divisor linearization
\cite[\S3]{AdlerVanMoerbeke2018} consequently restricts the motion
to a translate of this Prym. This proves containment of the reduced
divisor motion; a global reconstruction of the entire reduced
isospectral manifold would require its normalization data as well.

For completeness, there are $s$ constant diagonal mode gauges.
Writing $\boldsymbol v=(v_{i1})_{i=1}^{s}$, their action is
\[
 v_{i1}\mapsto e^{\epsilon_i}v_{i1},\qquad
 u_{ij}\mapsto e^{-\epsilon_i-\epsilon_j}u_{ij},
 \qquad 1\leq i,j\leq s,
\]
with inverse factors on the partners and the same constraints on
$\boldsymbol w$. They preserve the normalized eigenvector's pole
divisor. The number of independent pairs after folding is
$s+s(s+1)/2$. Indeed, the pulled-back symplectic form has factor two
on each vector-mode pair and on each off-diagonal symmetric
matrix-mode pair. Multiplying both members of each such pair by
$\sqrt2$, and leaving the diagonal pairs unchanged, gives canonical
coordinates. This accounts for the gauge and Prym dimensions in
\cref{sec:geometry-prym}.

\section{The quartic spectral Hamiltonian}
\label{sec:quartic-expansion}

The trace hierarchy of \cref{sec:spectral-flows} generates higher
interaction orders without enlarging the mode list.
The fourth member $H_4=\tr\mathsf X^4/4$ adds quartic
interactions to this same mode list. Its fixed diagonal entries
also supply constant, quadratic and cubic terms. Taking four mode factors in
the trace gives its homogeneous quartic part:
\begin{align}\label{eq:hierarchy-quartic-part}
 H_4^{[4]}={}&2\tr\!\left[(\boldsymbol u\bar{\boldsymbol u}^{\trans})^2\right]
 +\tfrac12\tr\!\left[(\bar{\boldsymbol w}^{\trans}\boldsymbol w)^2\right]
 +\tfrac12\tr\!\left[(\boldsymbol v\bar{\boldsymbol v}^{\trans})^2\right]
 \nonumber\\
 &-2\tr\!\left(\boldsymbol u\bar{\boldsymbol u}^{\trans}
                         \bar{\boldsymbol w}^{\trans}\boldsymbol w\right)
  -2\tr\!\left(\bar{\boldsymbol u}^{\trans}\boldsymbol u
                         \boldsymbol v\bar{\boldsymbol v}^{\trans}\right)
  +\tr\!\left(\boldsymbol w\bar{\boldsymbol w}^{\trans}
                         \bar{\boldsymbol v}^{\trans}\boldsymbol v\right).
\end{align}
For example, when $L,M\geq2$, this includes the four-mode exchange
\[
 4\bigl(u_{11}u_{22}\bar{u}_{12}\bar{u}_{21}
       +u_{12}u_{21}\bar{u}_{11}\bar{u}_{22}\bigr).
\]
Thus higher-order mixing terms come with specific mode-power terms
fixed by the spectral trace.

\section{Constrained-KP normalization and canonical flow matching}
\label{app:KP-normalization}

We justify the matrix wave normalization used in
\cref{prop:KP-hierarchy} and explain how its times acquire canonical
Hamiltonians on a fixed birational BA level.

\subsection{Wave normalization, commutativity and source evolution}

At each puncture, the ordinary column has leading vector $e_k$ in
\cref{eq:general-BA-asymptotics}. Collecting its $N$ local Laurent
expansions gives a matrix wave function with identity leading matrix.
It determines a formally invertible dressing operator
$\mathscr W=\Id_N+\mathsf W_1\partial^{-1}+\mathsf W_2\partial^{-2}+\cdots$.
Before the gauge change, the $N$ Laurent expansions of $h_A$ form a
time-independent diagonal formal operator $\mathsf H_A(\partial)$.
The spectral equations at all punctures imply
$\mathcal A_A\mathscr W=\mathscr W\mathsf H_A$. Since the diagonal
operators commute, so do $\mathcal A_L$ and $\mathcal A_M$.
This proves \cref{eq:KP-matrix-commutation} without inferring
commutativity from only one eigenfunction column.

Now use the local coordinates $s'_k=1/h_L(P)$. The gauge
$\mathsf G^{-1}=e^{x\mathsf C_L}$ changes the leading ordinary
factor from $e^{x/s_k}e_k$ to $e^{xh_L(P)}e_k$; the remaining
factor has identity leading matrix. In these coordinates the
renormalized dressing operator therefore satisfies
\[
 \mathcal L_{\rm KP}=\mathscr W\partial\mathscr W^{-1}.
\]
For a higher time, the principal parts of $h_L^n$ prescribe the BA
exponentials in the original $s_k$ coordinates. Multiplying component
$k$ by $\exp(x_n\operatorname{reg}_{R_k}h_L^n)$ supplies the
required regular constant. The dual component receives the inverse
factor. Hence the normalization includes both the Abel velocity and
the component phases.

The BA construction with these times gives the Sato equation
\[
 \partial_{x_n}\mathscr W
 =-(\mathcal L_{\rm KP}^{\,n})_-\mathscr W,
\]
where $-$ retains the negative differential powers. Equivalently,
subtracting $\mathcal B_n=(\mathcal L_{\rm KP}^{\,n})_+$ from the
ordinary BA derivative leaves zero leading values at every puncture.
The fixed divisor and BA uniqueness give the first equation in
\cref{eq:KP-source-flows}.

For the dual, the residue pairing uses $d\Omega_0$.
Converting to the local spectral measure $dh_L$ multiplies the dual
wave by the fixed scalar factor $d\Omega_0/dh_L$. This ratio has
leading value one at every puncture. The residue pairing identifies
the resulting dual wave with the formal adjoint wave operator,
giving $-\mathcal B_n^{\rm ad}$ as its time generator. Since the
factor is independent of all BA times, the same differential
equation holds for $\widetilde{\boldsymbol\psi}^{\vee}$ itself.
These are the standard BA and adjoint-wave arguments
\cite{Krichever1977,KricheverShiota2011} in the present normalization.

For a matrix differential operator $B$ and column functions $F,G$,
the finite-rank identity is
\[
 [B,F\partial^{-1}G^{\trans}]_-
 =B(F)\partial^{-1}G^{\trans}
  -F\partial^{-1}(B^{\rm ad}(G))^{\trans}
\]
\cite[equations (7)--(8)]{ChvartatskyiSydorenko2013}.
It proves preservation of the negative parts of both operators in
\cref{eq:KP-operator-flows}. Their nonnegative parts remain fixed:
for $\mathcal L_{\rm KP}$ this follows from
$[\mathcal L_{\rm KP}^{\,n},\mathcal L_{\rm KP}]=0$; for
$\mathcal M_{\rm KP}$ it follows from
$[\mathcal L_{\rm KP}^{\,n},\mathcal M_{\rm KP}]=0$ and its constant
differential part $\Id_N\partial+\mathsf C_M-\mathsf C_L$.
The commutator of a negative-order operator with either differential
part has negative order. This completes the operator and source-flow
proof for arbitrary $N$.

\subsection{Selecting the canonical generator on a fixed level}

Fix a birational prescribed BA level of genus $g$. By
\cref{thm:BA-higher-flows}, select $g$ spectral Hamiltonians
$H_1,\ldots,H_g$ with independent Abel vectors
$\boldsymbol U^{(1)},\ldots,\boldsymbol U^{(g)}$.
For comparison with the physical Hamiltonian convention write
$x_n=-it_n$, and let $\boldsymbol U_{{\rm KP},n}$ be defined by
$z(t_n)=z_0-it_n\boldsymbol U_{{\rm KP},n}$.
The coefficients of the required spectral combination solve
\begin{equation}\label{eq:KP-Abel-linear-match}
 \sum_{\alpha=1}^{g}\beta_{n\alpha}\boldsymbol U^{(\alpha)}
 =\boldsymbol U_{{\rm KP},n}.
\end{equation}
The matrix of these vectors is invertible. When $g=0$, no spectral
combination is required.

The remaining difference of the puncture time representatives has
zero residue pairing with every holomorphic differential. The
Mittag--Leffler criterion therefore makes it the principal part of
a meromorphic function supported at the punctures, up to an
additive constant. Its evaluations at the ordinary points and its
regular puncture constants give fixed vertex phase differences
$\delta\varphi_\ell^L$, $\delta\varphi_m^M$, $\delta\varphi_k^N$.
The BA theta translations already agree; these vertex phases account
for the remaining exponential factors in the mode arrays.

The power charges generate them through
\begin{equation}\label{eq:KP-canonical-generator}
\begin{aligned}
 \widehat H_n={}&\sum_{\alpha=1}^{g}\beta_{n\alpha}H_\alpha
 -\sum_{\ell=1}^{L}\delta\varphi_\ell^L Q_\ell^L
 +\sum_{m=1}^{M}\delta\varphi_m^M Q_m^M
 +\sum_{k=1}^{N}\delta\varphi_k^N Q_k^N.
\end{aligned}
\end{equation}
The resulting phase rates of $u_{\ell m},v_{mk},w_{k\ell}$ are
respectively $\delta\varphi_\ell^L-\delta\varphi_m^M$,
$\delta\varphi_m^M-\delta\varphi_k^N$ and
$\delta\varphi_k^N-\delta\varphi_\ell^L$; partner rates are opposite.
Thus \cref{eq:KP-canonical-generator} matches every mode, including
its phase. All coefficients are constants after choosing the level,
and the Hamiltonians and charges in the combination commute.
Allowing these coefficients to vary with spectral integrals inside
one Hamiltonian would introduce additional derivatives and is not
the construction used here.

\bibliographystyle{plainurl}
\bibliography{references}
\end{document}